\documentclass[12pt, a4paper]{article}

\usepackage[T1, T2A]{fontenc}
\usepackage[cp1251]{inputenc}
\usepackage[english]{babel}
\usepackage{amsmath, amssymb, xcolor}
\usepackage[left=3cm, right=1.5cm, top=2cm, bottom=2cm]{geometry}
\usepackage[displaymath]{lineno}
\usepackage{graphicx}

\newcommand\rem[1]{}

\def\la{\mathrel{\mathchoice {\vcenter{\offinterlineskip\halign{\hfil
$\displaystyle##$\hfil\cr<\cr\sim\cr}}}
{\vcenter{\offinterlineskip\halign{\hfil$\textstyle##$\hfil\cr
<\cr\sim\cr}}}
{\vcenter{\offinterlineskip\halign{\hfil$\scriptstyle##$\hfil\cr
<\cr\sim\cr}}}
{\vcenter{\offinterlineskip\halign{\hfil$\scriptscriptstyle##$\hfil\cr
<\cr\sim\cr}}}}}
\def\ga{\mathrel{\mathchoice {\vcenter{\offinterlineskip\halign{\hfil
$\displaystyle##$\hfil\cr>\cr\sim\cr}}}
{\vcenter{\offinterlineskip\halign{\hfil$\textstyle##$\hfil\cr
>\cr\sim\cr}}}
{\vcenter{\offinterlineskip\halign{\hfil$\scriptstyle##$\hfil\cr
>\cr\sim\cr}}}
{\vcenter{\offinterlineskip\halign{\hfil$\scriptscriptstyle##$\hfil\cr
>\cr\sim\cr}}}}}

\def\utw{\smash{\rlap{\lower5pt\hbox{$\sim$}}}}
\def\udtw{\smash{\rlap{\lower6pt\hbox{$\approx$}}}}

\def\diameter{{\ifmmode\mathchoice
{\ooalign{\hfil\hbox{$\displaystyle/$}\hfil\crcr
{\hbox{$\displaystyle\mathchar"20D$}}}}
{\ooalign{\hfil\hbox{$\textstyle/$}\hfil\crcr
{\hbox{$\textstyle\mathchar"20D$}}}}
{\ooalign{\hfil\hbox{$\scriptstyle/$}\hfil\crcr
{\hbox{$\scriptstyle\mathchar"20D$}}}}
{\ooalign{\hfil\hbox{$\scriptscriptstyle/$}\hfil\crcr
{\hbox{$\scriptscriptstyle\mathchar"20D$}}}}
\else{\ooalign{\hfil/\hfil\crcr\mathhexbox20D}}%
\fi}}

\begin{document}

\begin{center}
{\Large\noindent
Physical Properties and Kinematics of Dense Cores
Associated with Regions of Massive Star Formation
from the Southern Sky
}

{\it L.\,E. Pirogov\,$^{a,*}$, P. M. Zemlyanukha\,$^a$, E. M. Dombek\,$^a$,
and M. A. Voronkov\,$^b$}

\scriptsize{$^a$ Gaponov-Grekhov Institute of Applied Physics, Russian Academy of Sciences, Nizhny Novgorod, Russia} \\
\scriptsize{$^b$ Commonwealth Scientific and Industrial Research Organization (CSIRO) Space and Astronomy, Epping NSW, Australia}\\
\scriptsize{$^*$ e-mail: pirogov@appl.sci-nnov.ru}
\end{center}

{\bf Abstract}\\
The results of spectral observations in the $\sim 84-92$~GHz frequency range of six objects in the southern
sky containing dense cores and associated with regions of massive stars and star clusters formation are
presented. The observations were carried out with the MOPRA-22m radio telescope. Within the framework
of the local thermodynamic equilibrium (LTE) approximation, the column densities and abundances of the
H$^{13}$CN, H$^{13}$CO$^+$, HN$^{13}$C, HC$_3$N, c-C$_3$H$_2$, SiO, CH$_3$C$_2$H and CH$_3$CN molecules are calculated. 
Estimates of kinetic temperatures ($\sim 30-50$~K), sizes of emission regions ($\sim 0.2-3.1$~pc) and virial masses
($\sim 70-4600~M_{\odot}$) are obtained. The linewidths in the three cores decrease with increasing distance from the
center. In four cores, asymmetry in the profiles of the optically thick lines HCO$^+$(1--0) and HCN(1--0) is
observed, indicating the presence of systematic motions along the line of sight. In two cases, the asymmetry
can be caused by contraction of gas. The model spectral maps of HCO$^+$(1--0) and H$^{13}$CO$^+$(1--0), obtained
within the framework of the non-LTE spherically symmetric model, are fitted into the observed ones. The
radial profiles of density ($\propto r^{-1.6}$), turbulent velocity ($\propto r^{-0.2}$), and contraction velocity ($\propto r^{0.5}$) 
in the G268.42--0.85 core have been calculated. The contraction velocity profile differs from that expected both in
the case of free fall of gas onto a protostar ($\propto r^{-0.5}$), and in the case of global core collapse (contraction velocity
does not depend on distance). A discussion of the obtained results is provided.

$Keywords$: star formation, molecular clouds, dense cores, molecular lines, modeling

\section{Introduction}

Studies of dense cores of molecular clouds are an
important tool for estimating initial conditions of
the star formation process. The cores associated with
the regions of formation of high-mass stars ($\ga 8~M_{\odot}$)
and star clusters attract special attention. Despite the
greatly increased number of observations of such
objects in recent years, the agreed scenario of their formation
is far from complete (see, for example, \cite{Tan14,Motte18}).
Compared to the regions of formation of stars with masses
on the order of or less than the solar one, they are more
rare, located at larger distances, evolve faster and have
a hidden phase of development before the main
sequence. In the early stages of evolution, massive
stars actively interact with the parent dense core,
increase turbulence and temperature, lead to shock
waves and outflows, cause fragmentation, further
compression and, under certain conditions, a new
phase of star formation. Chemical composition of
the gas in the regions of massive star formation
is enriched due to evaporation of molecules from the
surface of dust particles. The radiation of individual
compact regions in water, methanol, hydroxyl and
some other molecular lines may have a maser nature.
The observed line profiles are significantly broadened
due to turbulent and systematic motions. These effects
often overlap, which complicates the interpretation of
observations.

There are various assumptions about the state of
the cores in which star formation occurs. Thus, in the
model of a singular isothermal sphere \cite{Shu,Shu87},
it is assumed that a quasi-equilibrium spherical core with a
radial Bonnor-Ebert density profile (a flat area near
the center and a dependence close to $r^{-2}$ in the shell)
evolves to a state with a singularity in the center (protostar),
after which a collapse begins, which spreads
inside-out. In the turbulent core model \cite{MT03}, proposed
to describe the formation of massive stars and star
clusters, the initial state is considered to be a sphere in
hydrostatic equilibrium, possessing supersonic turbulence
and a density profile $r^{-3/2}$ \cite{MT03,zhangtan18},
with no systematic motions.
Both in the model of a singular
isothermal sphere and in the model of a turbulent core
the density and velocity radial profiles
in the region where gas collapses onto a protostar
in free fall are $r^{-3/2}$ and $r^{-1/2}$, respectively.

The global hierarchical collapse model \cite{vazquez} proceeds
from the fact that cores, like parent clouds, are non-equilibrium
objects that are in the process of global
collapse even before the formation of a protostar,
and their observed proximity to the state of virial equilibrium
occurs, in particular, due to the proximity of the
free fall velocity to the virial one. In this model, based
on the classical works of Larson and Penston \cite{Lar69,Pen69},
after the formation of a protostar, the density profile in
the envelope has a form $r^{-2}$, and the contraction
velocity is constant and does not depend on the radial
distance (see, for example, \cite{vazquez,naranjo}).
Since both in the quasi-equilibrium and non-equilibrium core models
the density profiles in the shell are close, it is the systematic
velocity profile that can serve as a factor allowing
one to make a choice between the models.

Due to the inhomogeneous distribution of gas density
in the cores and the presence of systematic gas
motions, the observed profiles of molecular lines with
large optical depth may differ from Gaussian and have
an asymmetry associated with differences in the conditions
affecting their excitation along the line of sight
and a velocity shift due to Doppler effect. To interpret
such spectra and to estimate parameters of the
structure and velocity field, it is necessary to calculate
excitation of molecules taking into account differences
from local thermodynamic equilibrium (LTE)
and compare the calculated spectra with observed ones.
Among the molecular lines that are indicators of
dense gas, the optically thick HCO$^+$(1--0) and
HCN(1--0) lines are among the most sensitive to kinematics
and density distribution.

An analysis of observational data of five massive star-forming regions
in the northern sky associated with methanol masers \cite{Pir16}
revealed an asymmetry in the profiles of the optically thick HCO$^+$(1--0)
and HCN(1--0) lines in four of them. For one of the
cores with signs of contraction considered in \cite{Pir16}
(L1287 or G121.28$+$0.65), we analyzed observational data in the HCO$^+$(1--0)
and HCN(1--0) lines and in the lines of rarer isotopes
using the non-LTE model \cite{PZ21}.
To fit model spectral maps into observed ones,
a specially developed algorithm was used, based on the
methods of principal components and k-nearest
neighbors \cite{PZ21}.
Estimates of the optimal parameters of
density, turbulent velocity and
contraction velocity radial profiles were obtained.
It turned out that
the power-law index of the contraction velocity profile
in L1287 is close to zero, which indicates the likelihood
of a global collapse of this core. To answer the
question of how typical this case is, it is necessary to
conduct further studies of cores associated with
regions of formation of massive stars and star clusters
and showing signs of contraction.

A sample of dense cores associated with the regions
of massive stars and star clusters formation in the
southern sky has previously been studied in various
molecular lines and in continuum
\cite{Zin95,Lap98,Harju98,Zin2000,Pir03,Pir07,Liu16,Pir22}.
A number of physical parameters of the cores were determined,
density profiles were calculated \cite{Pir09}, estimates of
the chemical composition were obtained.
Also, effects of
chemical differentiation were found in several objects \cite{Pir07}.

This paper presents the results of observations of six
objects from this sample in various molecular lines in
the 3-mm wavelength range and the estimates of their physical
characteristics and chemical composition. For two
cores, where the observed HCO$^+$(1--0) and
HCN(1--0) profiles have indications of
gas contraction, an analysis with model calculations is given.
For one of them, estimates of the parameters
of the spatial distribution of density and velocity are given.

\section{Sources}
\label{sec:sources}

The objects of the study are six regions
of massive star and star cluster formation in the southern sky,
which we had previously observed in the CS(2--1),
CS(5--4) \cite{Zin95,Pir07}, and N$_2$H$^+$(1--0) lines \cite{Pir03},
and in the continuum at the 1.2~mm and 350~$\mu$m wavelengths
\cite{Pir07,Pir22}.
The list of the sources is given in Table~\ref{table:list}.
Distances to the objects are also given there and associations with other
objects are indicated.

In addition to IRAS point sources and maser
sources, the cores under the study include ultracompact
H II regions, near-infrared sources, submillimeter
and radio sources, and molecular outflows.
We calculated bolometric
luminosities of the IRAS sources
as an integral of the fitted curve of a ``gray” body
emission into the dependence of the flux on frequency
\cite{Pir07}, taking into account the distances from Table~\ref{table:list}.
Information about the presence of maser sources is
taken from the $maserdb.net$ database \cite{Lad19}.

In the region G268.42--0.85, the IRAS
09002--4732 source ($L\sim 8\times 10^4~L_{\odot}$) is associated with a
young star cluster immersed in a dense gas and dust
cloud, the dominant member of which is a star of
spectral type O7 \cite{Getman19}. Measurements by the parallax
method using the $Gaia$ satellite data give a distance
to the cluster of 1.7--1.8~kpc \cite{Getman19}. The ultracompact
H II zone is associated with the cluster \cite{Apai05,Tremblay22}.
A water maser is observed in the core \cite{Urquhart09}.

The source IRAS 09018--4816 in the region
G269.11--1.12 ($L\sim 6\times 10^4~L_{\odot}$), associated with the
ultracompact H II zone \cite{QUAD}, is located south of the
dense core. Water masers have been detected both in
the direction of the core center and near the IRAS
source \cite{Breen11,Urquhart09}. A class II methanol maser is observed
near the center of the core at a frequency of 6.7~GHz \cite{Caswell09}.
Class I methanol maser lines are also observed in the core
\cite{Voron14,Breen19,Slysh94,Valtts00,Yang17}.

The dense core of G270.26$+$0.83 is associated with
the RCW41 H~II region which consists of two separate
compact regions, the sources of which are stars
of spectral types O9~V and B0~V, respectively \cite{Tremblay22}.
The IRAS 09149--4743 source with a luminosity of $L\sim 5\times 10^3~L_{\odot}$
is located at the periphery of the core.
The core contains a cluster of young stellar objects
observed as near-IR sources \cite{Ortiz07,Neichel15}, as well as water
\cite{Breen11}, class I \cite{Voron14,Breen19,Slysh94,Valtts00,Yang17}
and class II methanol masers \cite{Caswell09}.

The region G285.26--0.05 consists of two cores.
Core 1 contains the IRAS 10295--5746 source
($L\sim 5\times 10^5~L_{\odot}$) and the ultracompact H~II region
\cite{QUAD}, as well as a cluster of infrared 2MASS sources \cite{Dutra03},
masers of OH \cite{Caswell98}, water \cite{Breen10} and class II methanol
\cite{Gay93}. Water maser is observed in core 2 \cite{Breen10}, and
there (as in core 1) an emission in the $Br$ line was detected
\cite{Barnes13}, which indicates an existence of the H~II region,
probably immersed in a dense dust shell.

The IRAS 11097--6102 source ($L\sim 3\times 10^5~L_{\odot}$)
is located in the central part of the extended gas and dust
clump G291.27--0.71. Here is the H II region \cite{Eswaraiah17},
the interaction of which with the parent cloud apparently
determines morphology of the clump, consisting
of several fragments and corresponding IR sources.
Water \cite{Breen10} and class II methanol masers \cite{Breen12,Caswell09}
are observed in the region.

The region G294.97--1.73 consists of two cores.
Core 1 is associated with the IRAS 11368--6312 source
($L\sim 6\times 10^3~L_{\odot}$) and the ultracompact H~II region
\cite{Walsh98}. Water masers are observed in both cores \cite{Breen10,Breen11}.
In core 1, lines of class I \cite{Voron14,Breen19,Yang17}
and class II \cite{Green12} methanol masers are observed.
Core 2 is associated with
a class II methanol maser \cite{Caswell09}.

\section{Observations}

Observations of six objects from Table~\ref{table:list} were carried
out in 2010 using the 22-m radio telescope MOPRA (Australia)
\footnote{The MOPRA telescope is part of the Australian Telescope
National Foundation (ATNF). At the time of observations, it
was funded by the Australian Government as a national facility
managed by the Commonwealth Scientific and Industrial
Research Organization (CSIRO).} (project M526).
During the observations, a 3-mm wavelength range receiver with
an SIS mixer at the input was used. The noise temperature
of the system during the observation period
varied in the range $\sim 140-200$~K depending on the
source and weather conditions. For spectral analysis, a
MOPS spectrum analyzer with a total bandwidth of 8 GHz
\footnote{The University of New South Wales digital filter bank was made
available for observations on the MOPRA telescope with support
from the Australian Research Council.}
was used, which made it possible to simultaneously
detect several molecular lines. The observation band was divided into
16 subbands with a width of 137.5~MHz, tuned to the
lines of the HCO$^+$, HCN, HNC molecules and their rarer
isotopes, as well as to the lines of the HC$_3$N,
SiO, CH$_3$OH, CH$_3$C$_2$H, CH$_3$CN and some other molecules.
Each subband consisted of 4096 channels, which gave
a velocity resolution of $\sim 0.11$~km/s. The width of the
main beam of the MOPRA telescope was 36$''$ in this
frequency range \cite{Ladd05}. The main beam efficiency at a
frequency of 86~GHz was 0.49 \cite{Ladd05}.

Mapping was carried out in the OTF mode. For
each object, maps of size $2'\times 2'$ were obtained.
Pointing errors were regularly checked using SiO maser
observations and did not exceed $5''\pm 2''$. The Orion~KL
source was used to calibrate spectral observations.
Processing of the obtained data was carried out
using the $Livedata$ and $Gridzilla$
packages\footnote{https://www.atnf.csiro.au/computing/software/livedata/
index.html}
with the help of which the 1st order baseline was subtracted,
the averaging and formation of data cubes (two coordinates
and velocity) with a fixed step along both coordinates
(15$''$, Nyquist sampling) were made.

\begin{table}[h]
\centering
\caption{Source List}
\vskip 2mm
\scriptsize
\begin{tabular}{l|c|c|r|l}
\noalign{\hrule}\noalign{\smallskip}
Source         &$\alpha$(2000)  &$\delta$(2000)     & $D$ (kpc) & Associations \\
               &                &                   &             & with other objects \\
\noalign{\hrule}\noalign{\smallskip}
G~268.42$-$0.85  &09 01 54.3   &$-$47 43 59  & 1.7(0.1) \cite{Getman19}   & IRAS 09002--4732 \\
G~269.11$-$1.12  &09 03 32.8   &$-$48 28 39  & 2.6 \cite{Zin95}      & IRAS 09018--4816 \\
G~270.26$+$0.83  &09 16 43.3   &$-$47 56 36  & 1.3(0.2) \cite{RL09}  & IRAS 09149--4743, RCW 41 \\
G~285.26$-$0.05  &10 31 30.0   &$-$58 02 07  & 4.7 \cite{Zin95}      & IRAS 10295--5746 \\
G~291.27$-$0.71  &11 11 49.9   &$-$61 18 14  & 2.8(0.3) \cite{Binder18}    & IRAS 11097--6102, NGC~3576, RCW 57 \\
G~294.97$-$1.73  &11 39 12.6   &$-$63 28 47  & 1.2 \cite{Zin95}      & IRAS 11368--6312 \\
\noalign{\hrule}\noalign{\smallskip}
\end{tabular}
\flushleft
{\scriptsize
The distances to G268.42, G291.27 were obtained by the parallax method using data from the $Gaia$ satellite, the distance
to G270.26 was estimated by the spectroscopic method, and kinematic distances are given for the remaining objects. Since the error values
of the latter are not given in the literature, in the calculations we arbitrarily took them equal to 0.3~kpc (G269.11 and G294.97)
and 0.5~kpc (G285.26). Here and below, abbreviated names of objects are used. The numbers in parentheses following the names
G285.26 and G294.97 correspond to individual cores.}
\label{table:list}
\end{table}

\normalsize

\section{The results of observations}

For all the observed objects, maps were obtained in the
CH$_3$OH(5$_{-1}$--4$_0$~E), c-C$_3$H$_2$(2$_{1,2}$--1$_{0,1}$),
H$^{13}$CN(1--0), H$^{13}$CO$^+$(1--0), SiO(2--1), HCN(1--0),
HCO$^+$(1--0), HNC(1--0) and HC$_3$N(10--9) lines.
In the direction of the central positions of the objects,
the HN$^{13}$C(1--0), CH$_3$C$_2$H(5--4) and CH$_3$CN(5--4) lines were detected.

\begin{figure}[h]

\centering


\begin{minipage}[b]{0.3\textwidth}
    \includegraphics[width=\textwidth,angle=-0]{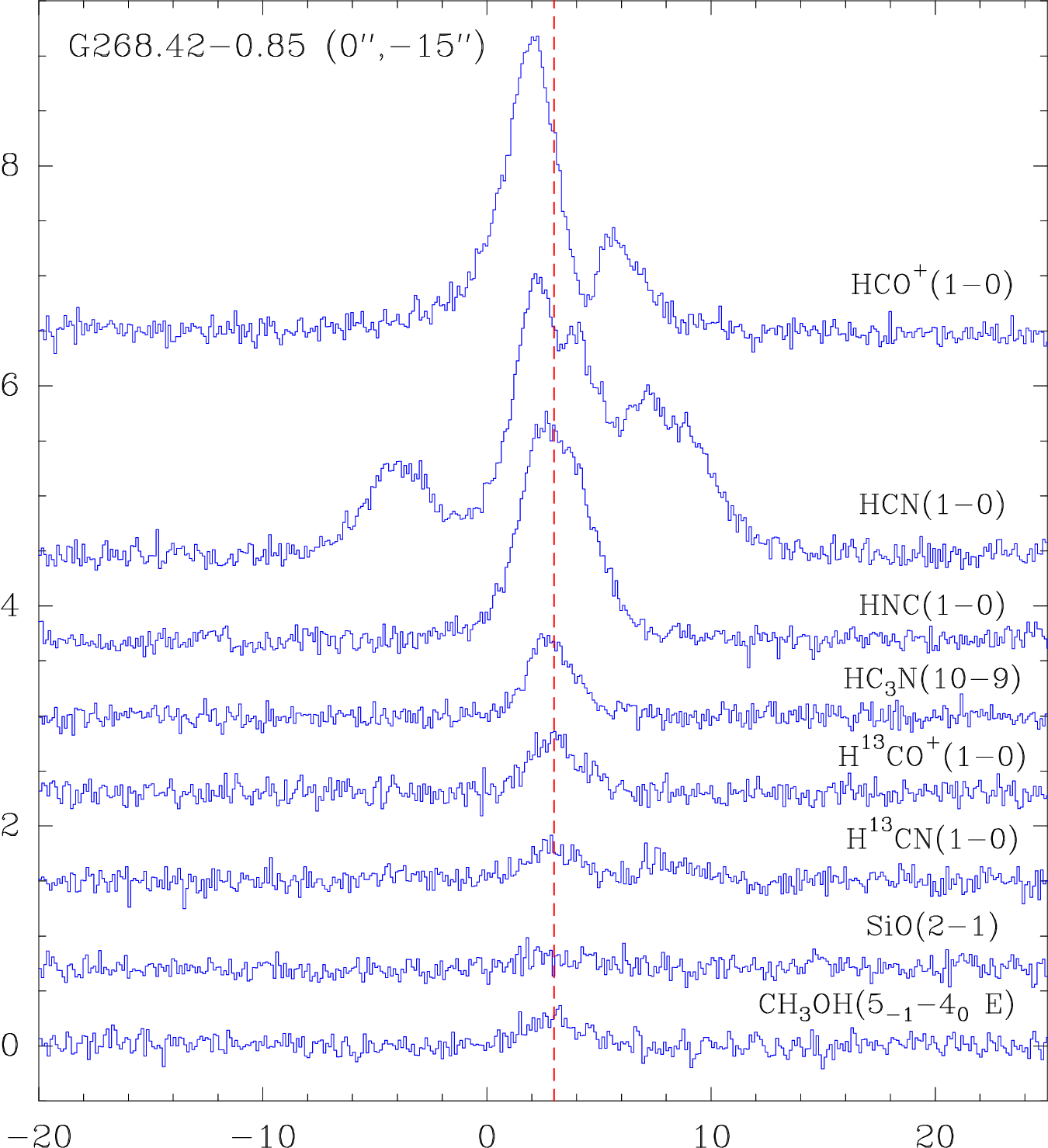}
\end{minipage}
\hspace{2mm}
\begin{minipage}[b]{0.3\textwidth}
    \includegraphics[width=\textwidth,angle=-0]{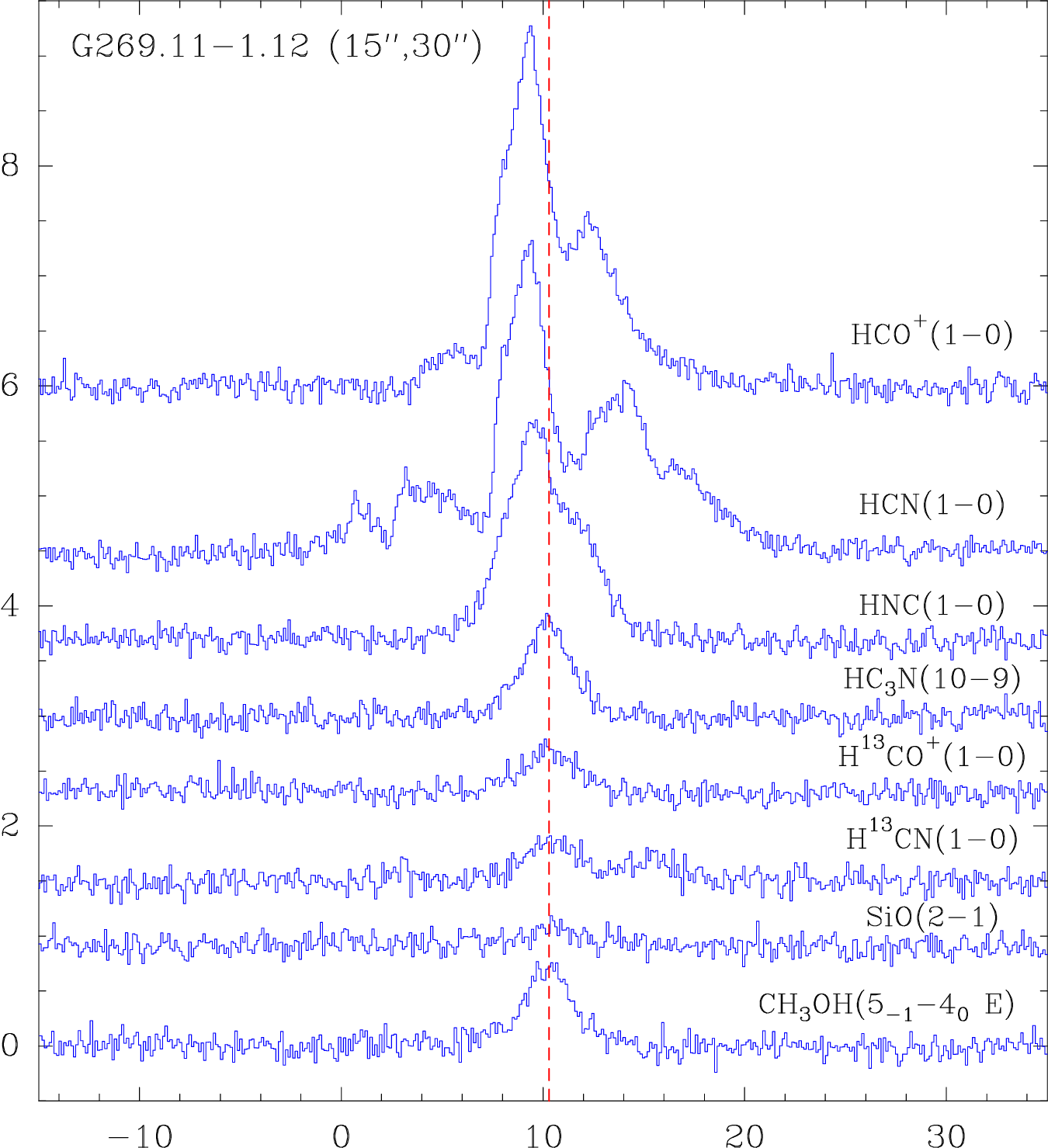}
\end{minipage}

\vspace{2mm}

\begin{minipage}[b]{0.3\textwidth}
    \includegraphics[width=\textwidth,angle=-0]{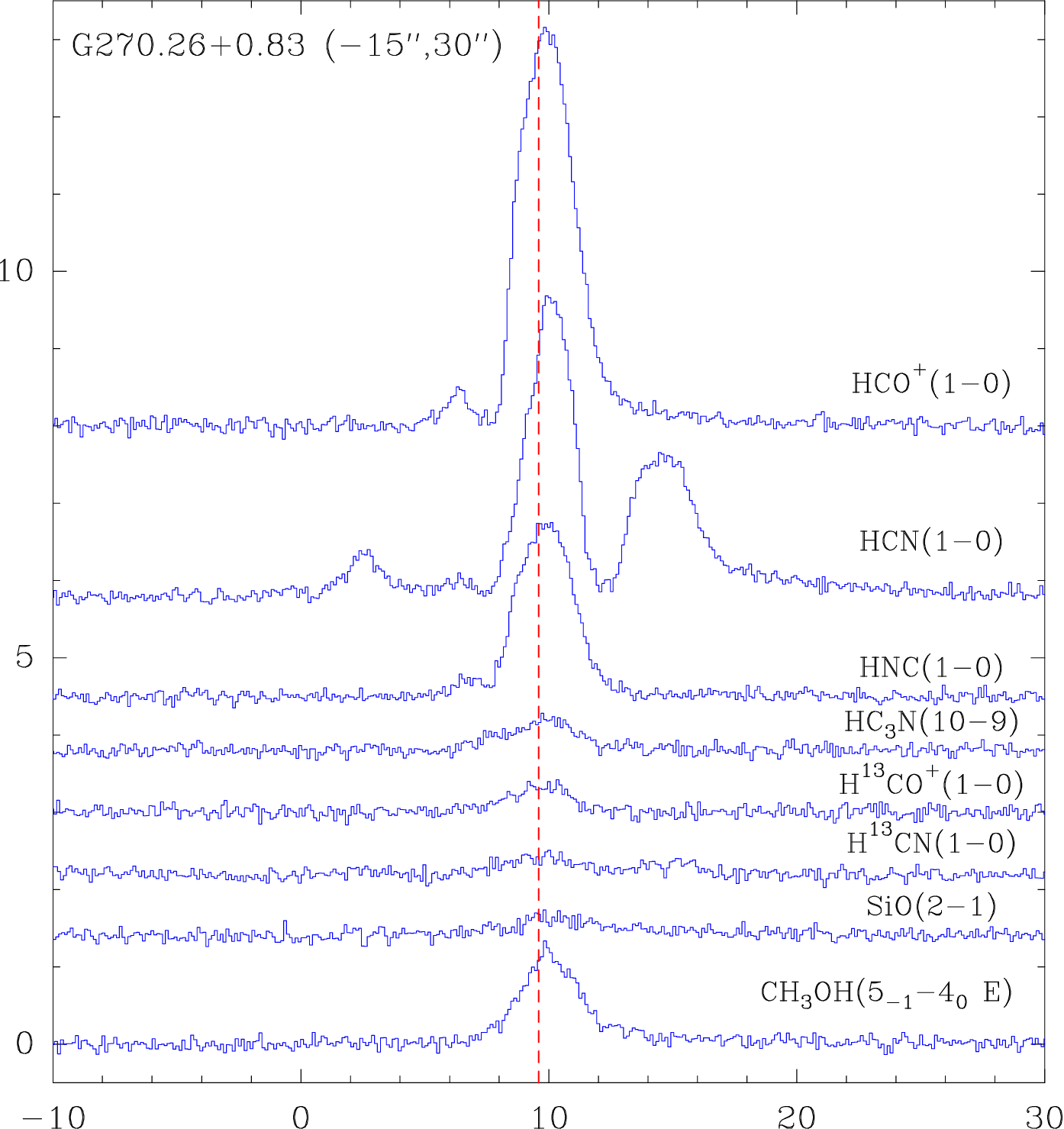}
\end{minipage}
\hspace{2mm}
\begin{minipage}[b]{0.3\textwidth}
    \includegraphics[width=\textwidth,angle=-0]{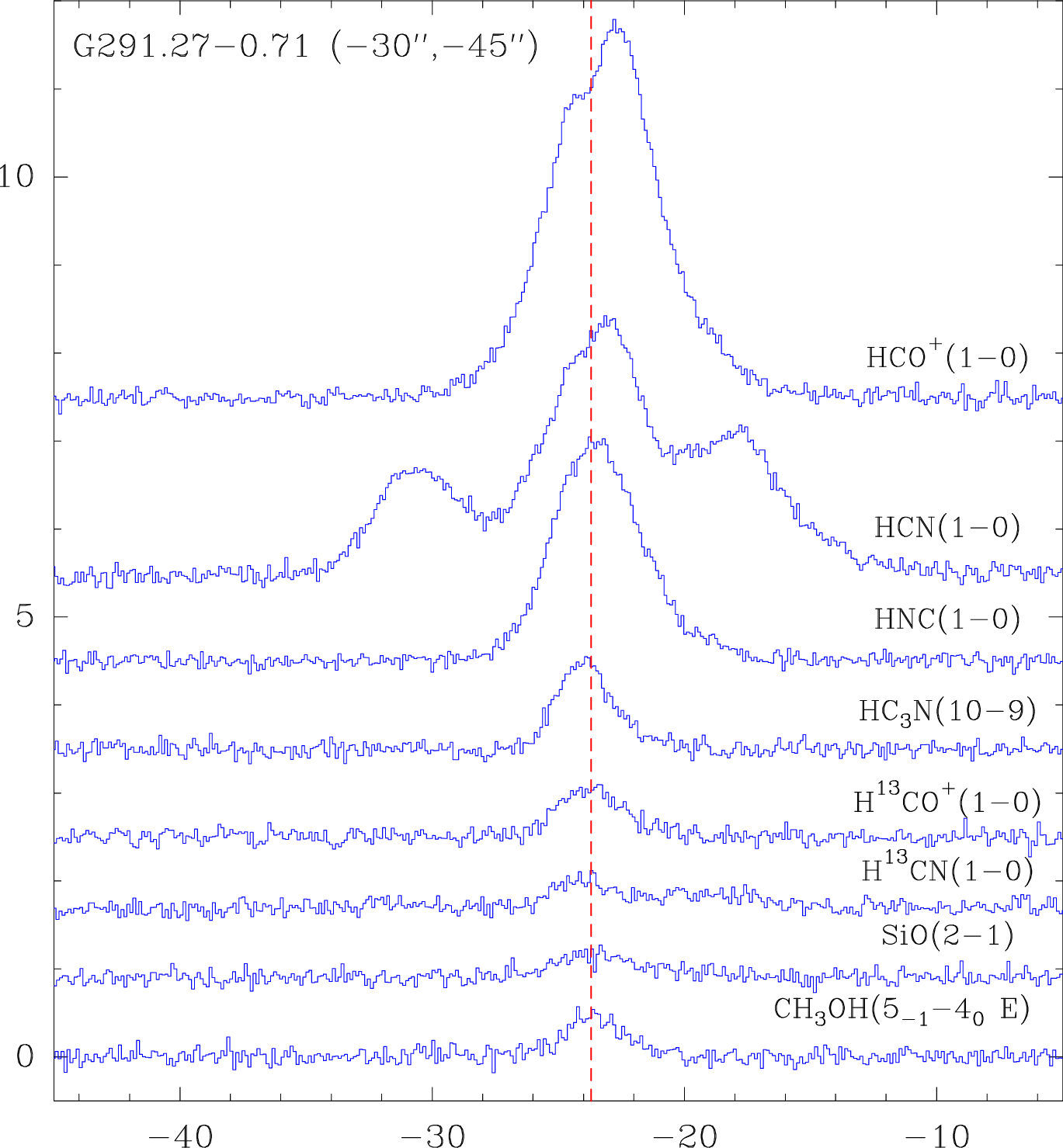}
\end{minipage}

\vspace{2mm}

\begin{minipage}[b]{0.3\textwidth}
    \includegraphics[width=\textwidth,angle=-0]{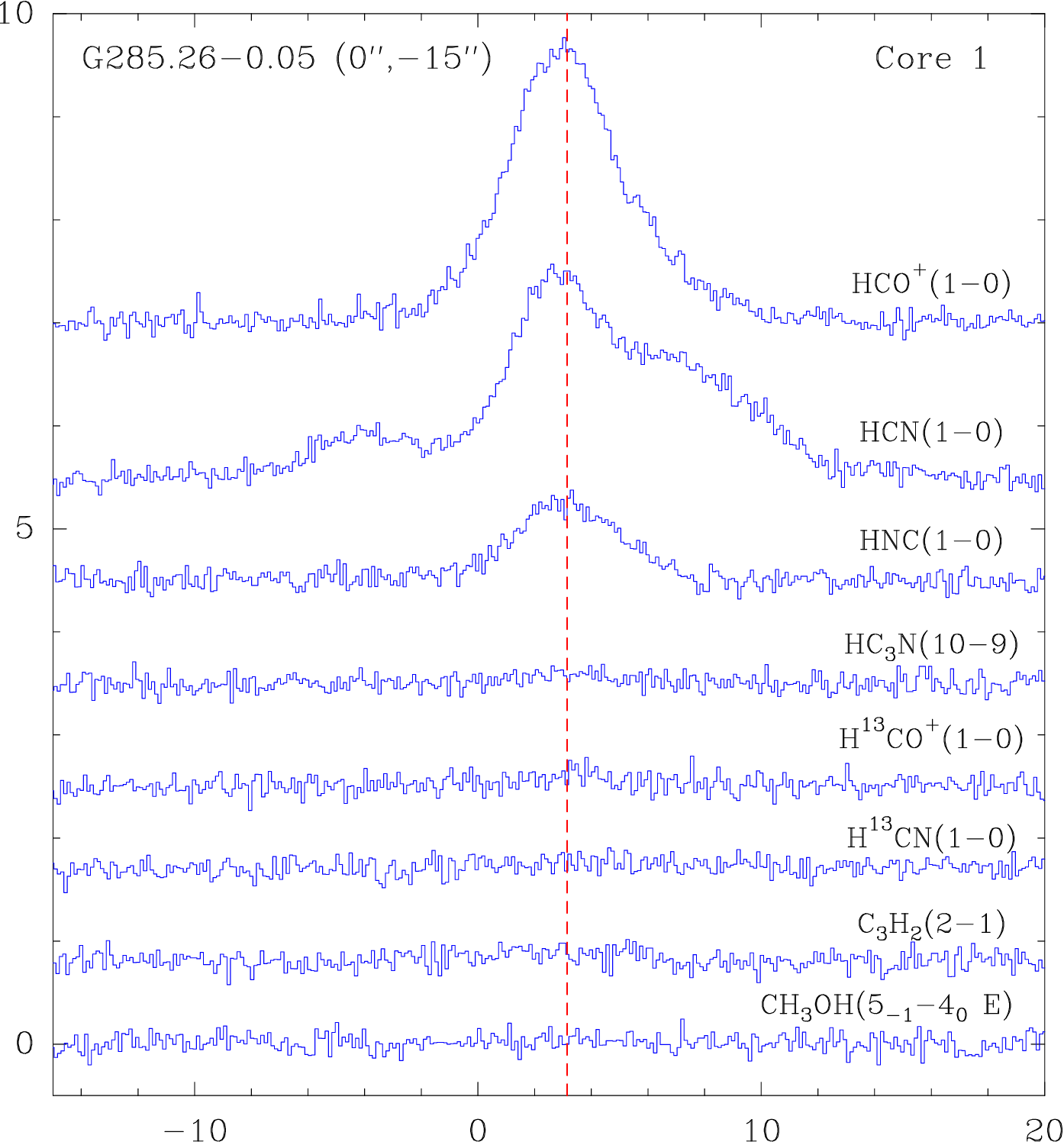}
\end{minipage}
\hspace{2mm}
\begin{minipage}[b]{0.3\textwidth}
    \includegraphics[width=\textwidth,angle=-0]{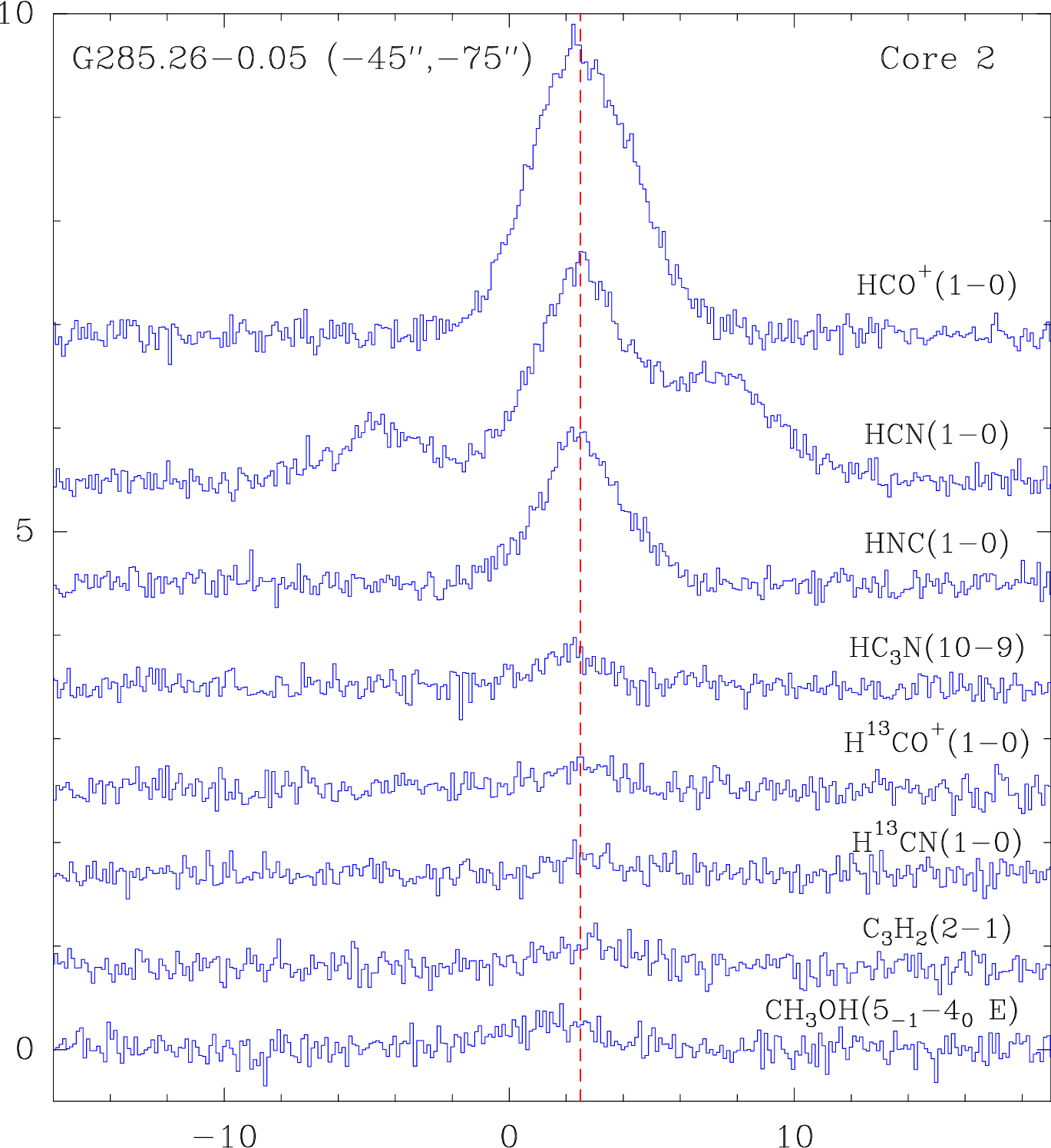}
\end{minipage}

\vspace{2mm}

\begin{minipage}[b]{0.3\textwidth}
    \includegraphics[width=\textwidth,angle=-0]{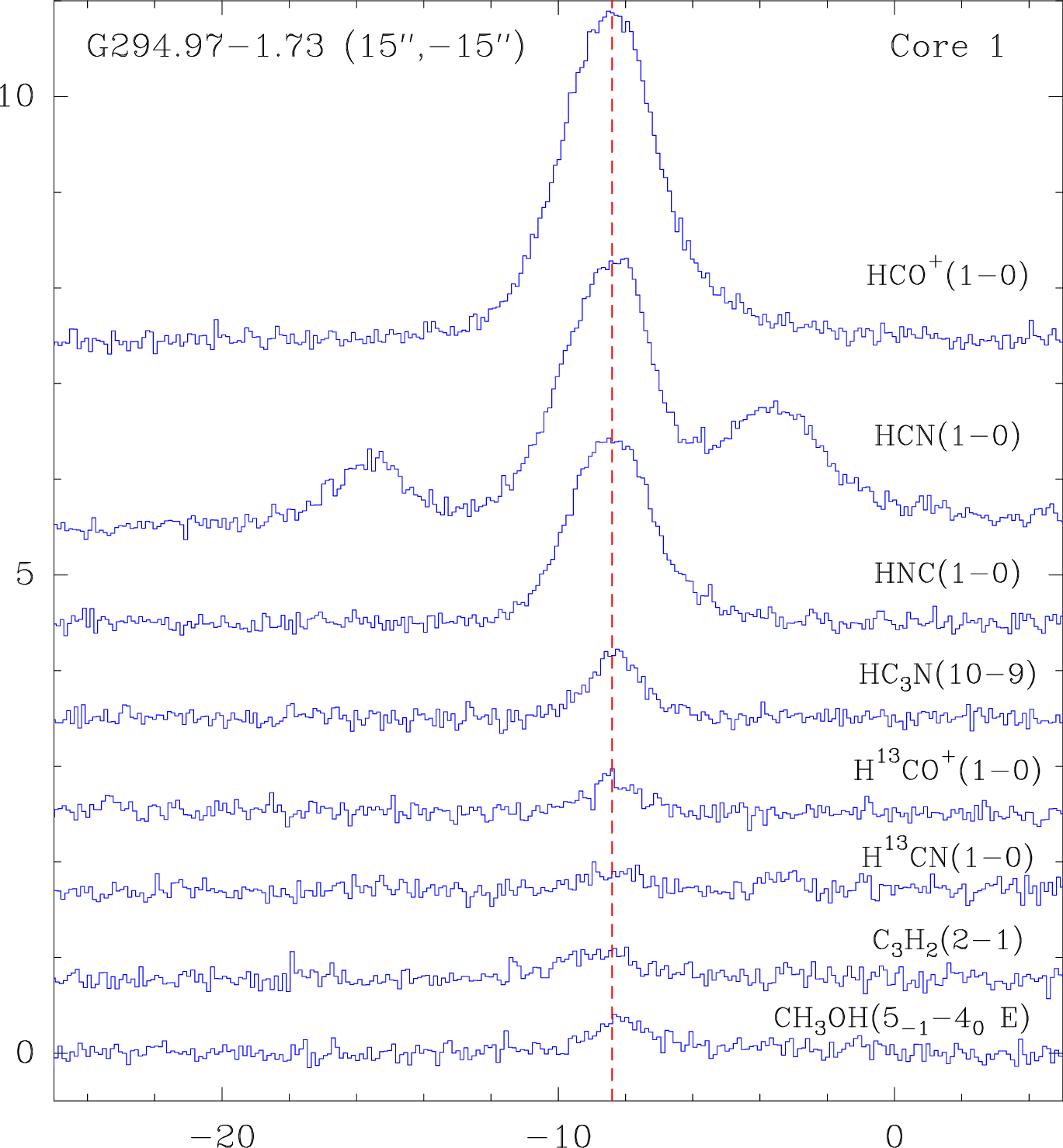}
\end{minipage}
\hspace{2mm}
\begin{minipage}[b]{0.3\textwidth}
    \includegraphics[width=\textwidth,angle=-0]{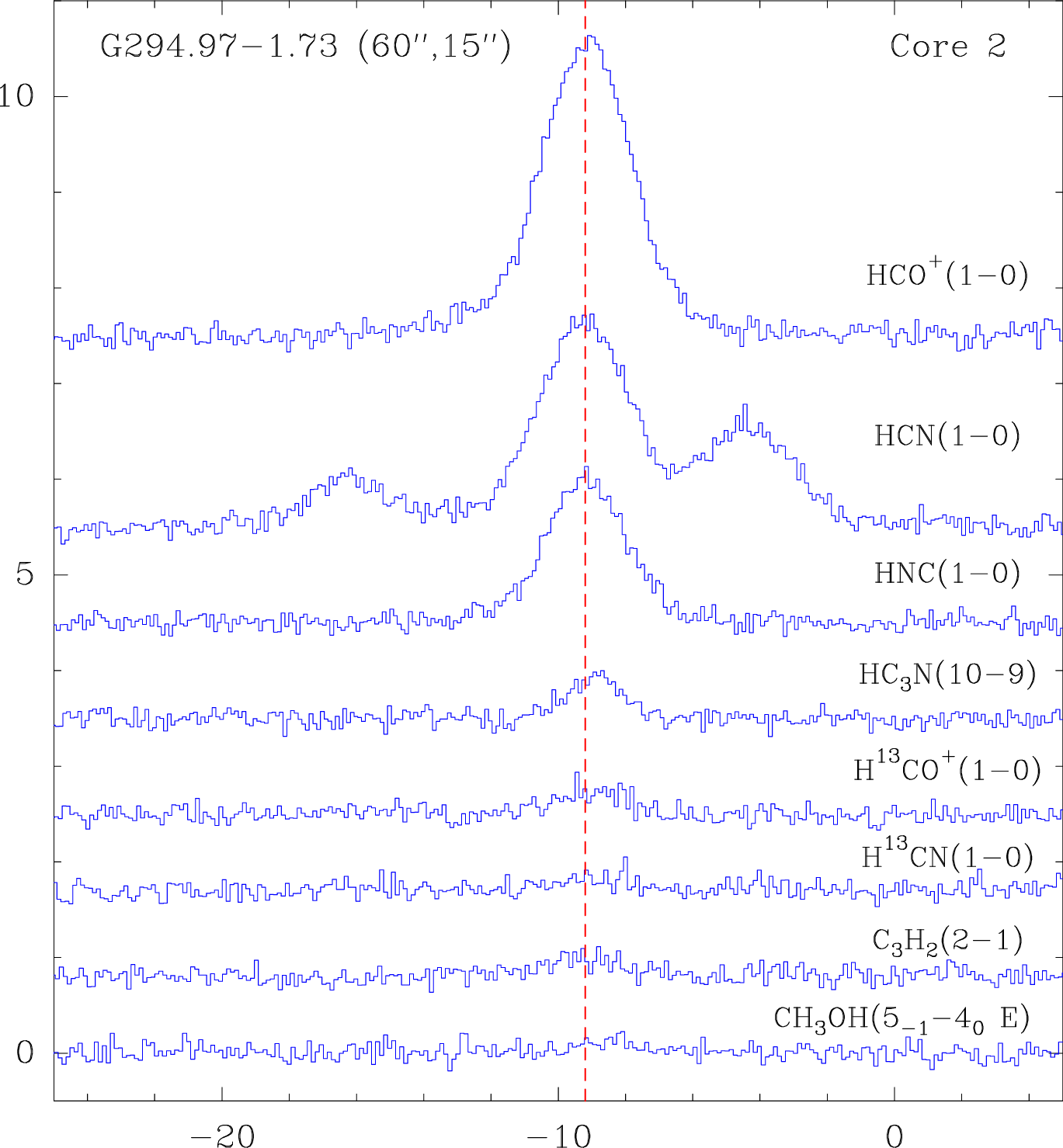}
\end{minipage}

\caption{\scriptsize
Spectra in the direction of emission peaks in objects.
The horizontal axis shows the line of sight velocity, and the vertical
axis shows the intensity in the telescope main beam temperature units
($T_{\rm MB}$).
For G285.26--0.05 and G294.97--1.73, the spectra are shown for two positions
close to the centers of different cores in these objects.
The dashed red lines correspond to the center of the optically thin
H$^{13}$CO$^+$(1--0) line.
}
\label{spectra}
\end{figure}

Table~\ref{table:lines} lists the detected molecular lines in order of
increasing frequency, indicating the transition, frequency,
and energy of the upper level (in temperature
units). Figure~\ref{spectra} shows the spectra in the direction of
positions close to the centers of the cores, where the
intensities of most lines are close to maximum. For the
regions G285.26 and G297.97, which
contain two compact cores, the spectra are shown in
the direction of the central positions of each of them.
The vertical dashed line in each diagram corresponds
to the center of the optically thin H$^{13}$CO$^+$(1--0) line.
It can be seen that the profiles of the optically thick lines
HCO$^+$(1--0), HCN(1--0) and HNC(1--0) in
G268.42, G269.11, G270.26 and G291.27 are asymmetric.
In G268.42 and G269.11 there is a pronounced
absorption dip between the blue and red peaks, and the
intensity of the blue peak exceeds the intensity of the red one.

\begin{table}[h]

\centering
\caption{List of observed molecular lines}
\vskip 2mm
\scriptsize
\begin{tabular}{l|c|r|r}
\noalign{\hrule}\noalign{\smallskip}
Molecule  &  Transition  & Frequency (MHz) & $E_u/k$ (K) \\
\noalign{\smallskip}\hline\noalign{\smallskip}

CH$_3$OH 	&	 5$_{-1}$--4$_0$~E    	&	 84521.206 	&	40.39 	\\
c-C$_3$H$_2$ 	&	 2$_{1,2}$--1$_{0,1}$ 	&	 85338.906 	&	6.45 	\\
CH$_3$C$_2$H 	&	 5$_3$--4$_3$     	&	 85442.600 	&	77.34 	\\
CH$_3$C$_2$H 	&	 5$_2$--4$_2$     	&	 85450.765 	&	41.21 	\\
CH$_3$C$_2$H 	&	 5$_1$--4$_1$     	&	 85455.665 	&	19.53 	\\
CH$_3$C$_2$H 	&	 5$_0$--4$_0$     	&	 85457.299 	&	12.30 	\\
H$^{13}$CN  	&	 1--0  $F$=1--1  	&	 86338.737 	&	4.14 	\\
H$^{13}$CN  	&	 1--0  $F$=2--1  	&	 86340.176 	&	4.14 	\\
H$^{13}$CN  	&	 1--0  $F$=0--1  	&	 86342.255 	&	4.14 	\\
H$^{13}$CO$^+$ 	&	 1--0         		&	 86754.288 	&	4.16 	\\
SiO         	&	 2--1           	&	 86846.995 	&	6.25 	\\
HN$^{13}$C  	&	 1--0  $F$=2--1   	&	 87090.859 	&	4.18 	\\
HCN         	&	 1--0  $F$=1--1  	&	 88630.416 	&	4.25 	\\
HCN         	&	 1--0  $F$=2--1  	&	 88631.847 	&	4.25 	\\
HCN         	&	 1--0  $F$=0--1  	&	 88633.936 	&	4.25 	\\
HCO$^+$     	&	 1--0           	&	 89188.526 	&	4.28 	\\
HNC     	&	 1--0           	&	 90663.564 	&	4.35 	\\
HC$_3$N     	&	 10--9           	&	 90978.989 	&       24.01 	\\
CH$_3$CN 	&	 5$_3$--4$_3$     	&	 91971.130 	&	77.55 	\\
CH$_3$CN 	&	 5$_2$--4$_2$     	&	 91979.994 	&	41.83 	\\
CH$_3$CN 	&	 5$_1$--4$_1$     	&	 91985.314 	&	20.39 	\\
CH$_3$CN 	&	 5$_0$--4$_0$     	&	 91987.088 	&	13.24 	\\

\noalign{\smallskip}\hline\noalign{\smallskip}
\end{tabular}
\label{table:lines}
\end{table}

\subsection{Maps in molecular lines}

\begin{figure}[h]

\begin{minipage}[b]{0.33\textwidth}
    \includegraphics[width=\textwidth,angle=-0]{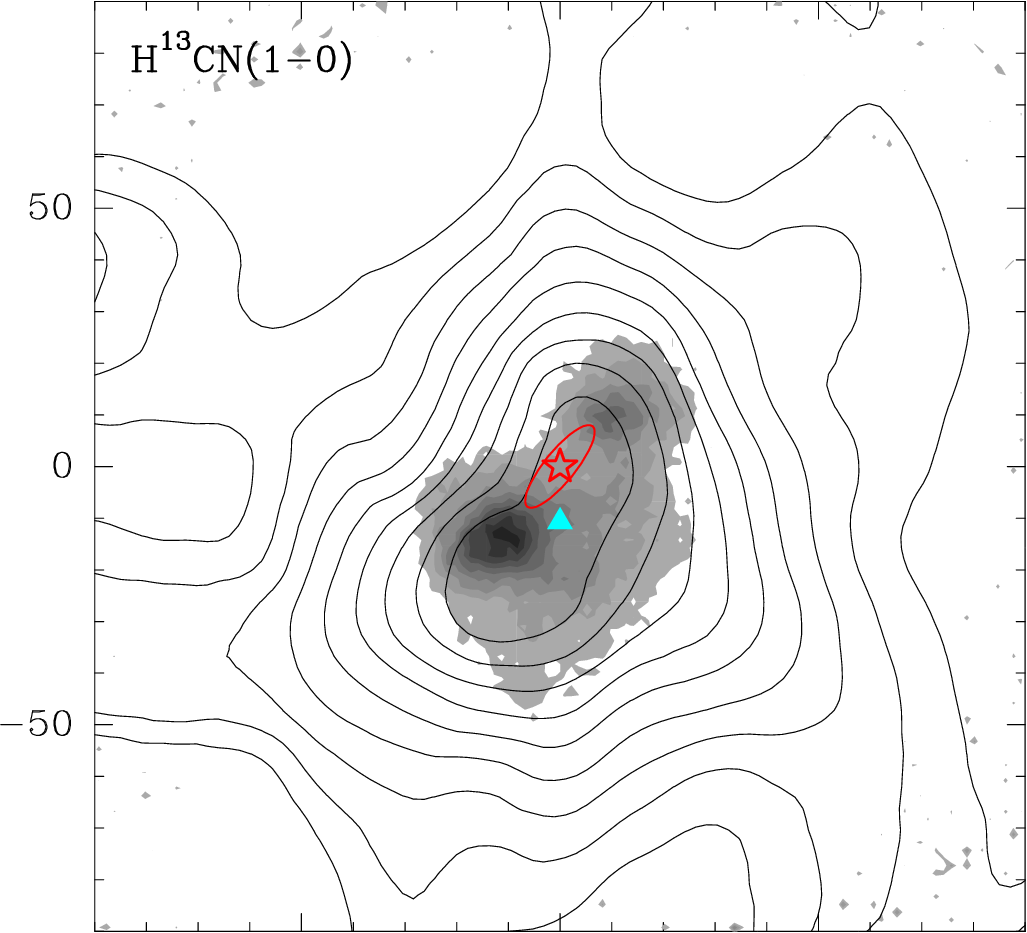}
\end{minipage}
\begin{minipage}[b]{0.3\textwidth}
    \includegraphics[width=\textwidth,angle=-0]{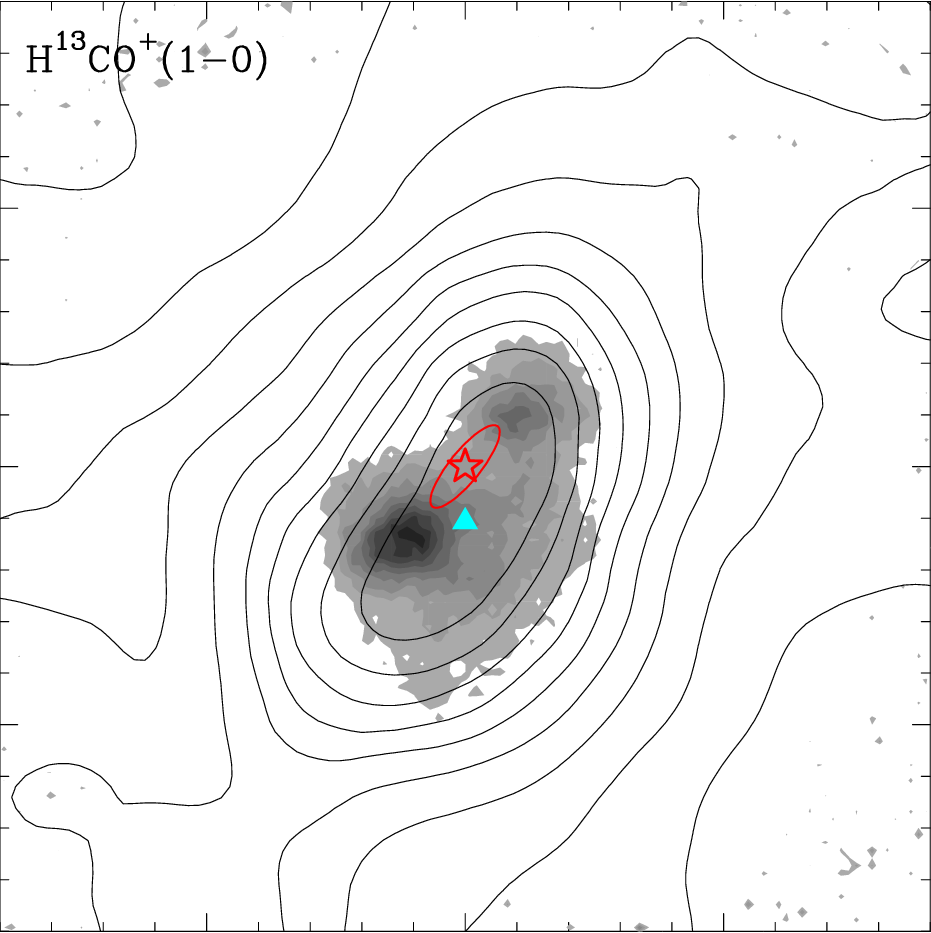}
\end{minipage}
\begin{minipage}[b]{0.3\textwidth}
    \includegraphics[width=\textwidth,angle=-0]{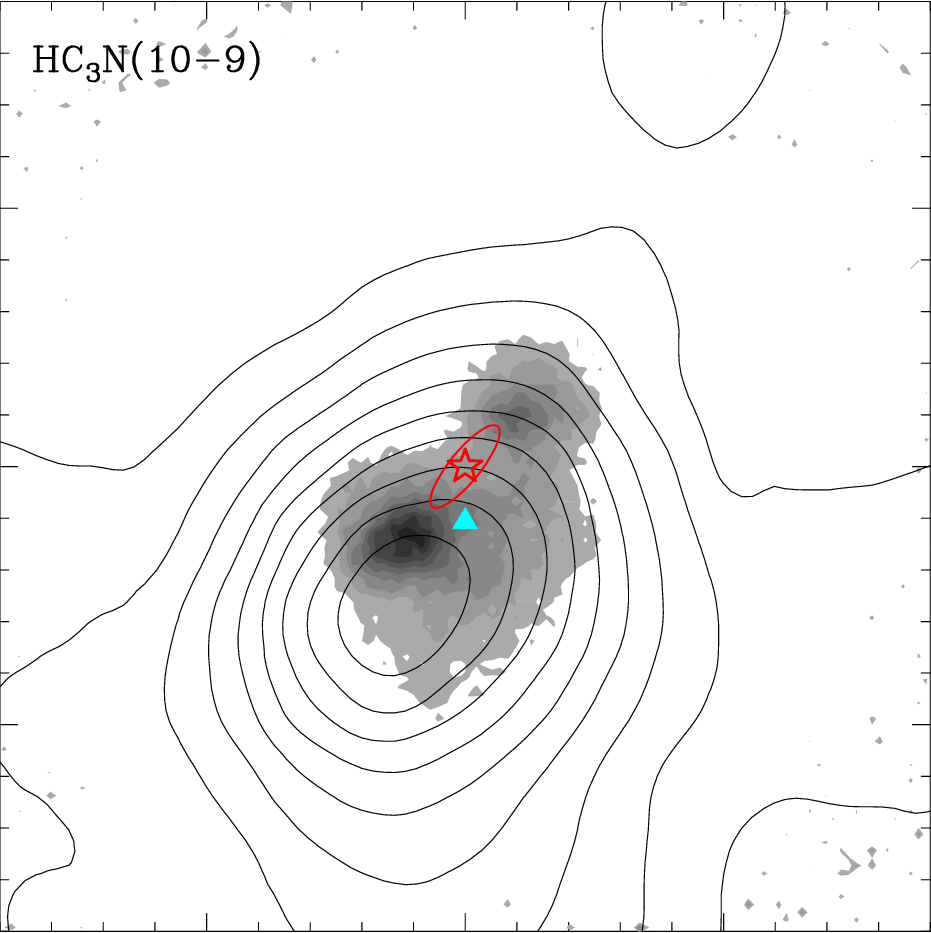}
\end{minipage}

\begin{minipage}[b]{0.33\textwidth}
    \includegraphics[width=\textwidth,angle=-0]{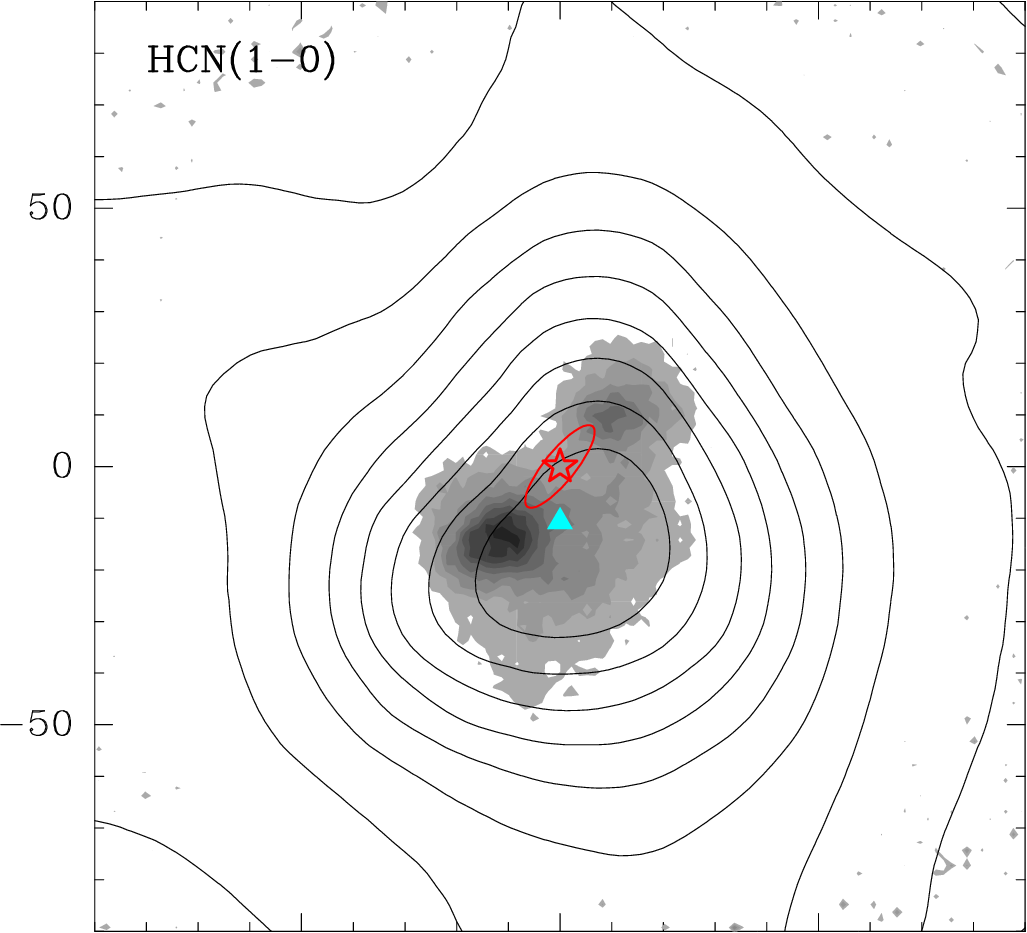}
\end{minipage}
\begin{minipage}[b]{0.3\textwidth}
    \includegraphics[width=\textwidth,angle=-0]{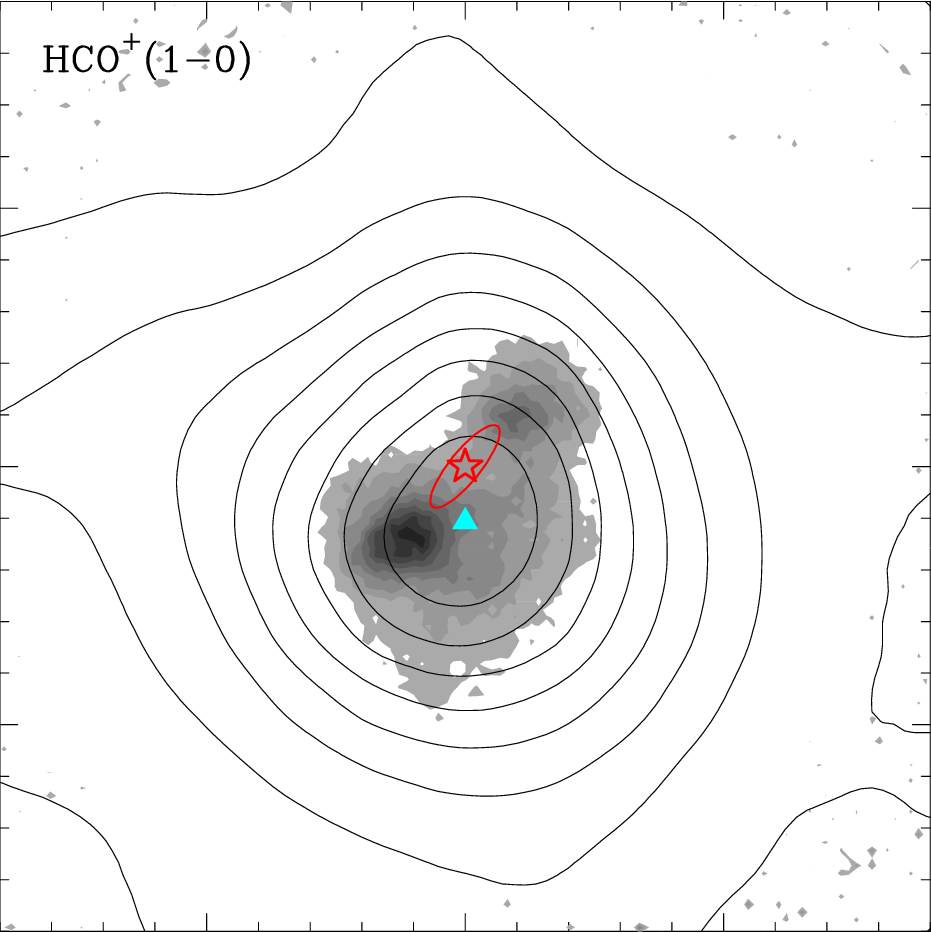}
\end{minipage}
\begin{minipage}[b]{0.3\textwidth}
    \includegraphics[width=\textwidth,angle=-0]{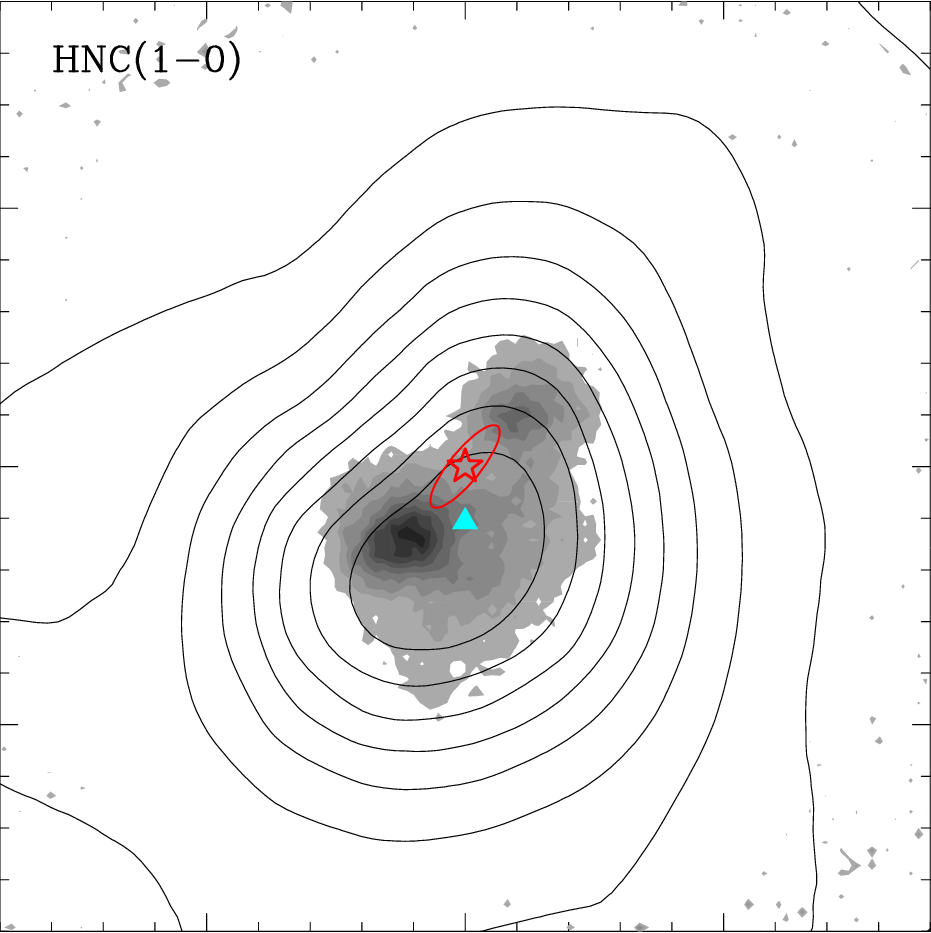}
\end{minipage}

\begin{minipage}[b]{0.33\textwidth}
    \includegraphics[width=\textwidth,angle=-0]{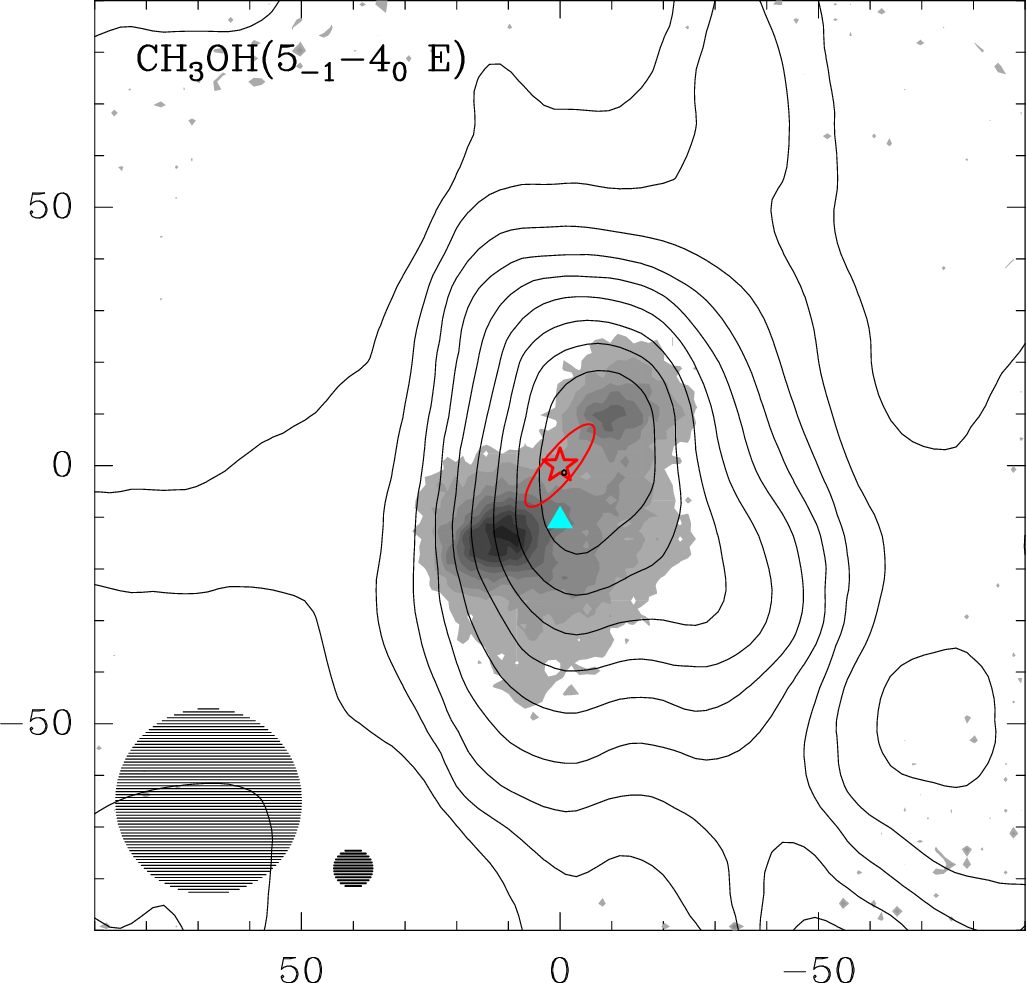}
\end{minipage}
\begin{minipage}[b]{0.3\textwidth}
    \includegraphics[width=\textwidth,angle=-0]{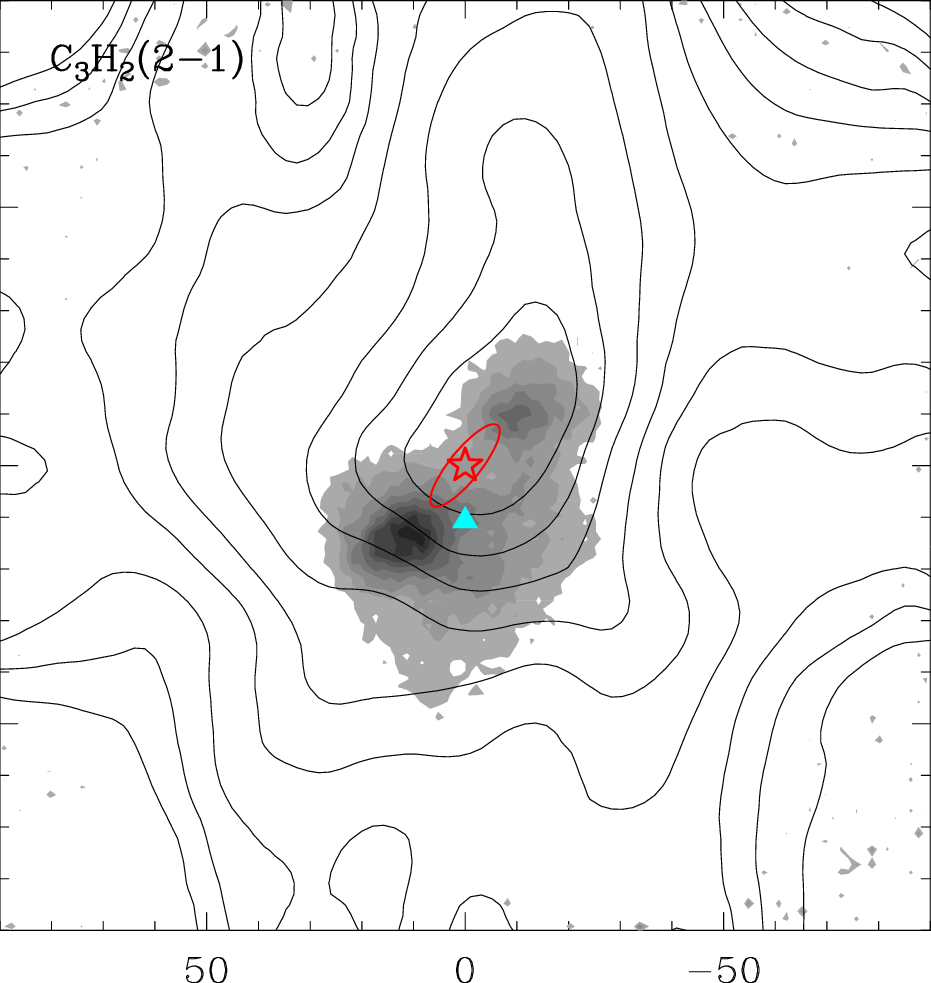}
\end{minipage}
\begin{minipage}[b]{0.3\textwidth}
    \includegraphics[width=\textwidth,angle=-0]{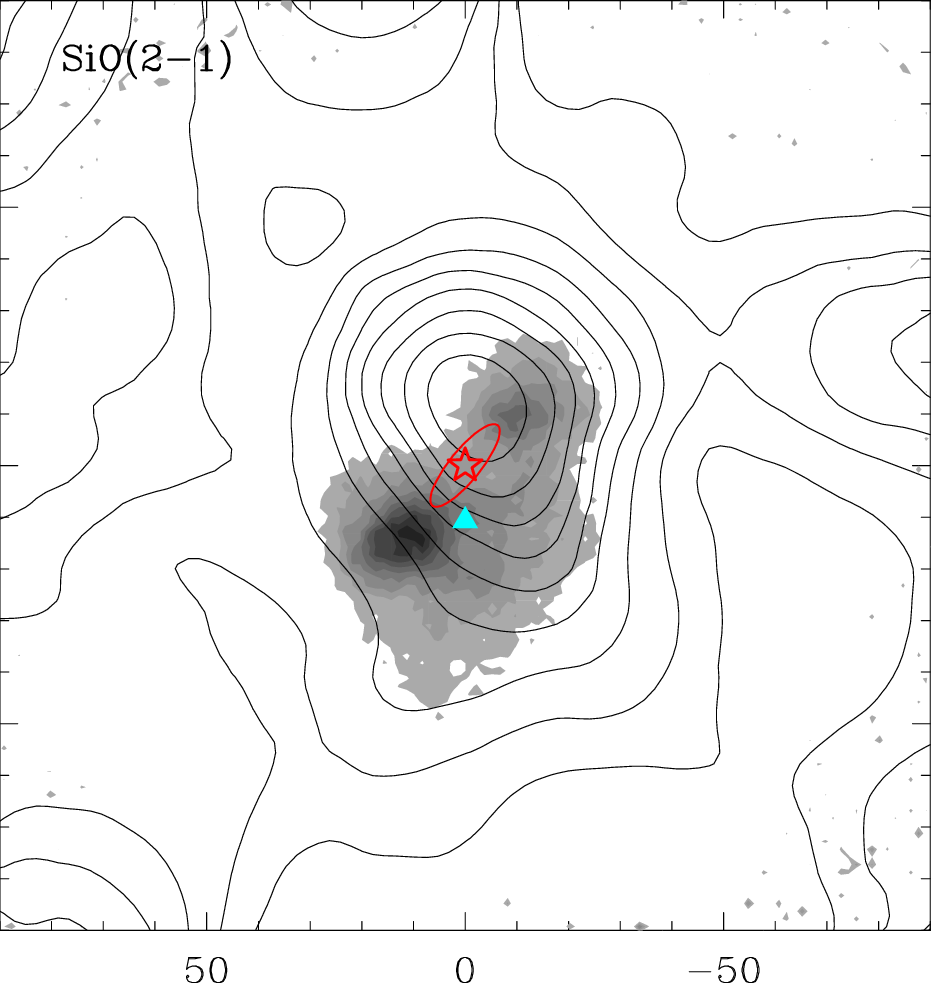}
\end{minipage}

\vskip 20mm

\caption{\scriptsize
Maps of molecular lines observed in the G268.42--0.85 region (contours)
and maps of dust emission in continuum at 350~$\mu$m \cite{Pir22}
(shades of gray).
The axes show the offsets relative to the coordinates given
in Table~\ref{table:list}. Isolines of integrated intensities lie
in the range from 10\% to 90\% of peak values (Table~\ref{table:list}).
The IRAS source is indicated
by a red star, water maser \cite{Urquhart09} is indicated by a blue triangle.
The uncertainty of the IRAS source position is shown
by an ellipse corresponding to the 95\% confidence level.
The lower left corner of the CH$_3$OH map shows
main beams of MOPRA-22m (36$''$) and APEX-12m (7.5$''$).}
\label{fig:G268}
\end{figure}

\begin{figure}[h]

\begin{minipage}[b]{0.33\textwidth}
    \includegraphics[width=\textwidth,angle=-0]{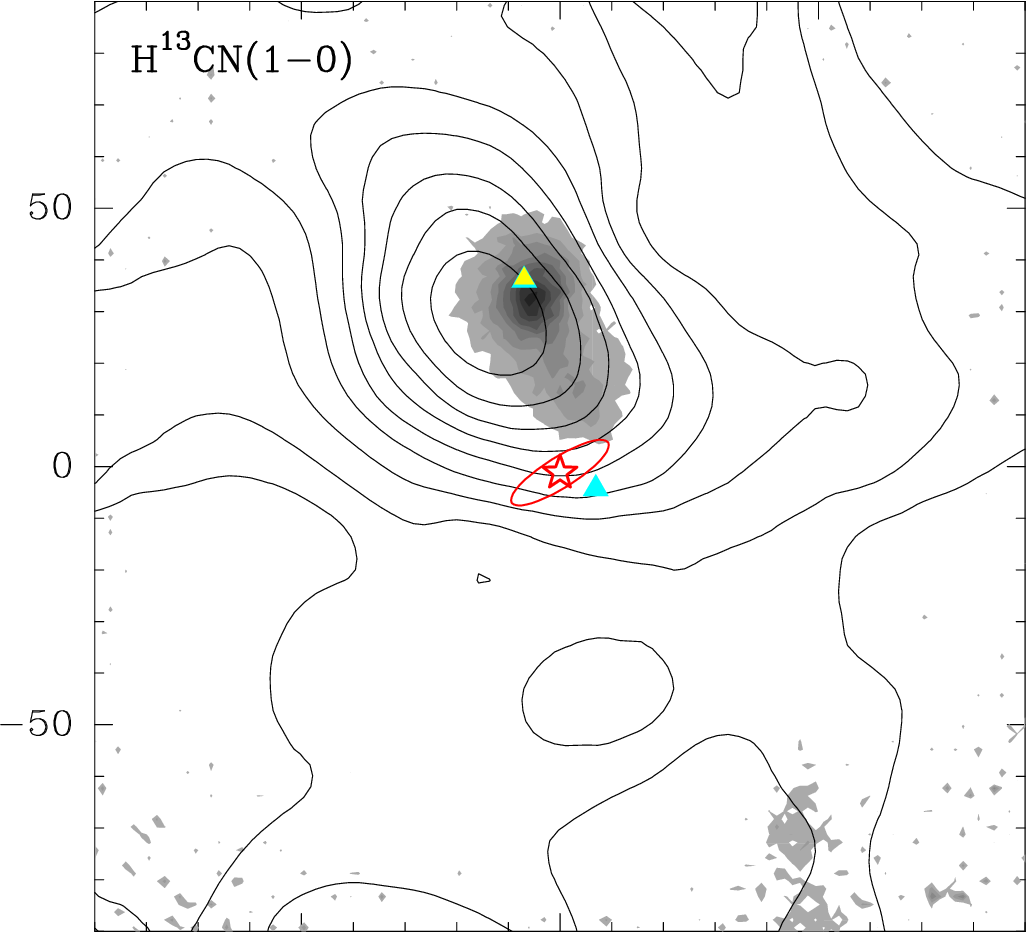}
\end{minipage}
\begin{minipage}[b]{0.3\textwidth}
    \includegraphics[width=\textwidth,angle=-0]{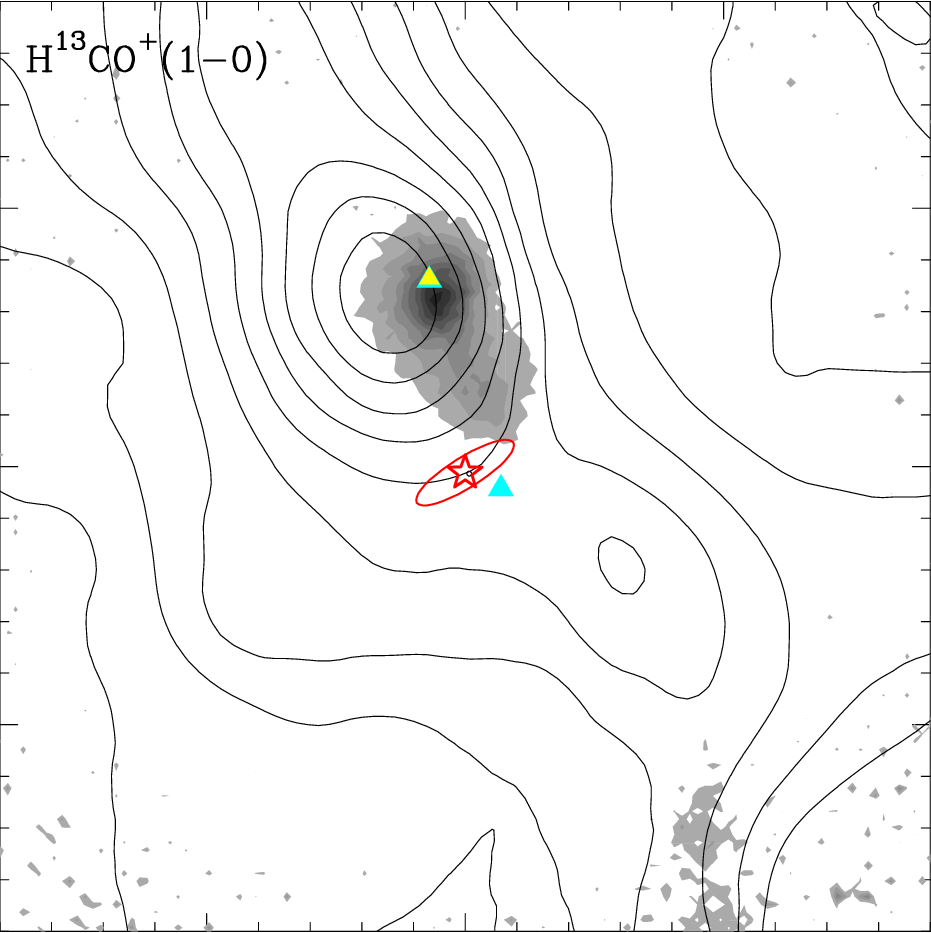}
\end{minipage}
\begin{minipage}[b]{0.3\textwidth}
    \includegraphics[width=\textwidth,angle=-0]{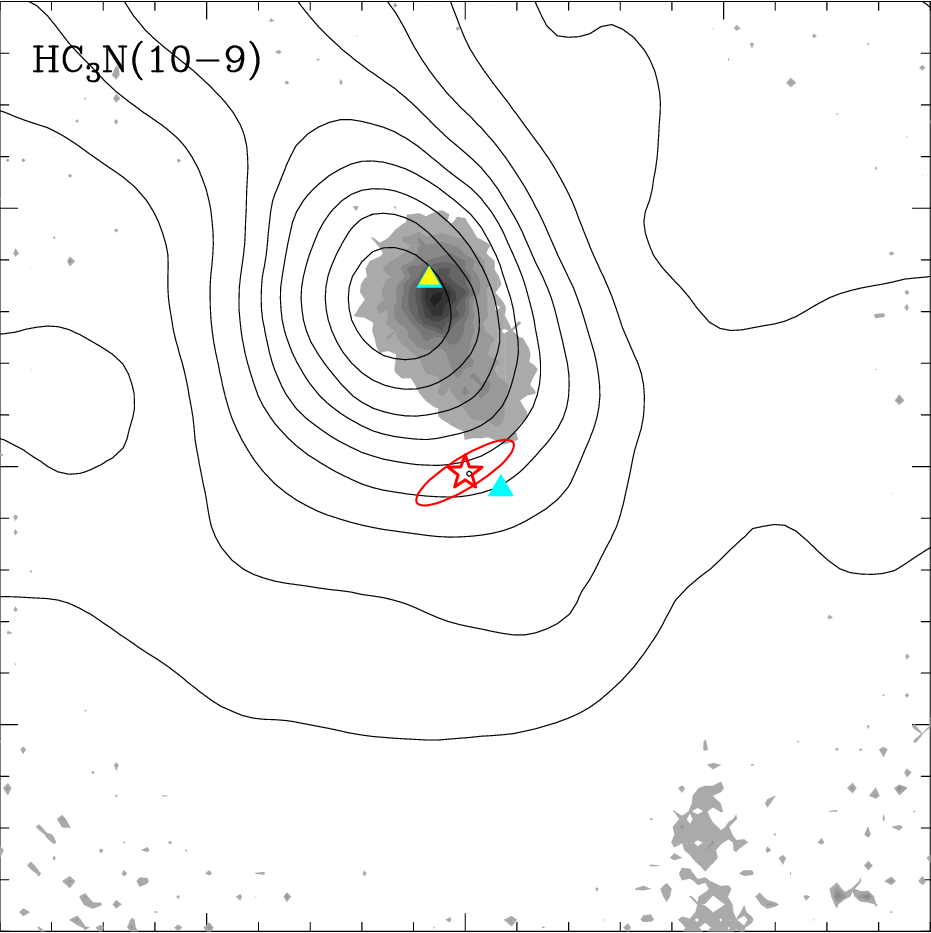}
\end{minipage}

\begin{minipage}[b]{0.33\textwidth}
    \includegraphics[width=\textwidth,angle=-0]{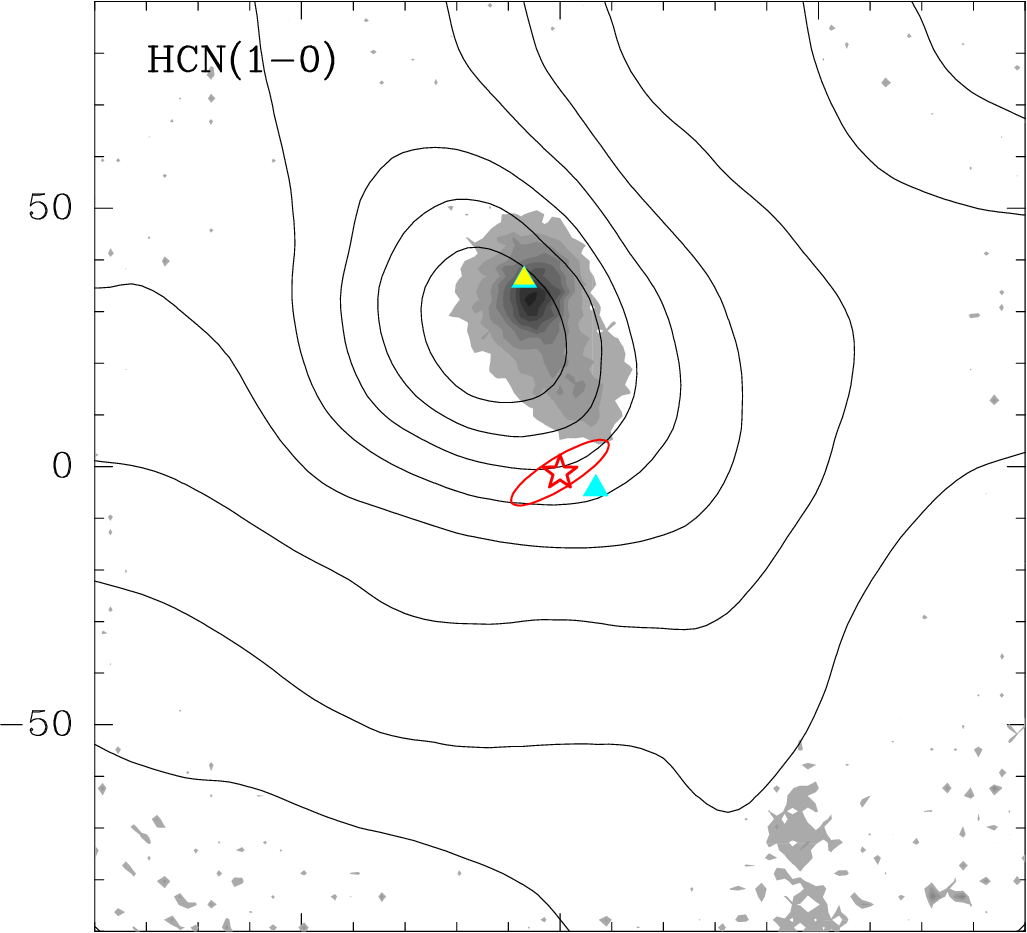}
\end{minipage}
\begin{minipage}[b]{0.3\textwidth}
    \includegraphics[width=\textwidth,angle=-0]{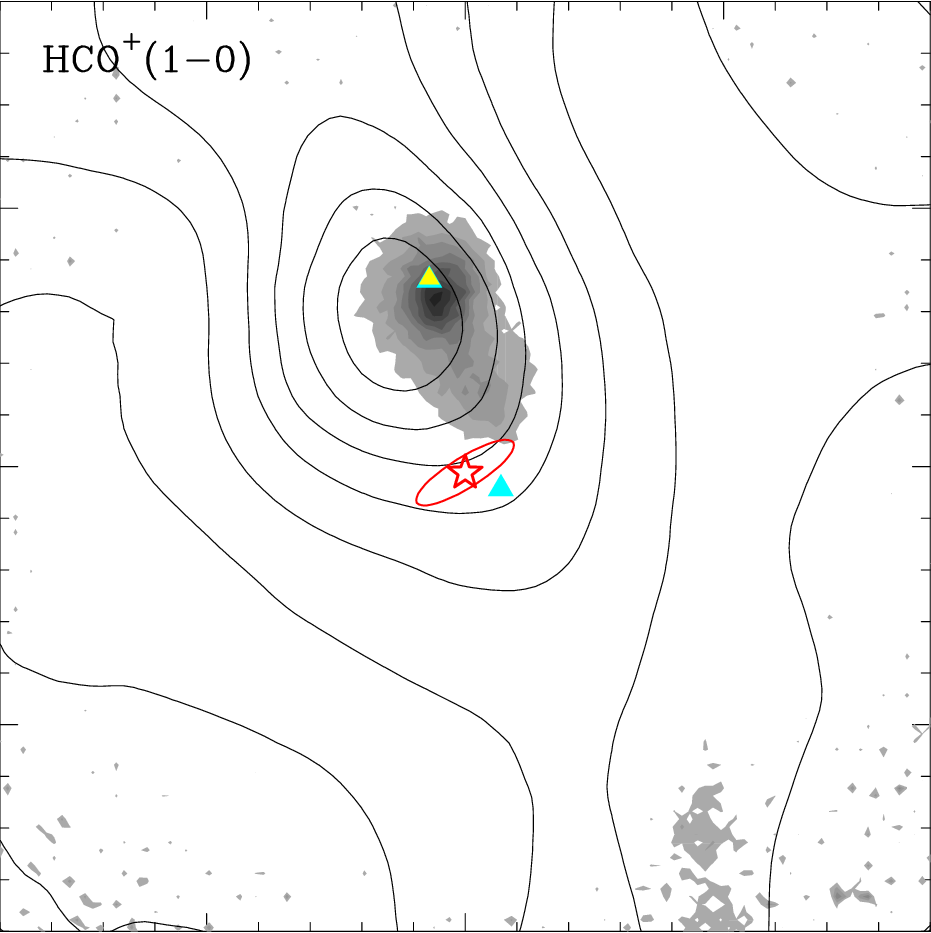}
\end{minipage}
\begin{minipage}[b]{0.3\textwidth}
    \includegraphics[width=\textwidth,angle=-0]{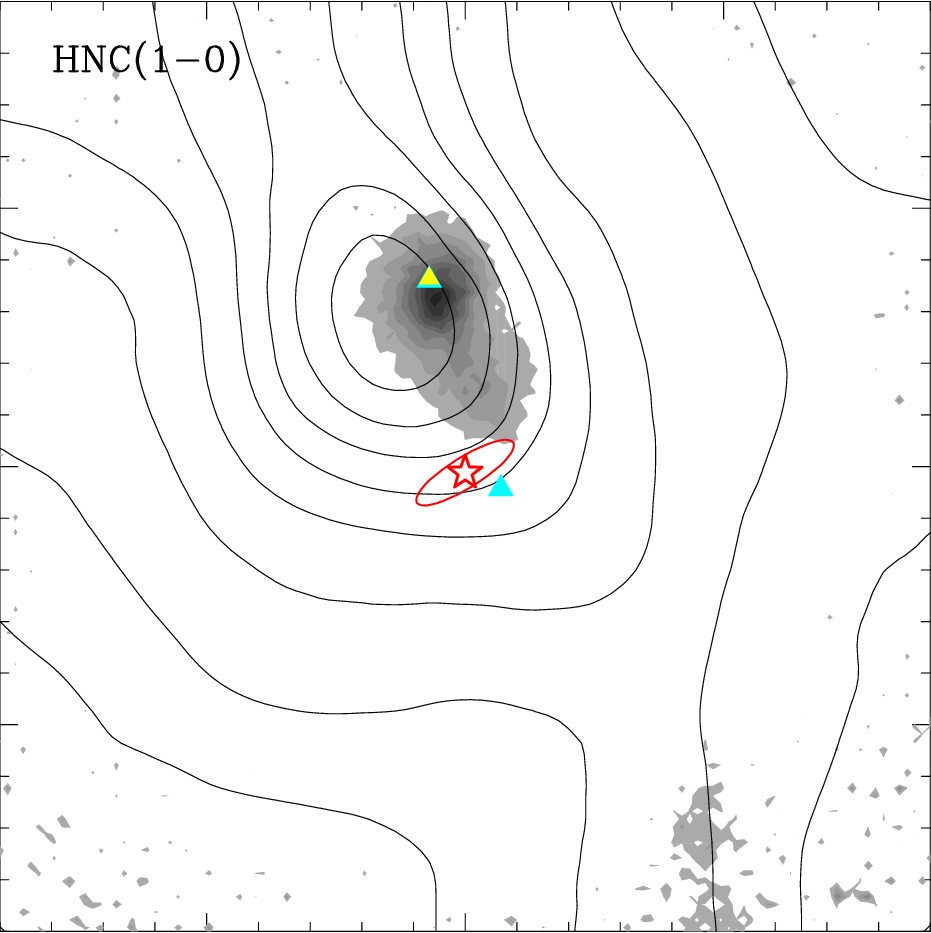}
\end{minipage}

\begin{minipage}[b]{0.33\textwidth}
    \includegraphics[width=\textwidth,angle=-0]{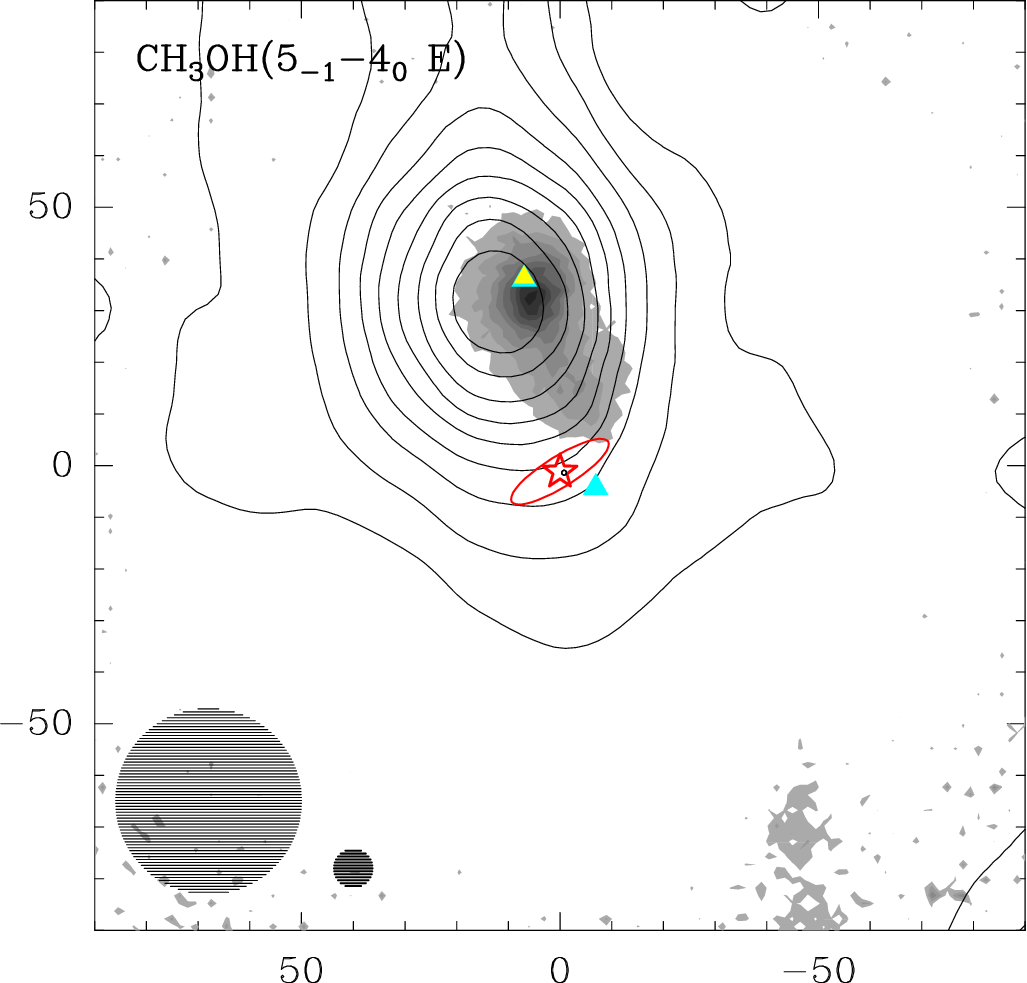}
\end{minipage}
\begin{minipage}[b]{0.3\textwidth}
    \includegraphics[width=\textwidth,angle=-0]{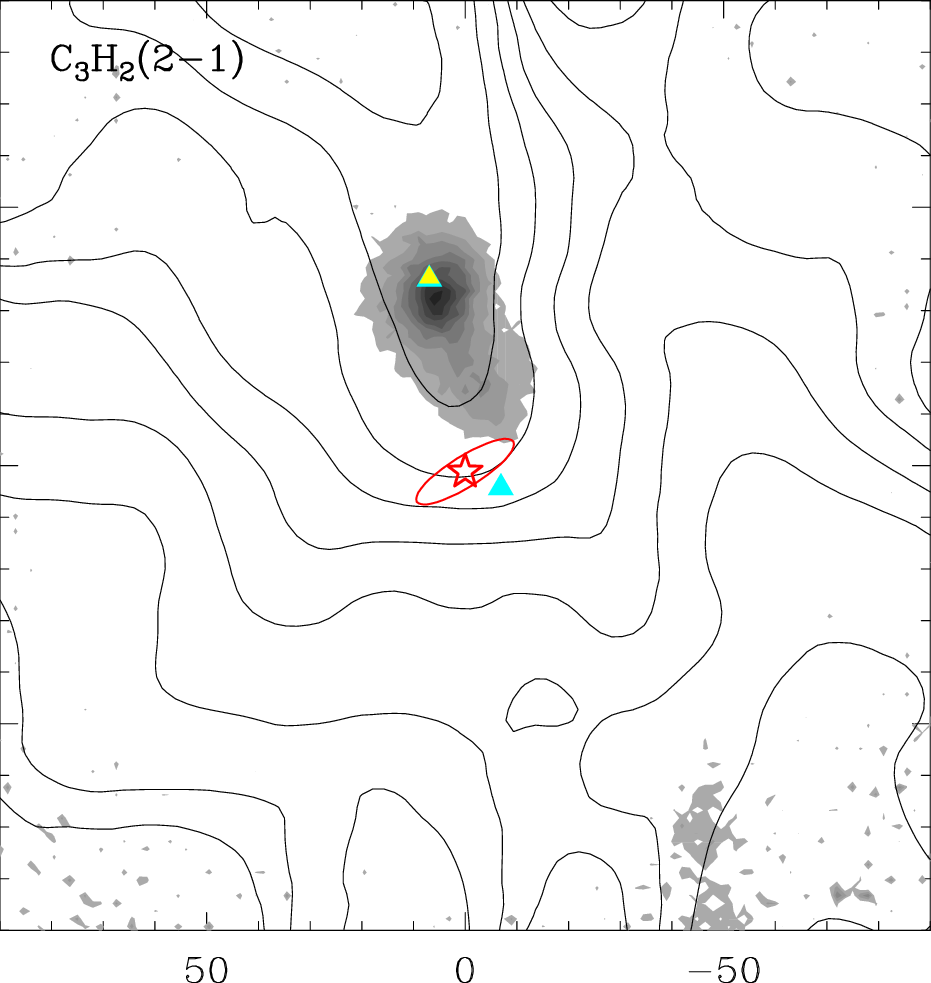}
\end{minipage}
\begin{minipage}[b]{0.3\textwidth}
    \includegraphics[width=\textwidth,angle=-0]{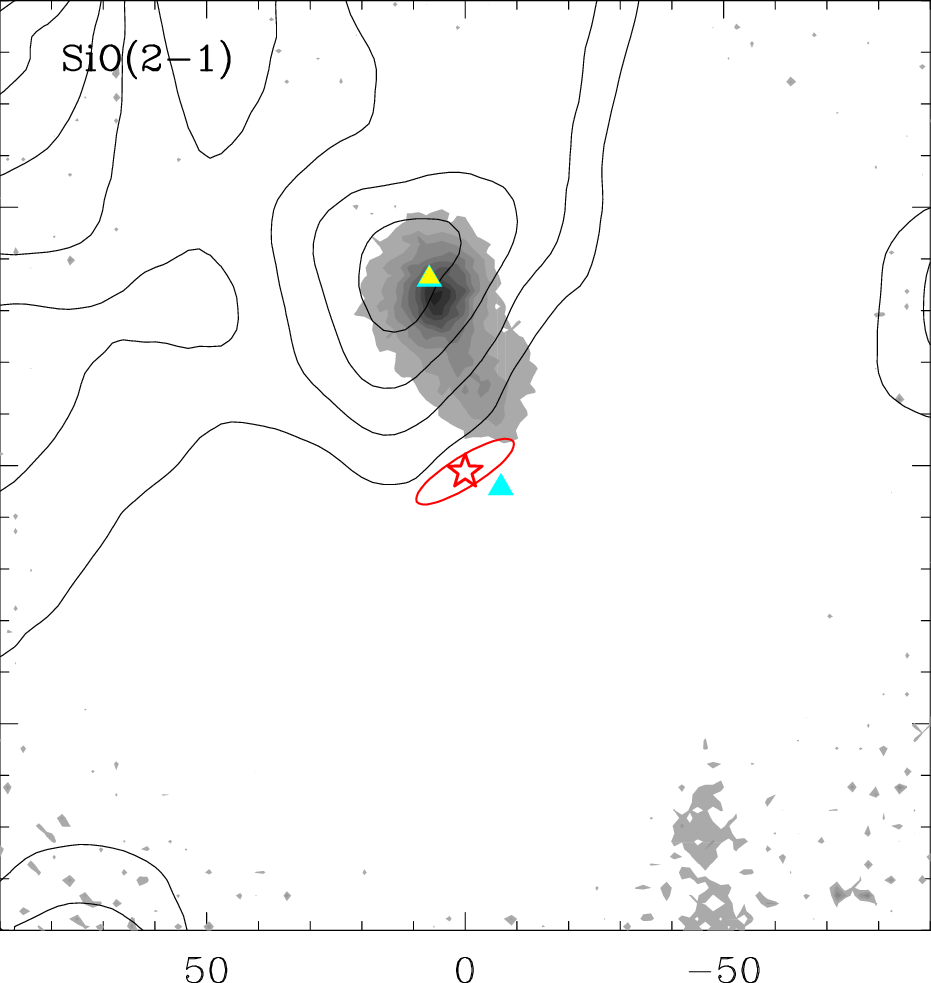}
\end{minipage}

\vskip 20mm

\caption{\scriptsize
Maps of molecular lines observed in G269.11--1.12.
A water maser is observed near the IRAS source \cite{Urquhart09}.
Near the center of the core there is a water maser \cite{Breen11}
and a class II methanol maser (6.7~GHz) \cite{Caswell09}.
Water and methanol masers are indicated by
blue and yellow triangles, respectively.
The remaining symbols are the same as in Fig.~\ref{fig:G268}.
}
\label{fig:G269}
\end{figure}

\begin{figure}[t!]

\begin{minipage}[b]{0.33\textwidth}
    \includegraphics[width=\textwidth,angle=-0]{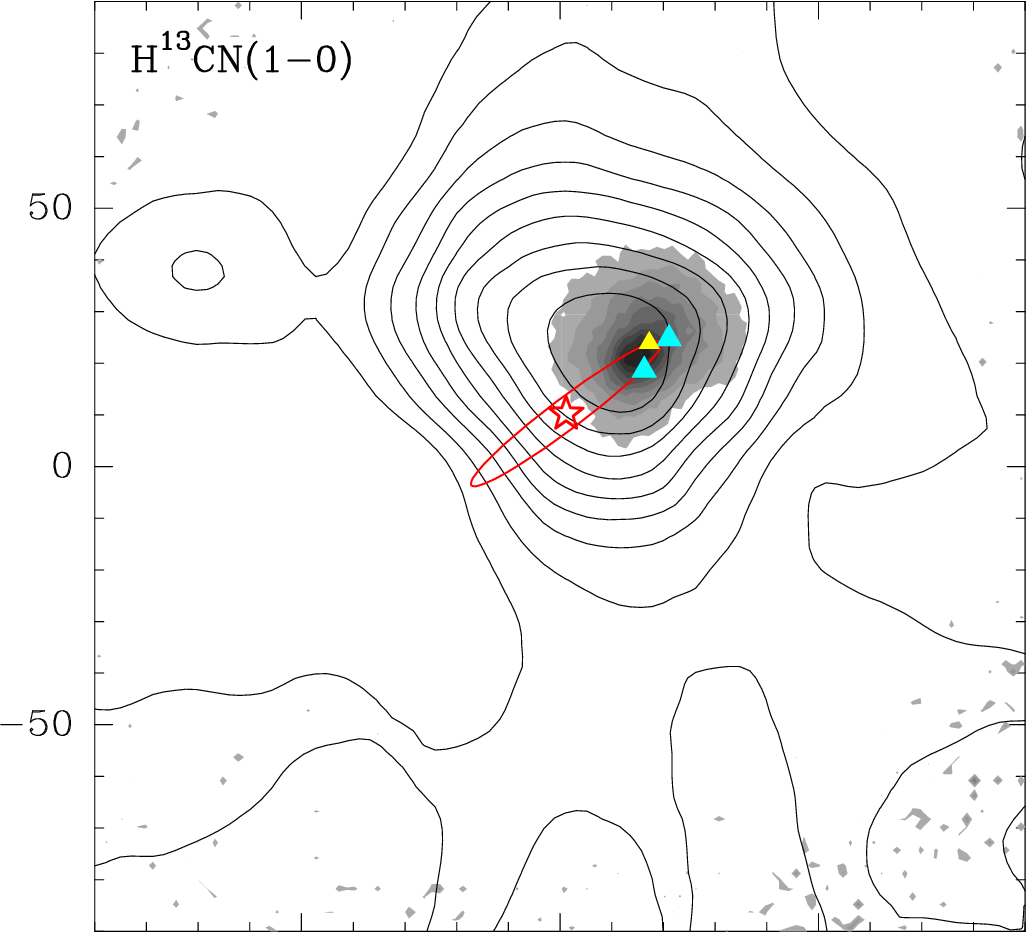}
\end{minipage}
\begin{minipage}[b]{0.3\textwidth}
    \includegraphics[width=\textwidth,angle=-0]{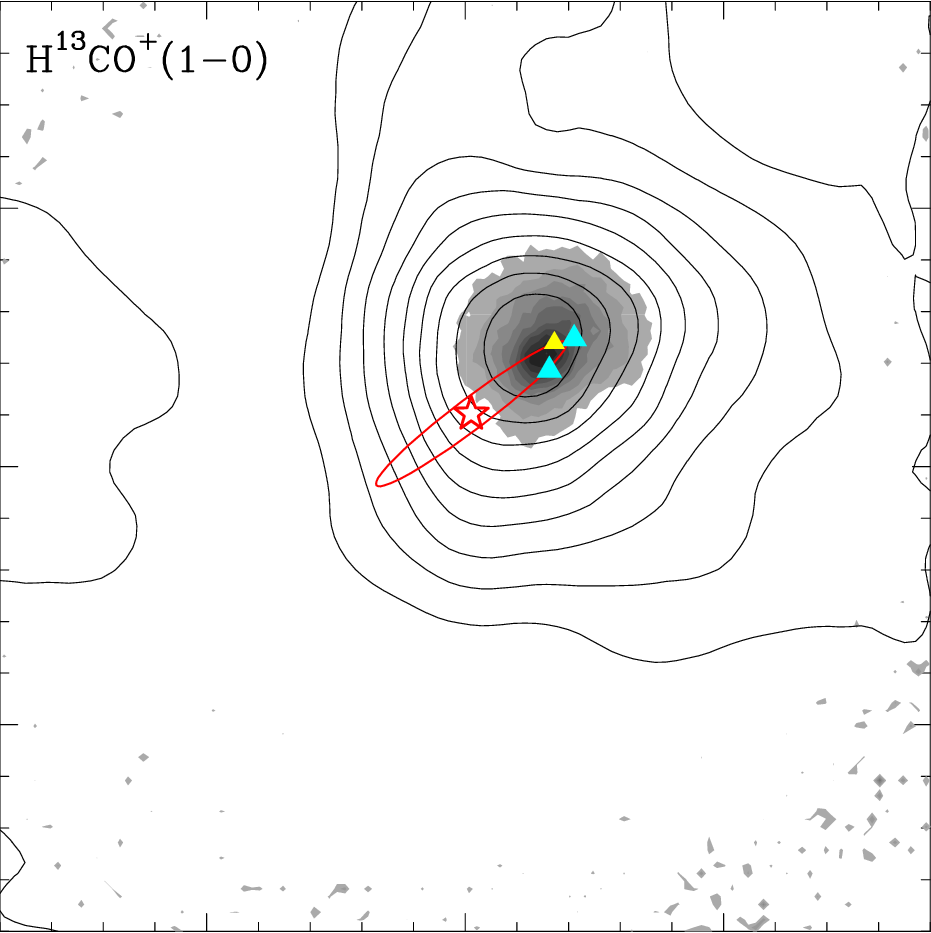}
\end{minipage}
\begin{minipage}[b]{0.3\textwidth}
    \includegraphics[width=\textwidth,angle=-0]{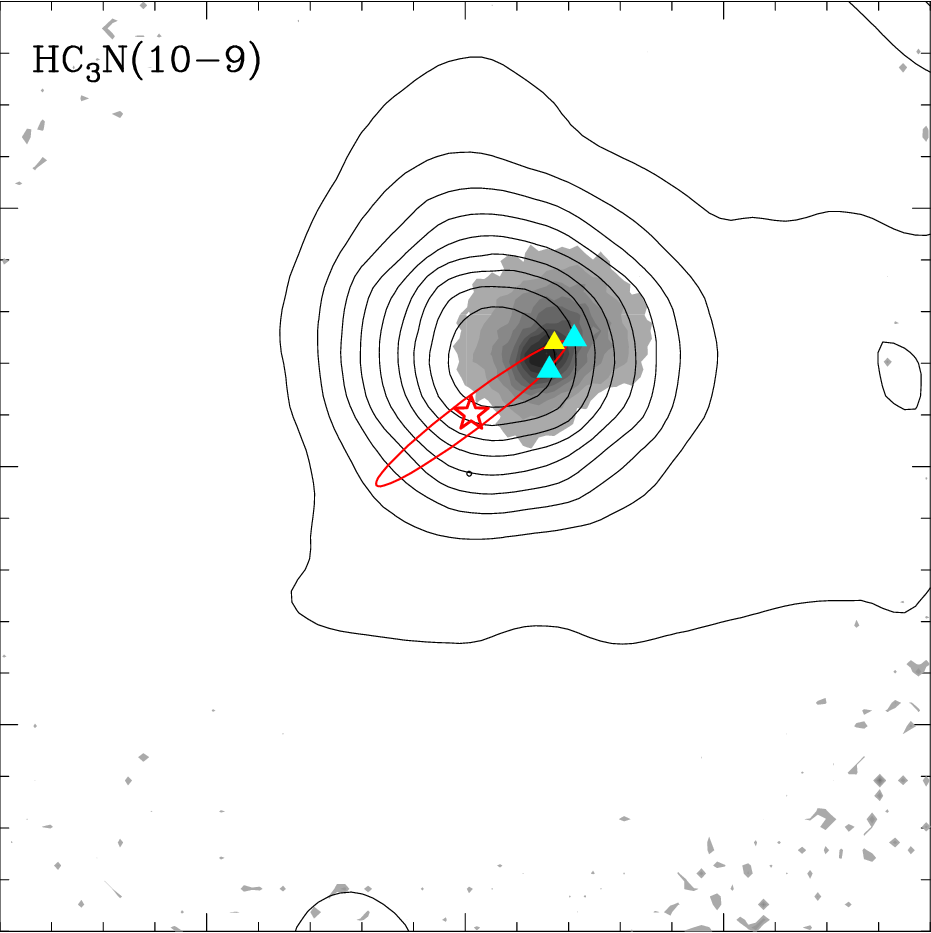}
\end{minipage}

\begin{minipage}[b]{0.33\textwidth}
    \includegraphics[width=\textwidth,angle=-0]{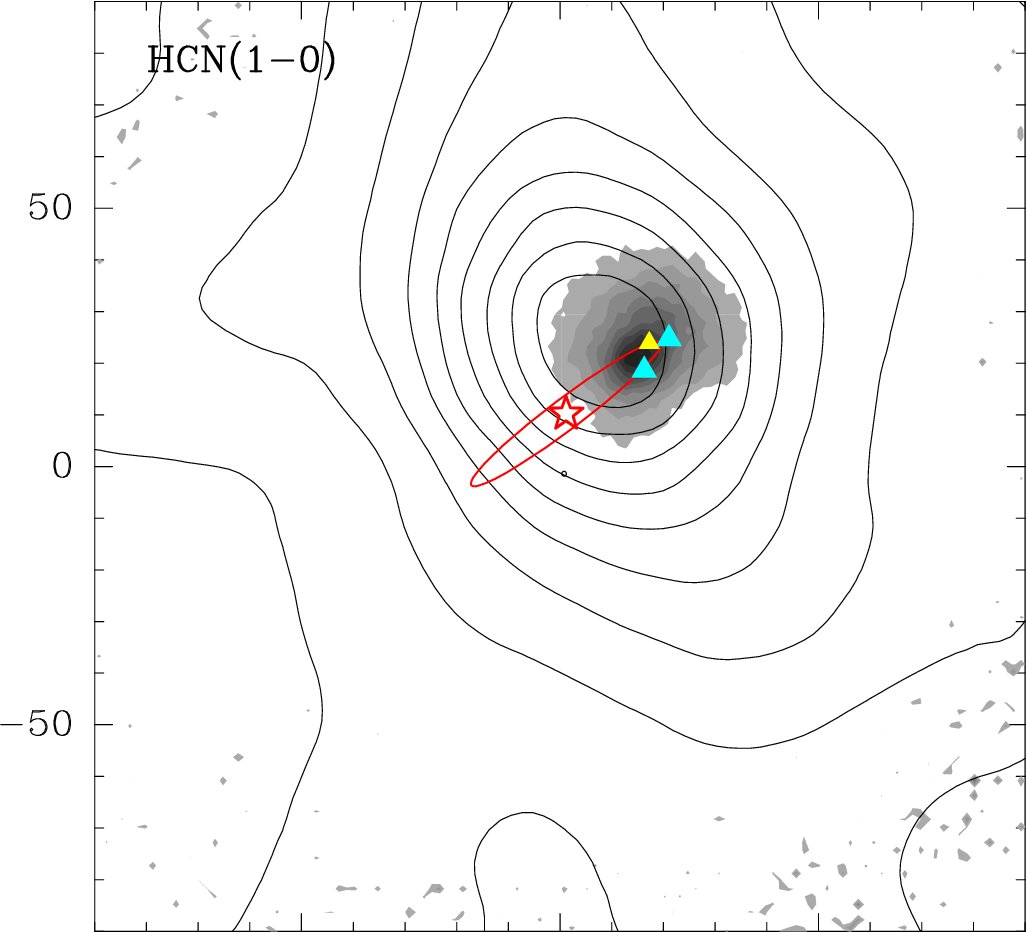}
\end{minipage}
\begin{minipage}[b]{0.3\textwidth}
    \includegraphics[width=\textwidth,angle=-0]{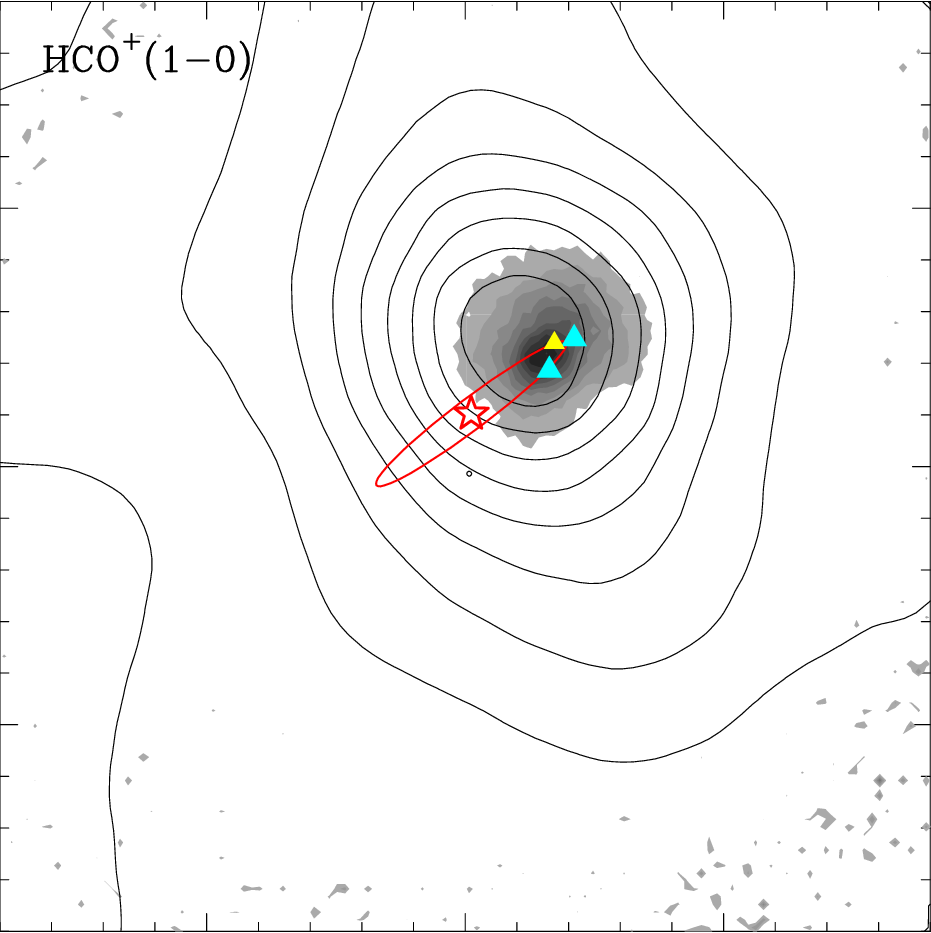}
\end{minipage}
\begin{minipage}[b]{0.3\textwidth}
    \includegraphics[width=\textwidth,angle=-0]{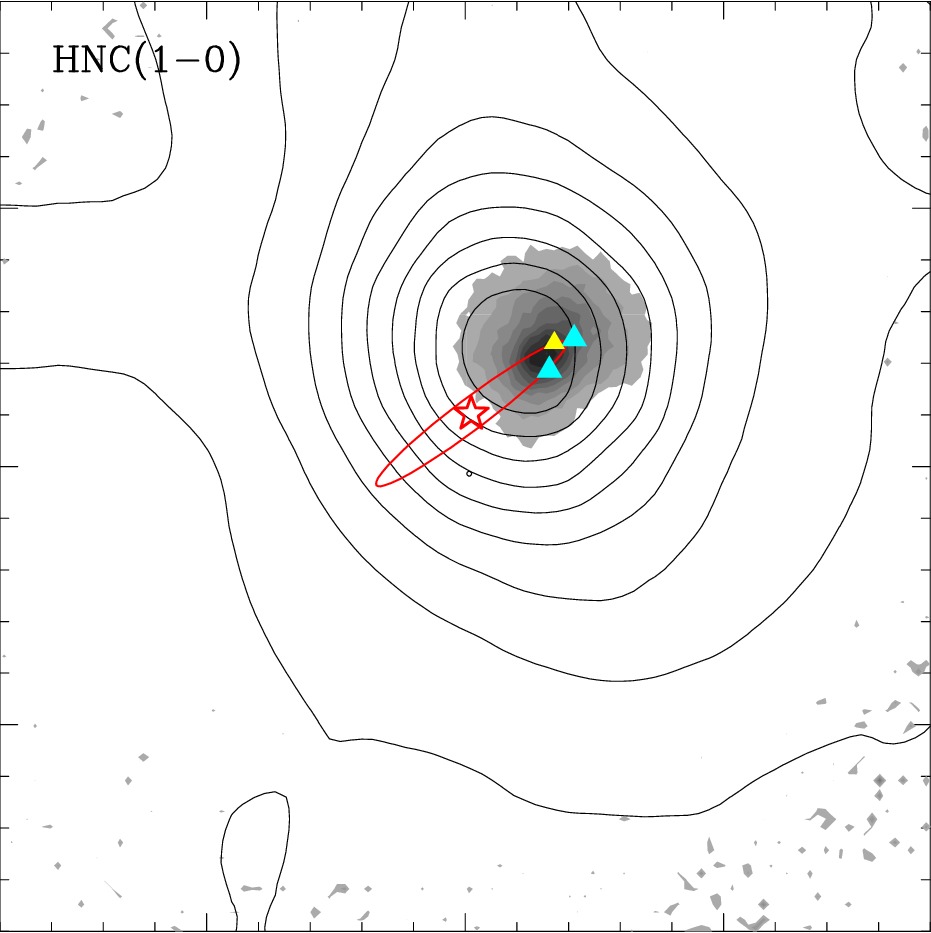}
\end{minipage}

\begin{minipage}[b]{0.33\textwidth}
    \includegraphics[width=\textwidth,angle=-0]{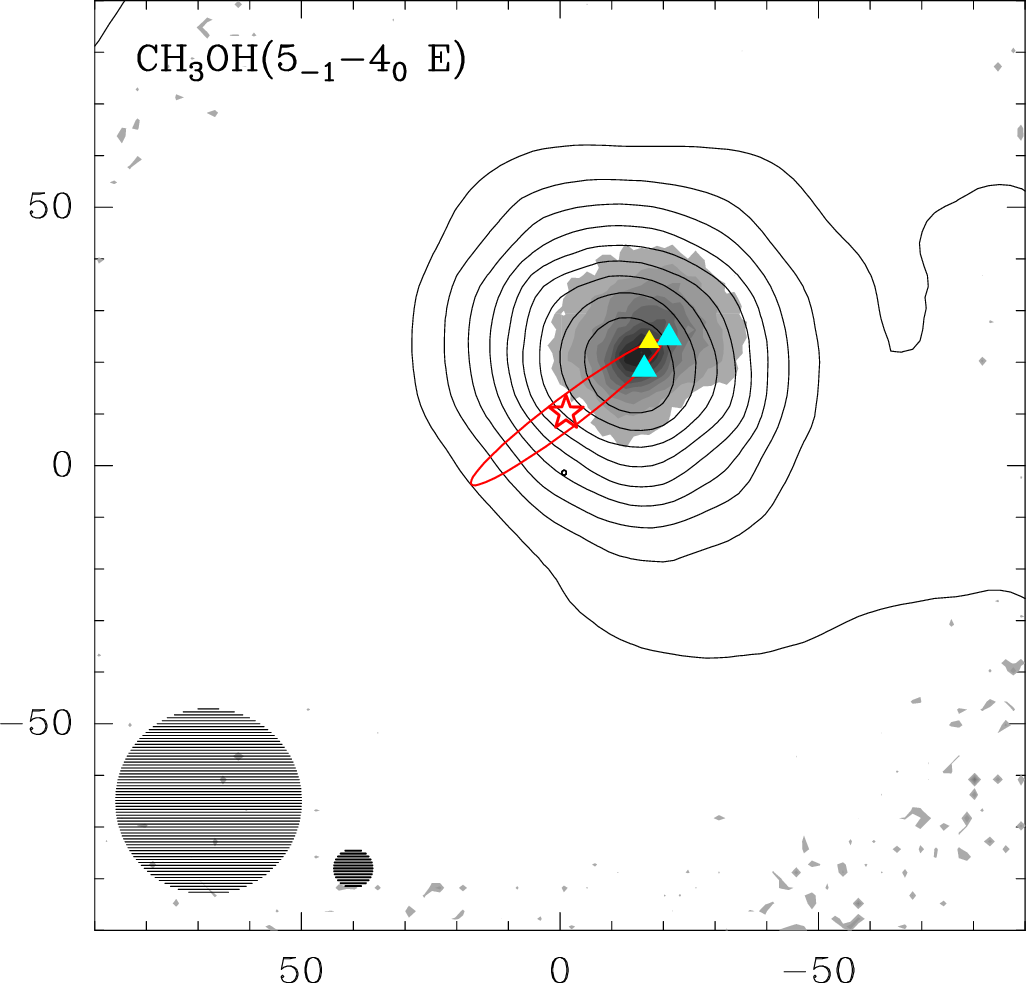}
\end{minipage}
\begin{minipage}[b]{0.3\textwidth}
    \includegraphics[width=\textwidth,angle=-0]{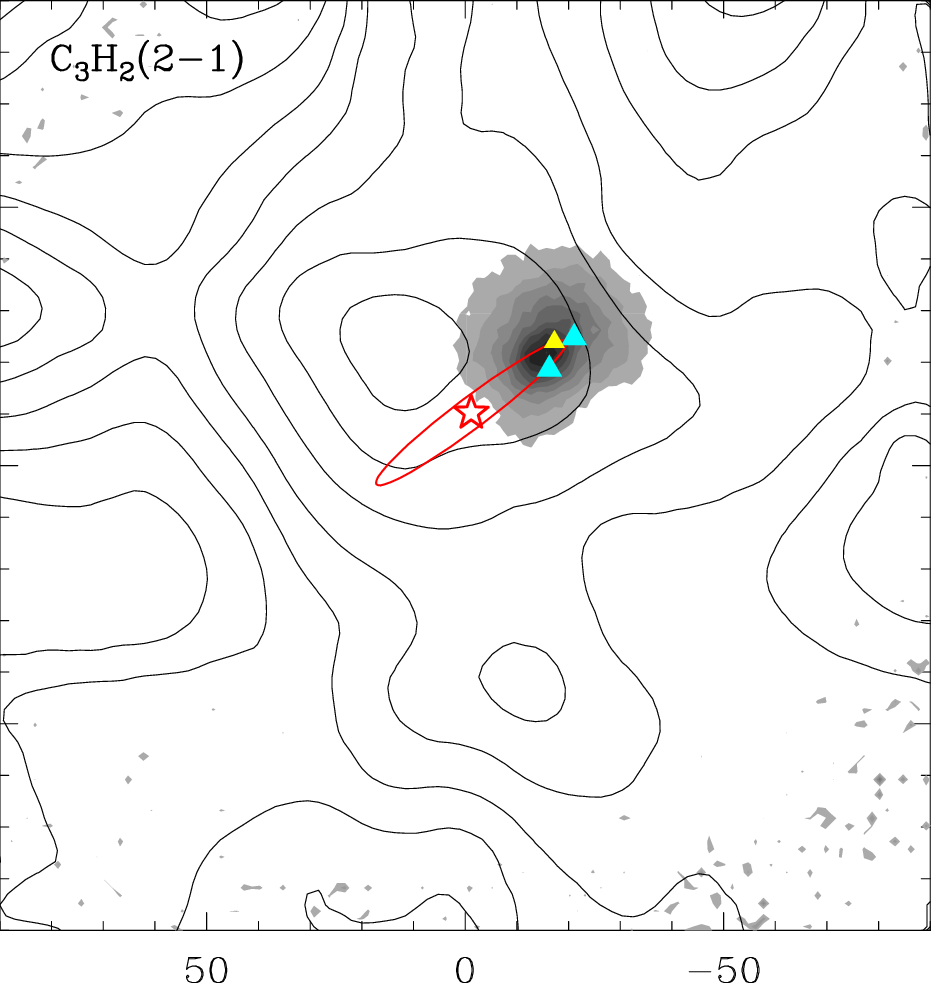}
\end{minipage}
\begin{minipage}[b]{0.3\textwidth}
    \includegraphics[width=\textwidth,angle=-0]{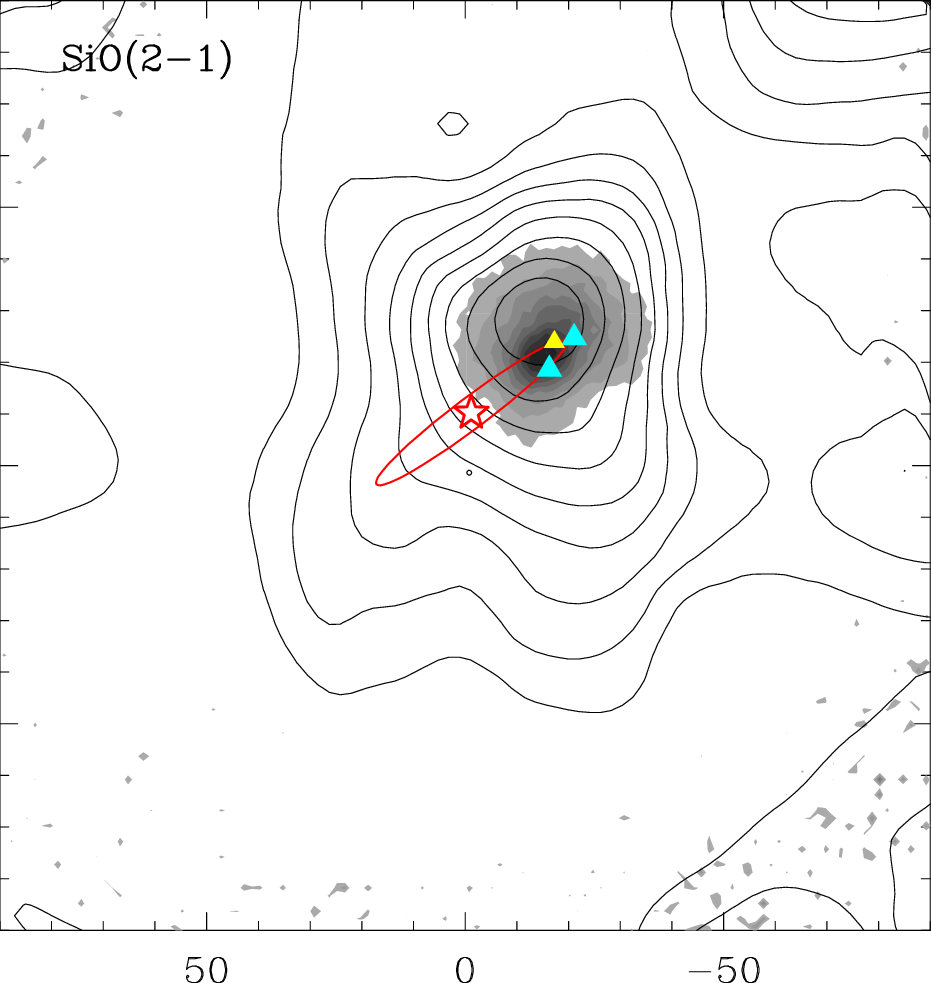}
\end{minipage}

\vskip 20mm

\caption{\scriptsize
Maps of molecular lines observed in G270.26$+$0.83.
Near the center of the core, water masers \cite{Breen11},
indicated by blue triangles, and a class II methanol maser (6.7~GHz) \cite{Caswell09},
indicated by a yellow triangle, are observed. The remaining symbols are the
same as in Fig.~\ref{fig:G268}.
}
\label{fig:G270}
\end{figure}

\begin{figure}[h]

\begin{minipage}[b]{0.33\textwidth}
    \includegraphics[width=\textwidth,angle=-0]{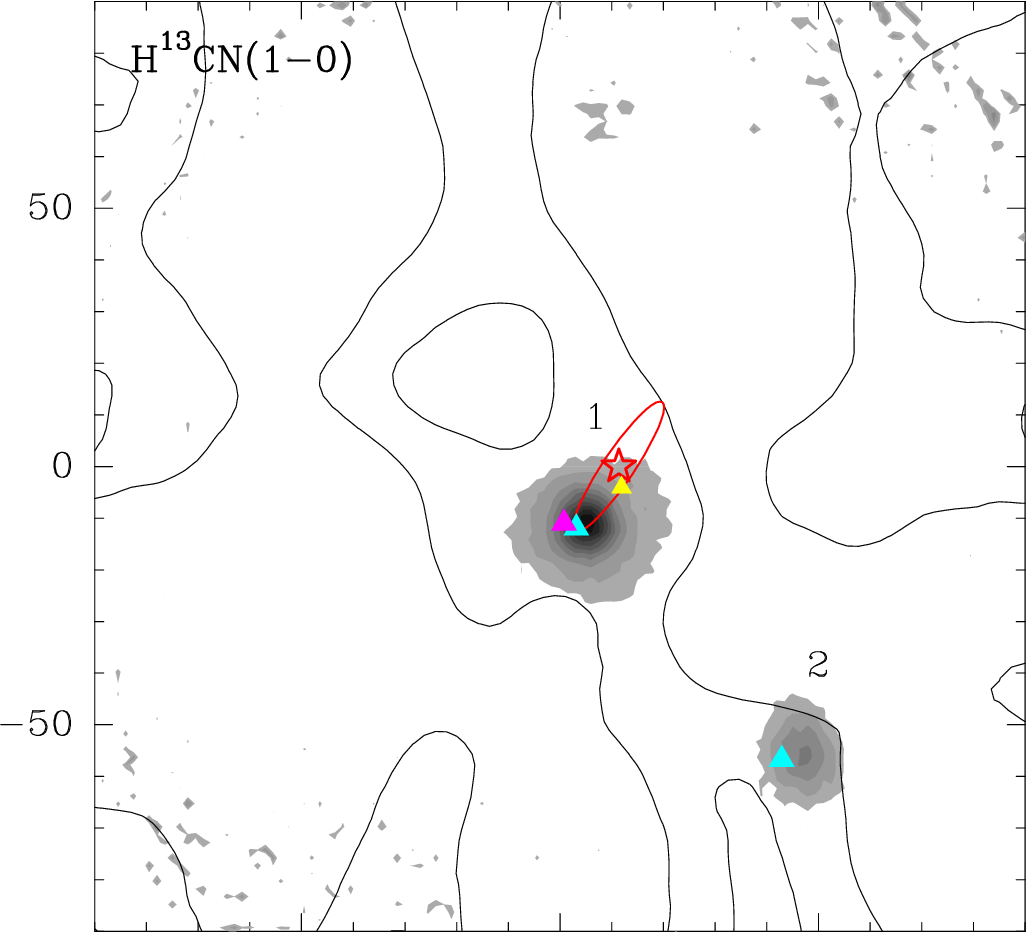}
\end{minipage}
\begin{minipage}[b]{0.3\textwidth}
    \includegraphics[width=\textwidth,angle=-0]{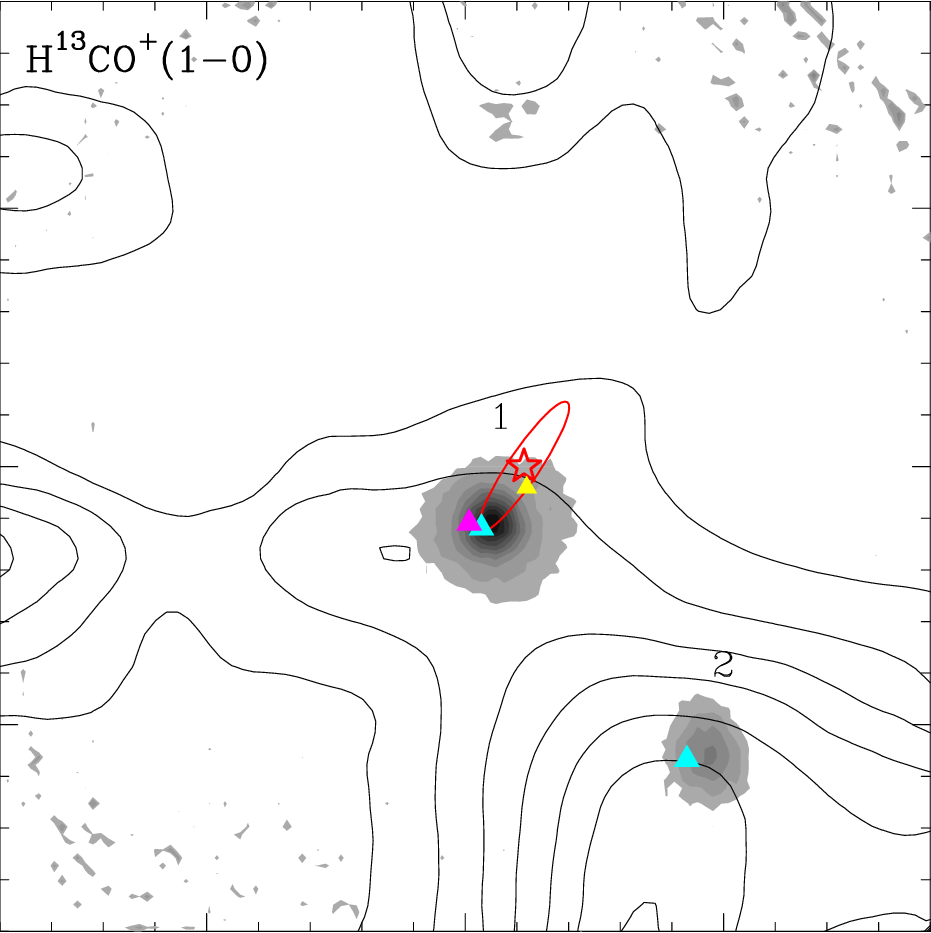}
\end{minipage}
\begin{minipage}[b]{0.3\textwidth}
    \includegraphics[width=\textwidth,angle=-0]{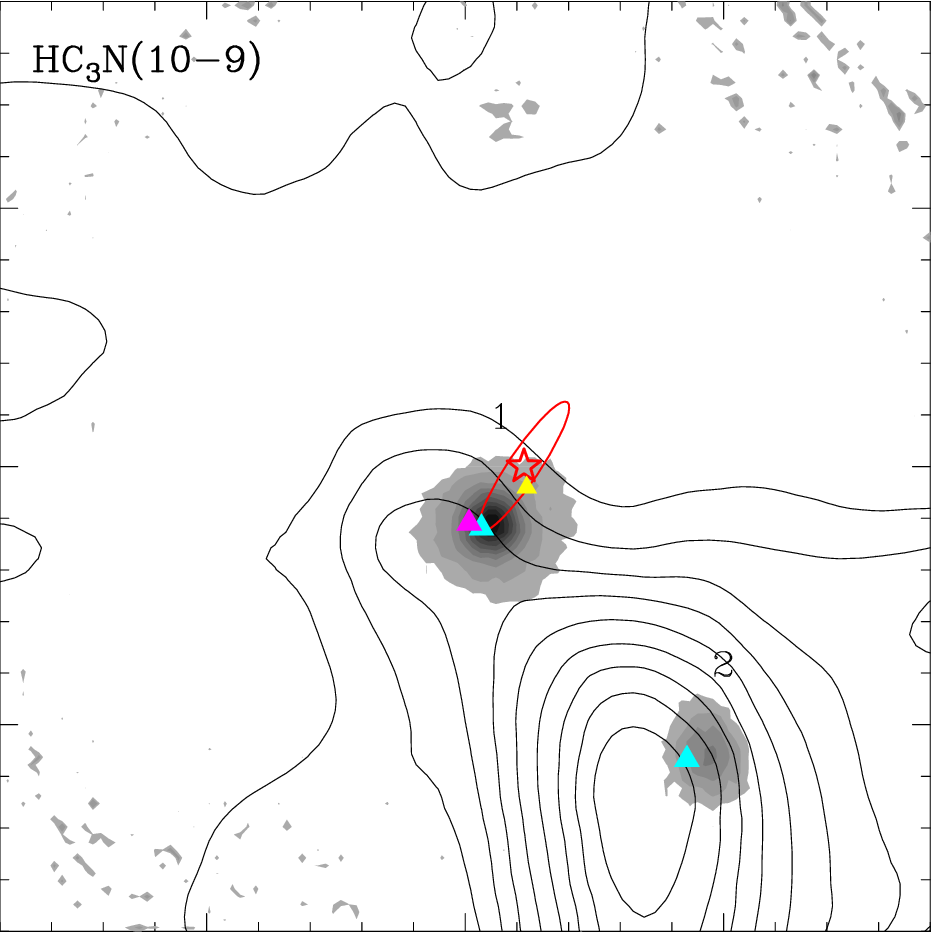}
\end{minipage}

\begin{minipage}[b]{0.33\textwidth}
    \includegraphics[width=\textwidth,angle=-0]{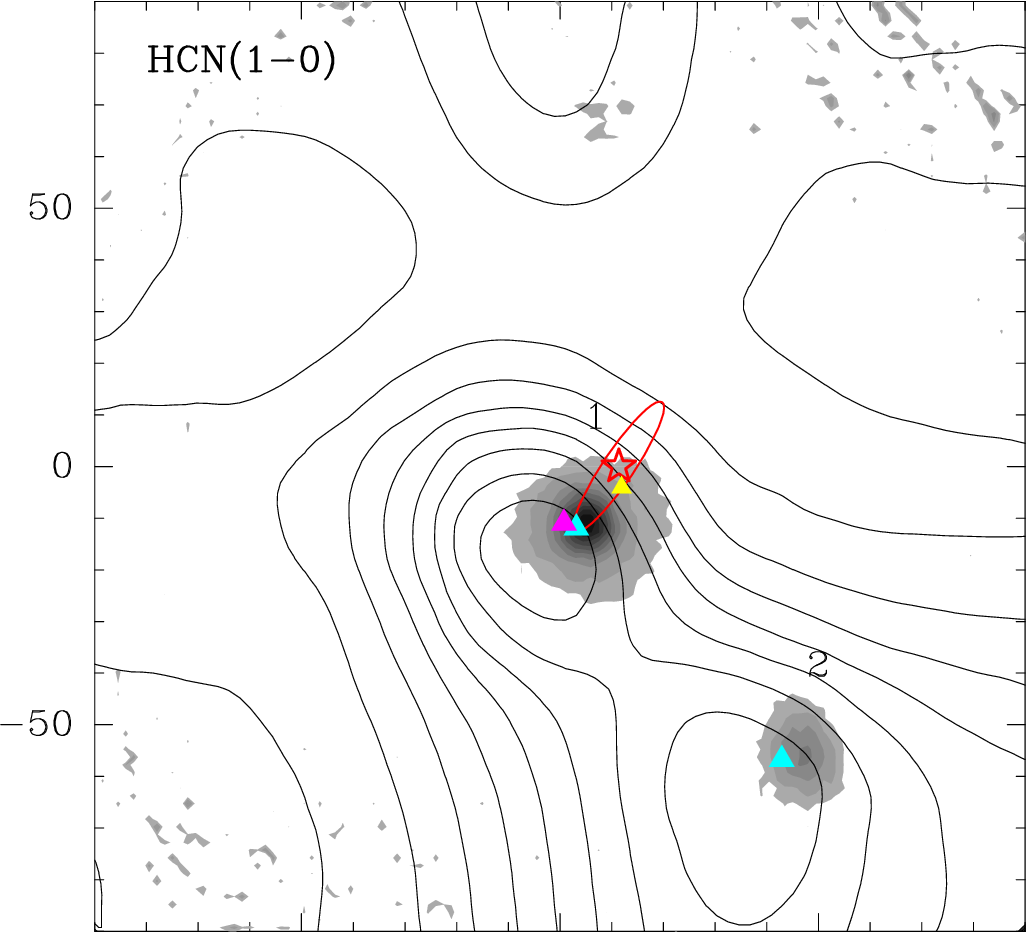}
\end{minipage}
\begin{minipage}[b]{0.3\textwidth}
    \includegraphics[width=\textwidth,angle=-0]{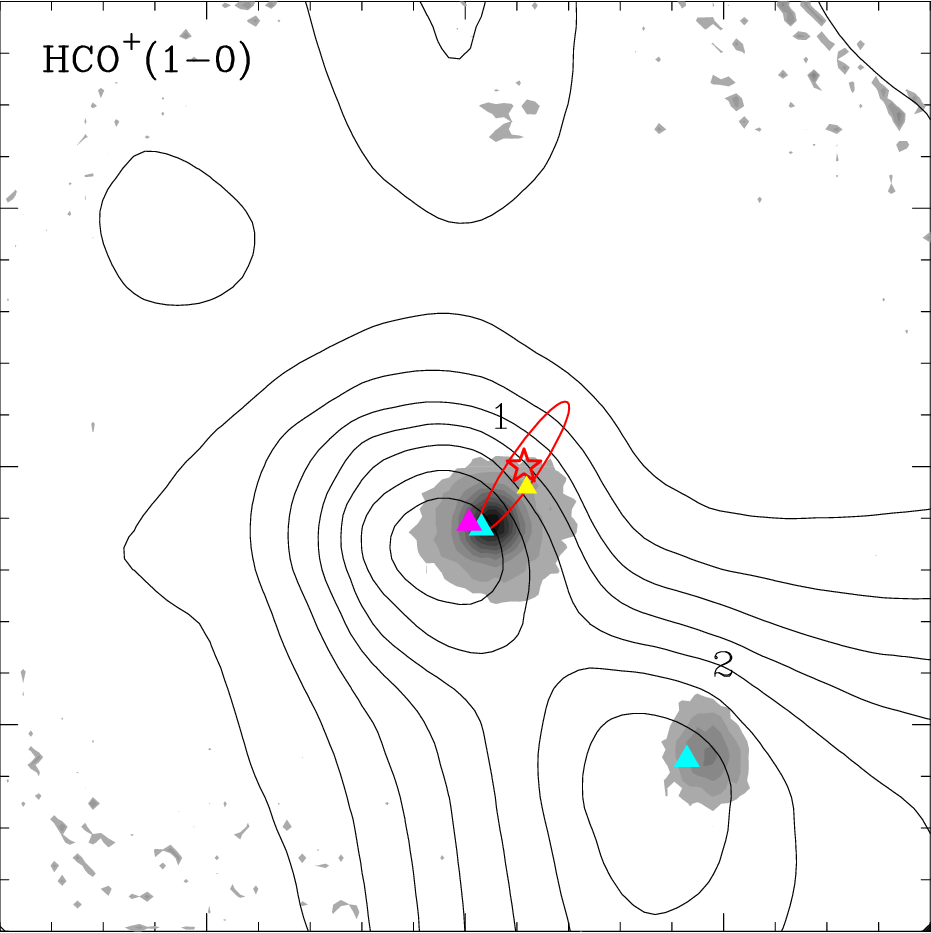}
\end{minipage}
\begin{minipage}[b]{0.3\textwidth}
    \includegraphics[width=\textwidth,angle=-0]{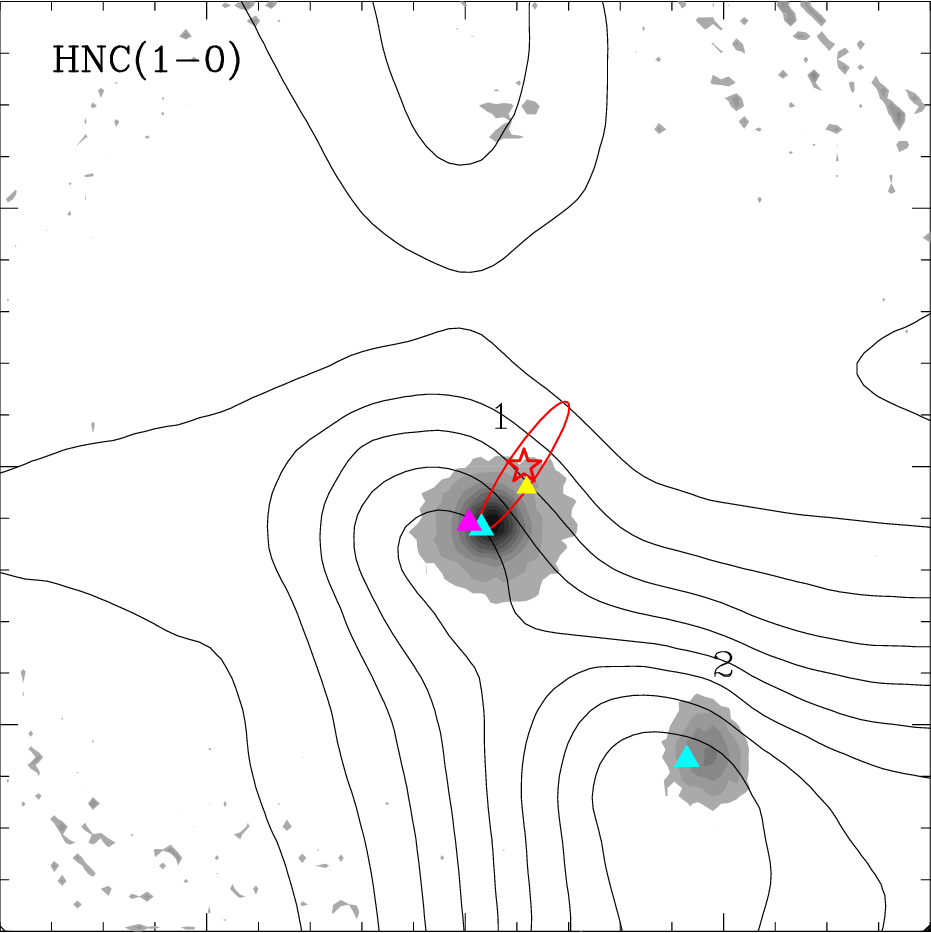}
\end{minipage}

\begin{minipage}[b]{0.33\textwidth}
    \includegraphics[width=\textwidth,angle=-0]{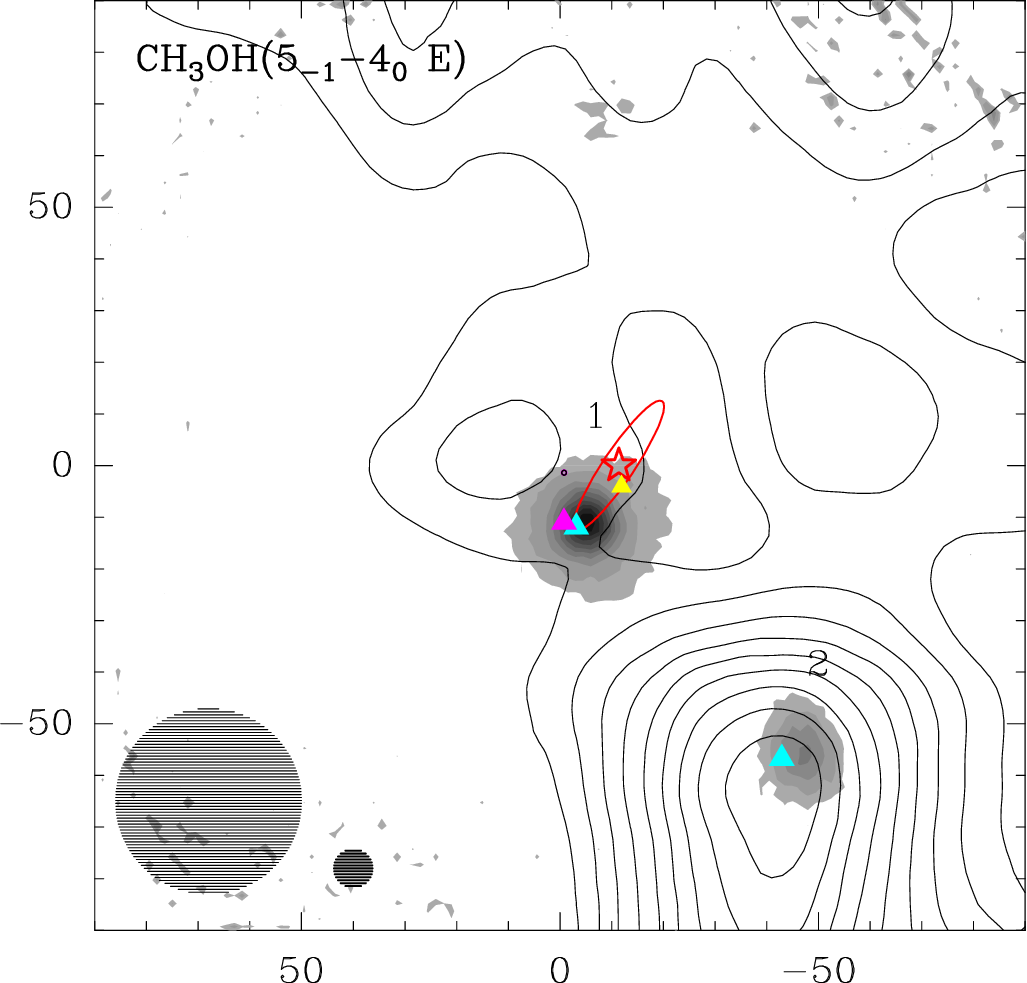}
\end{minipage}
\begin{minipage}[b]{0.3\textwidth}
    \includegraphics[width=\textwidth,angle=-0]{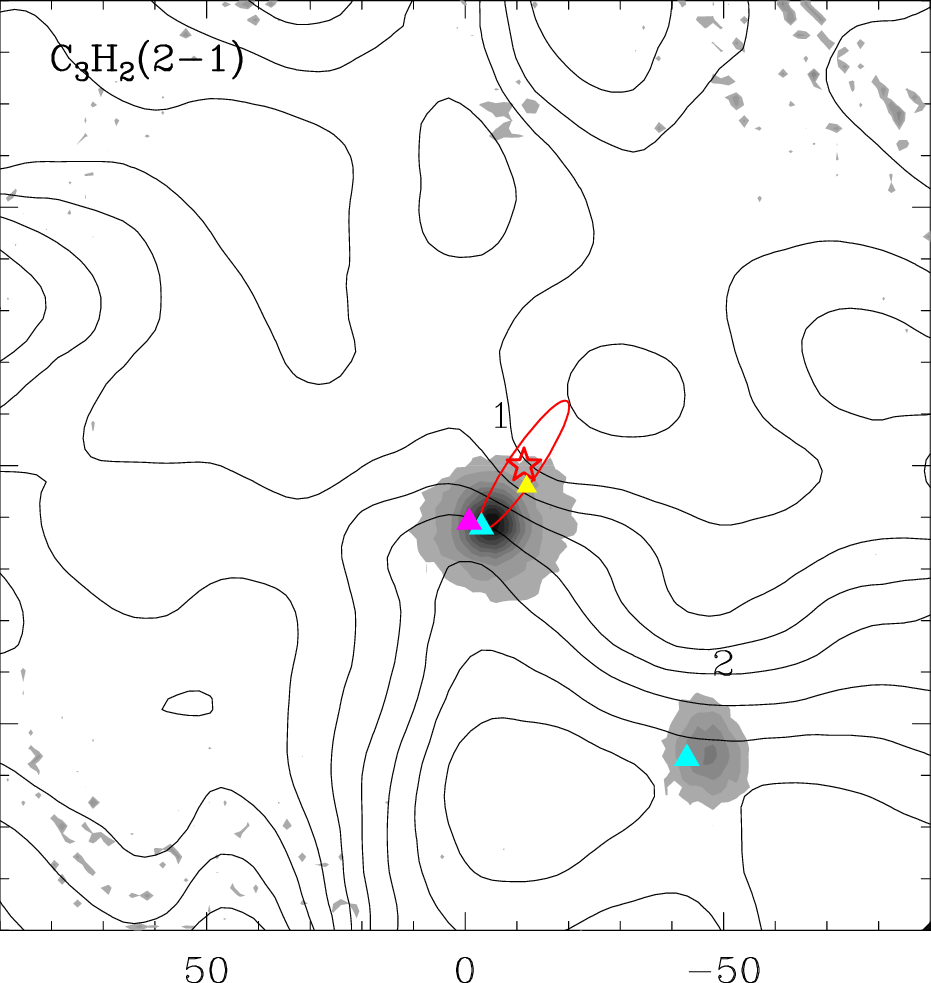}
\end{minipage}
\begin{minipage}[b]{0.3\textwidth}
    \includegraphics[width=\textwidth,angle=-0]{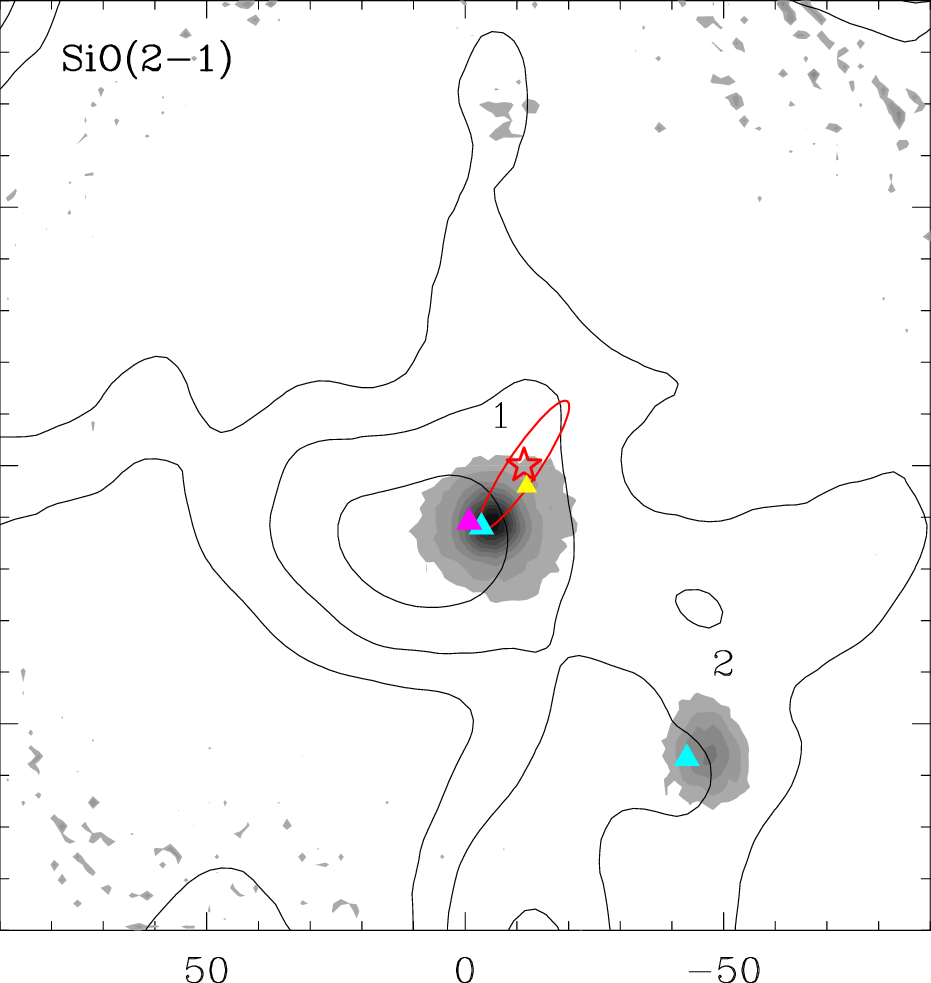}
\end{minipage}

\vskip 20mm

\caption{\scriptsize
Maps of molecular lines observed in G285.26--0.05.
Water masers \cite{Breen10} are indicated by blue triangles.
A class II methanol maser (6.7~GHz) \cite{Gay93}
and OH maser \cite{Caswell98} are indicated by yellow and pink triangles,
respectively. The remaining symbols are the same as in Fig.~\ref{fig:G268}.
The numbers indicate cores whose parameters are calculated separately.
}
\label{fig:G285}
\end{figure}

\begin{figure}[h]

\begin{minipage}[b]{0.33\textwidth}
    \includegraphics[width=\textwidth,angle=-0]{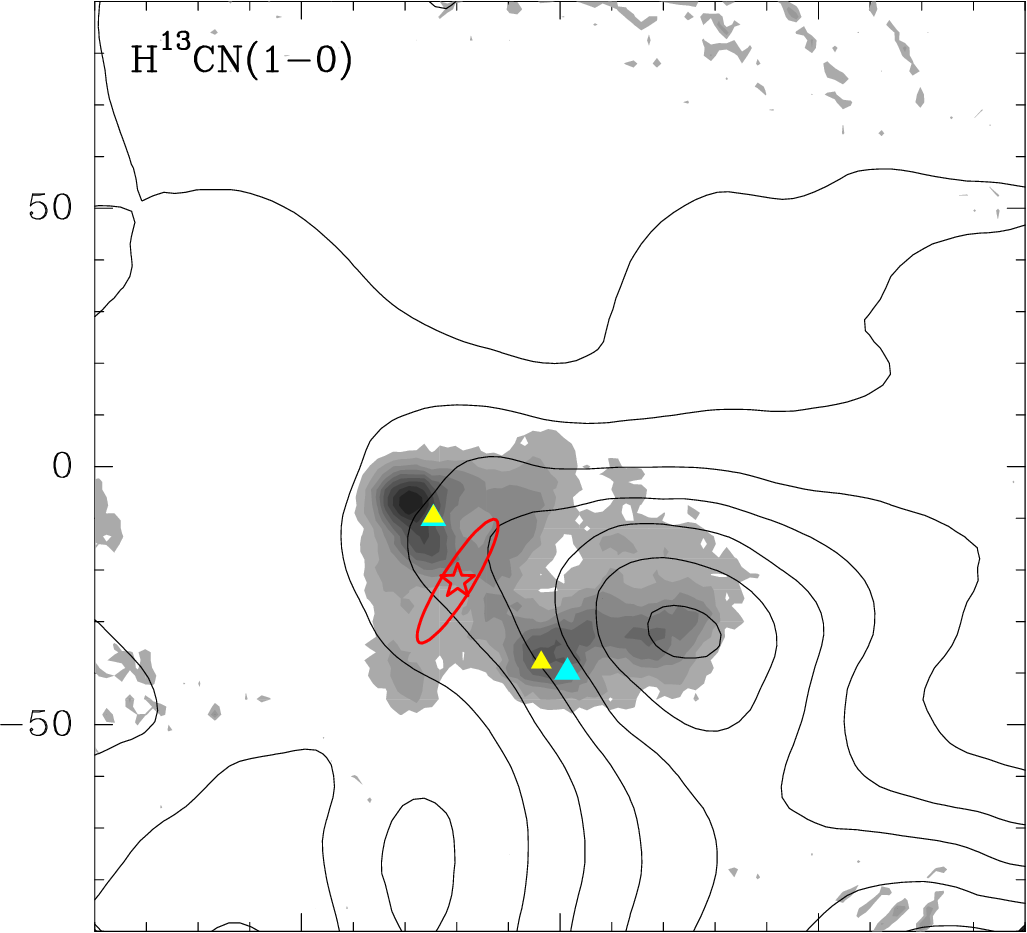}
\end{minipage}
\begin{minipage}[b]{0.3\textwidth}
    \includegraphics[width=\textwidth,angle=-0]{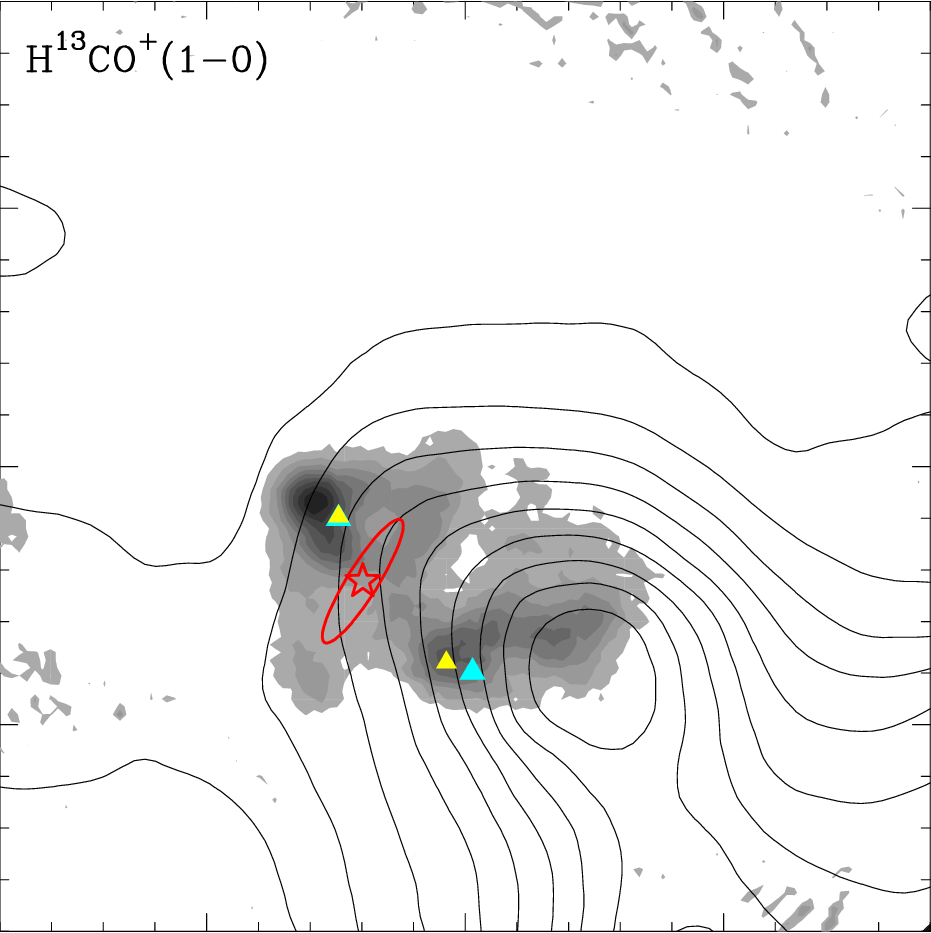}
\end{minipage}
\begin{minipage}[b]{0.3\textwidth}
    \includegraphics[width=\textwidth,angle=-0]{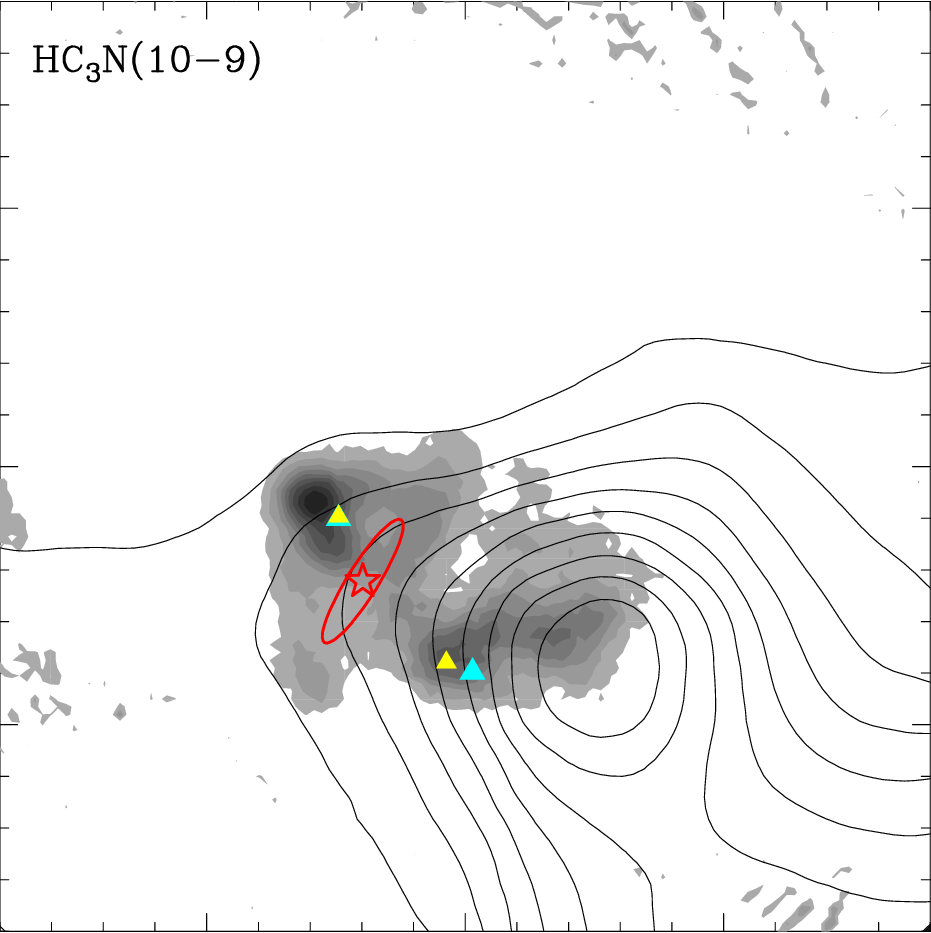}
\end{minipage}

\begin{minipage}[b]{0.33\textwidth}
    \includegraphics[width=\textwidth,angle=-0]{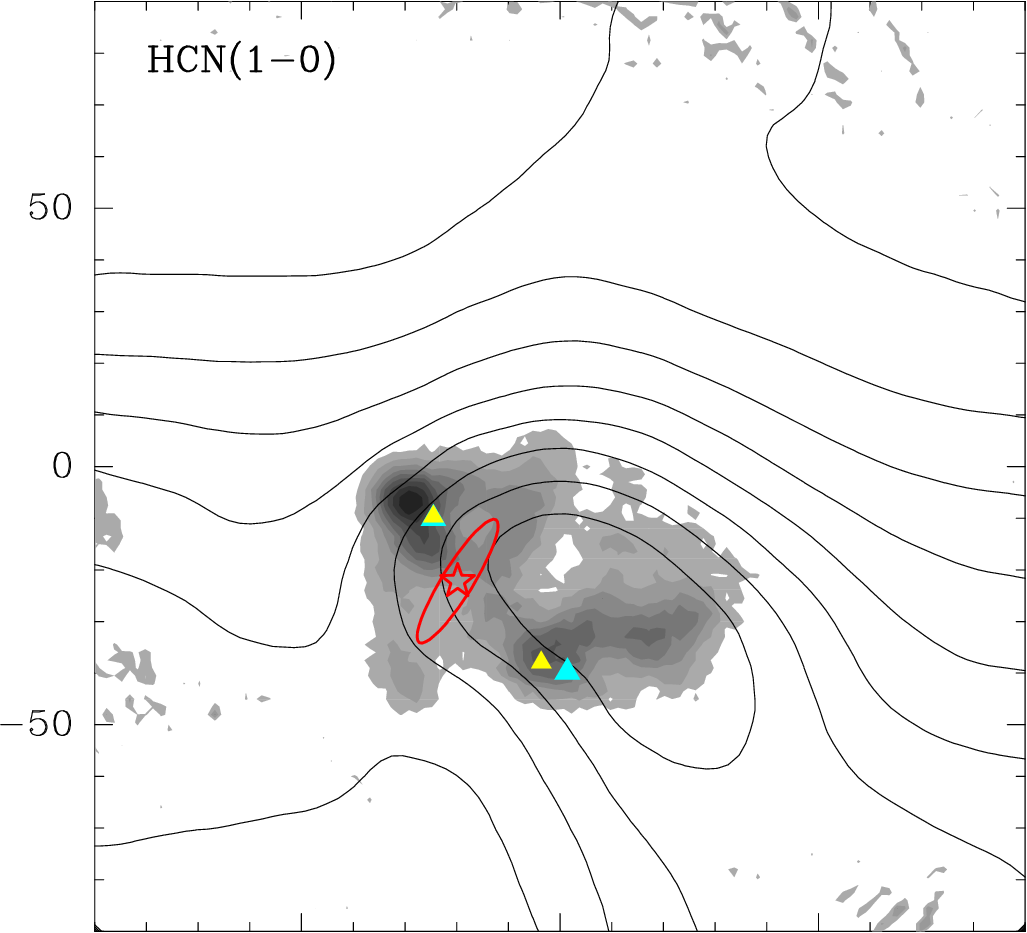}
\end{minipage}
\begin{minipage}[b]{0.3\textwidth}
    \includegraphics[width=\textwidth,angle=-0]{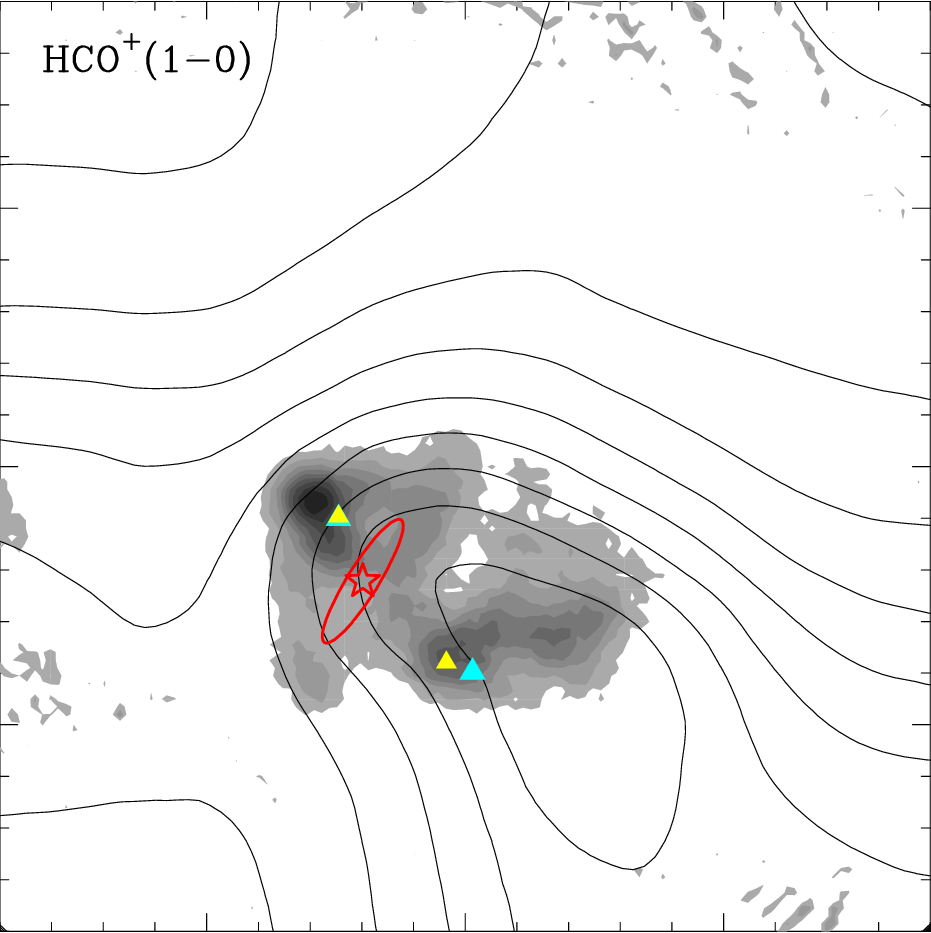}
\end{minipage}
\begin{minipage}[b]{0.3\textwidth}
    \includegraphics[width=\textwidth,angle=-0]{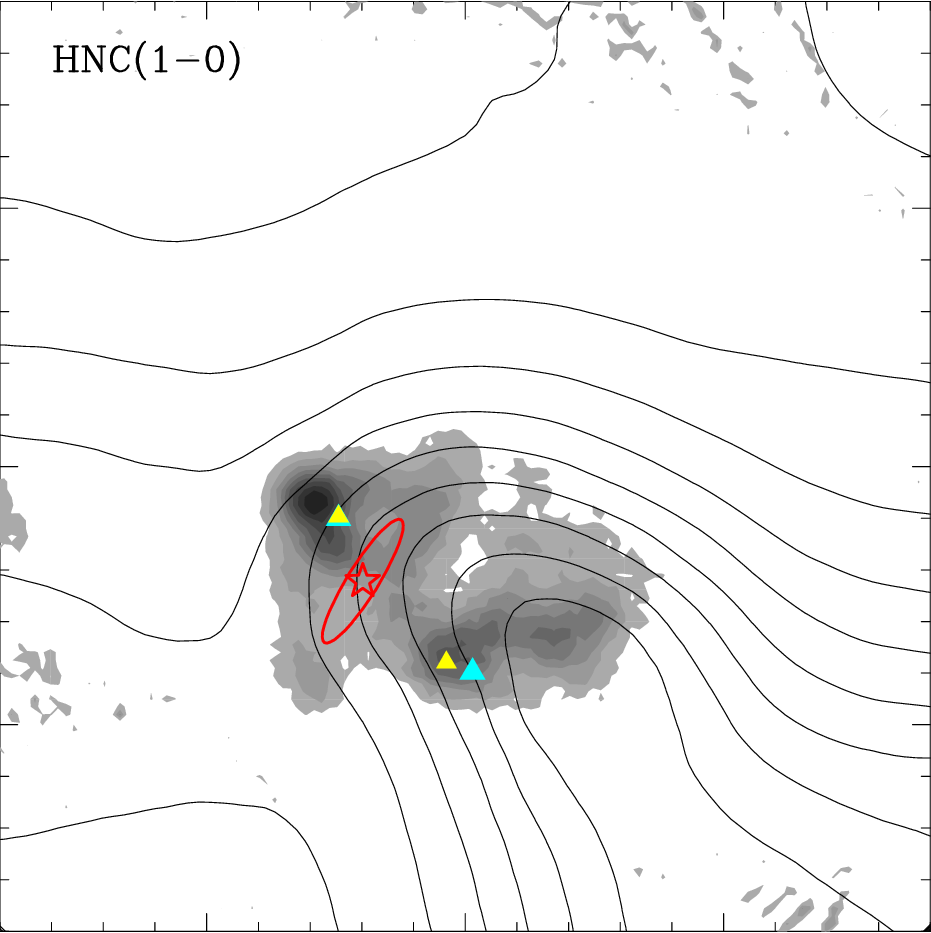}
\end{minipage}

\begin{minipage}[b]{0.33\textwidth}
    \includegraphics[width=\textwidth,angle=-0]{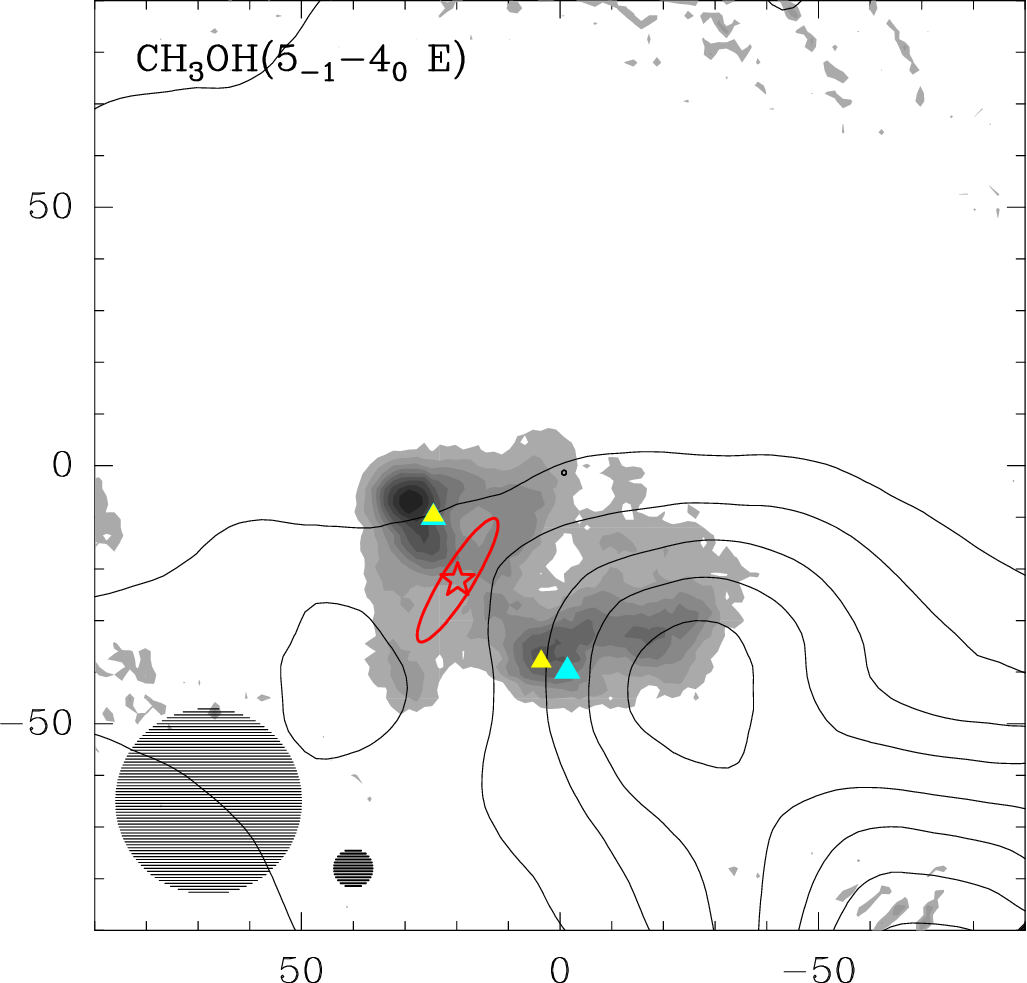}
\end{minipage}
\begin{minipage}[b]{0.3\textwidth}
    \includegraphics[width=\textwidth,angle=-0]{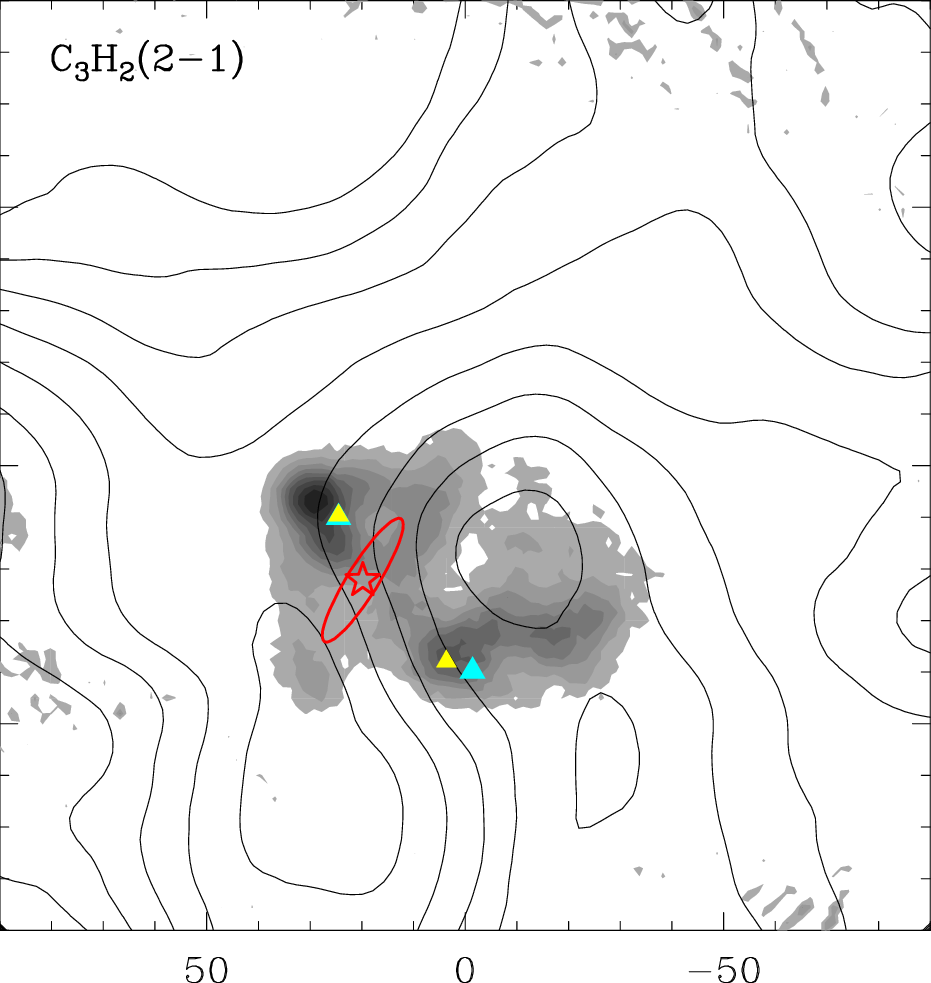}
\end{minipage}
\begin{minipage}[b]{0.3\textwidth}
    \includegraphics[width=\textwidth,angle=-0]{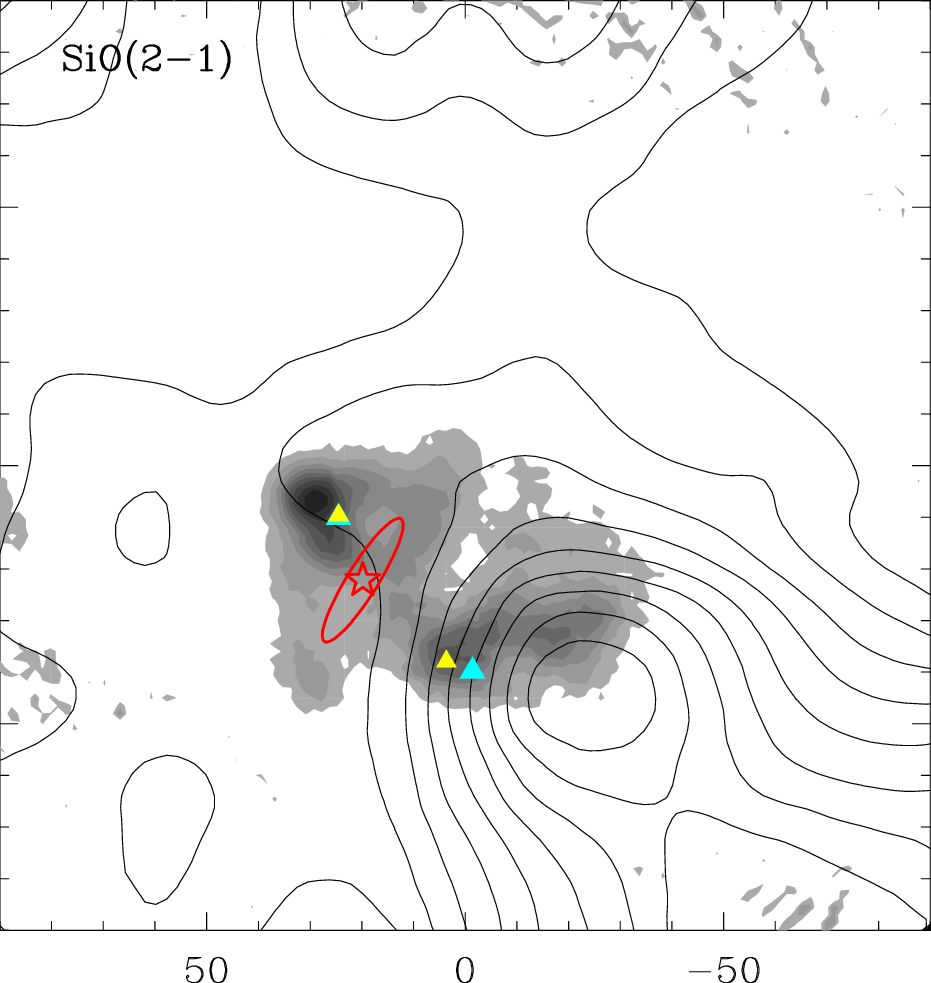}
\end{minipage}

\vskip 20mm

\caption{\scriptsize
Maps of molecular lines observed in G291.27--0.71.
Water masers \cite{Breen10} are indicated by blue triangles, class II methanol
masers (6.7~GHz) \cite{Caswell09} are indicated by yellow triangles.
The remaining symbols are the same as in Fig.~\ref{fig:G268}.
}
\label{fig:G291}
\end{figure}

\begin{figure}[h]

\begin{minipage}[b]{0.33\textwidth}
    \includegraphics[width=\textwidth,angle=-0]{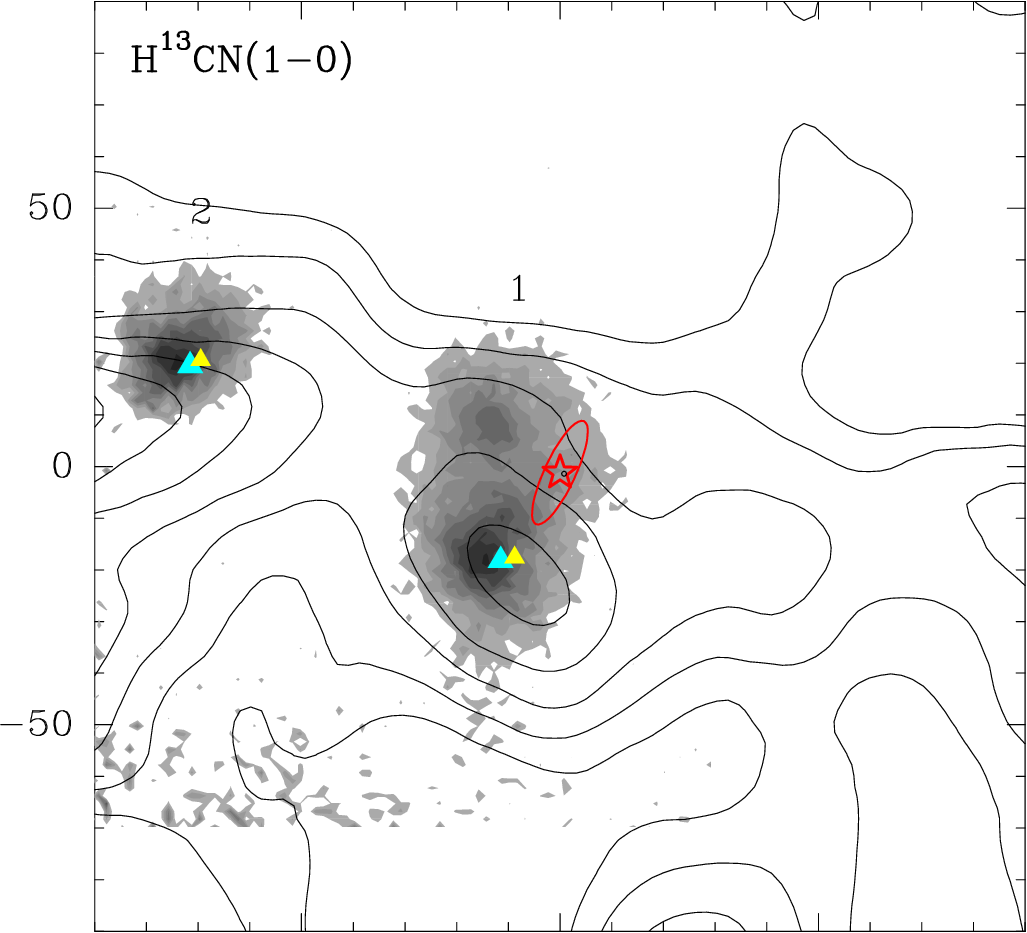}
\end{minipage}
\begin{minipage}[b]{0.3\textwidth}
    \includegraphics[width=\textwidth,angle=-0]{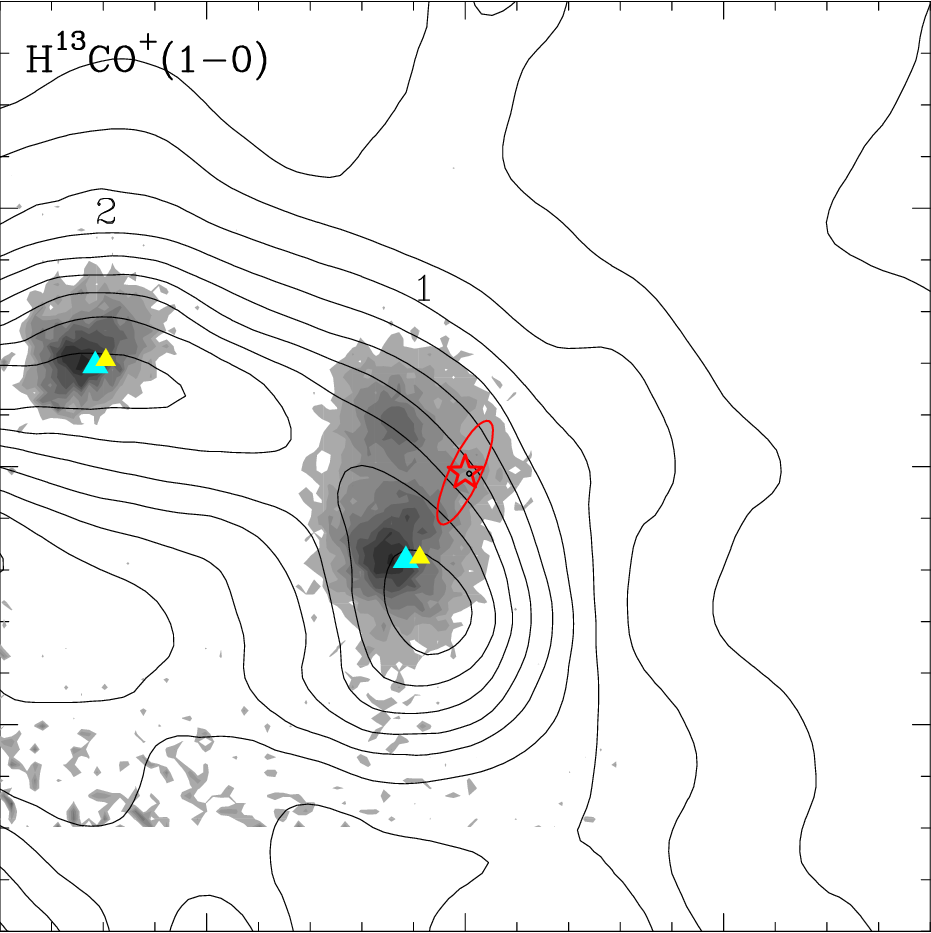}
\end{minipage}
\begin{minipage}[b]{0.3\textwidth}
    \includegraphics[width=\textwidth,angle=-0]{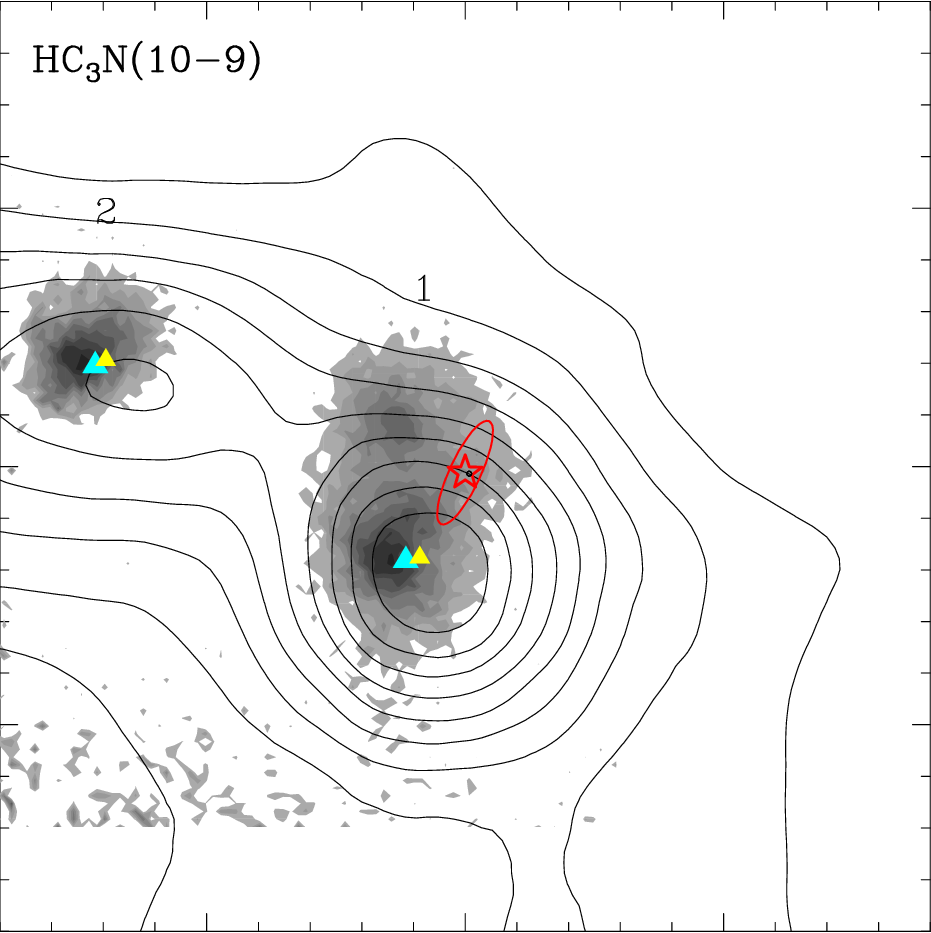}
\end{minipage}

\begin{minipage}[b]{0.33\textwidth}
    \includegraphics[width=\textwidth,angle=-0]{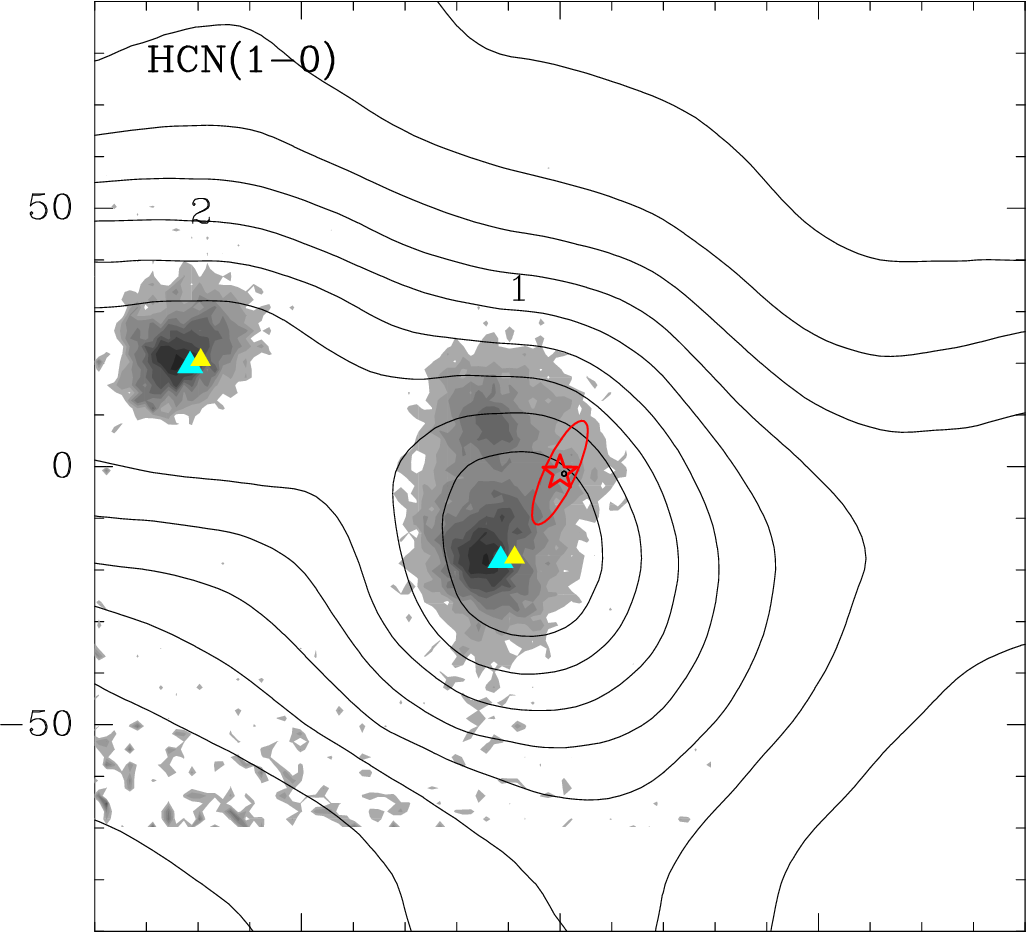}
\end{minipage}
\begin{minipage}[b]{0.3\textwidth}
    \includegraphics[width=\textwidth,angle=-0]{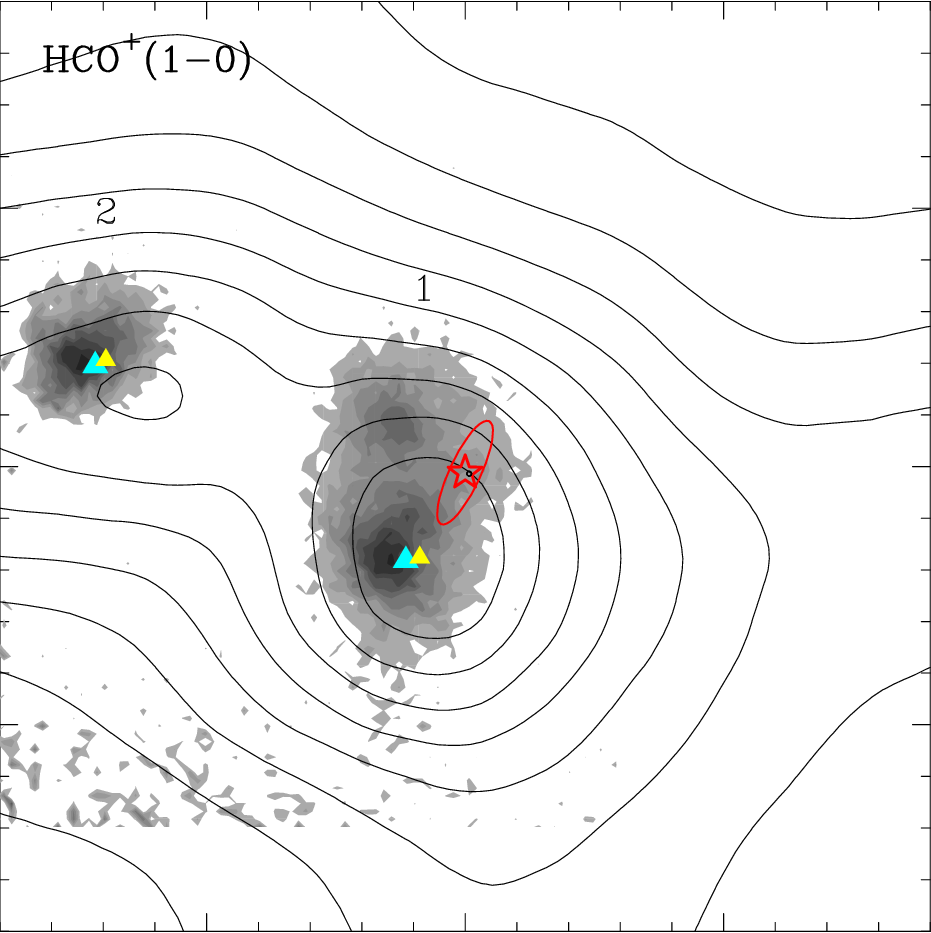}
\end{minipage}
\begin{minipage}[b]{0.3\textwidth}
    \includegraphics[width=\textwidth,angle=-0]{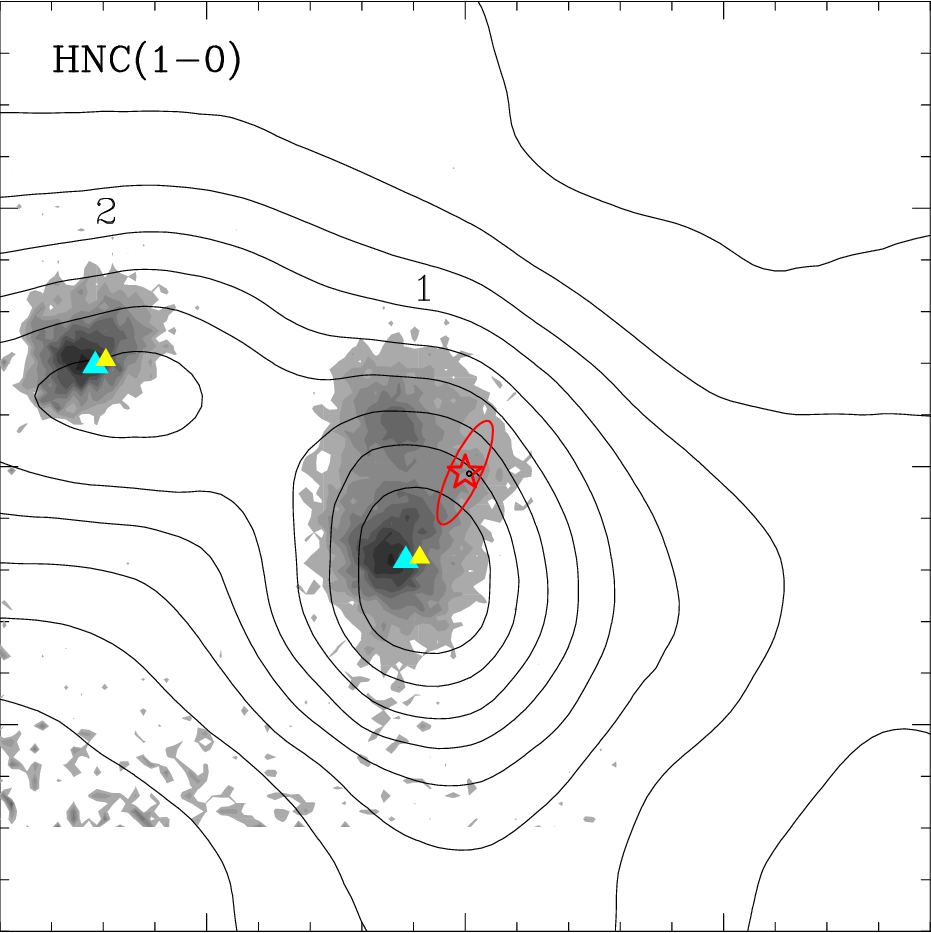}
\end{minipage}

\begin{minipage}[b]{0.33\textwidth}
    \includegraphics[width=\textwidth,angle=-0]{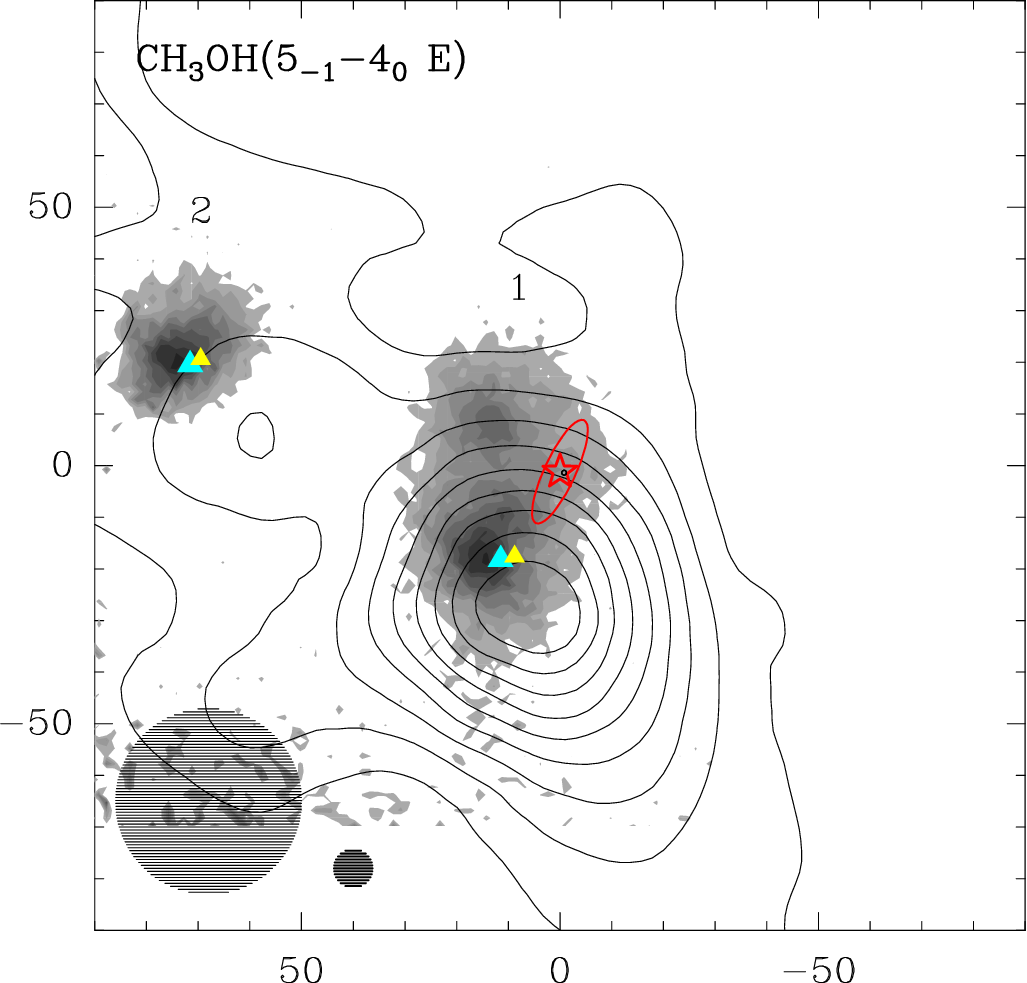}
\end{minipage}
\begin{minipage}[b]{0.3\textwidth}
    \includegraphics[width=\textwidth,angle=-0]{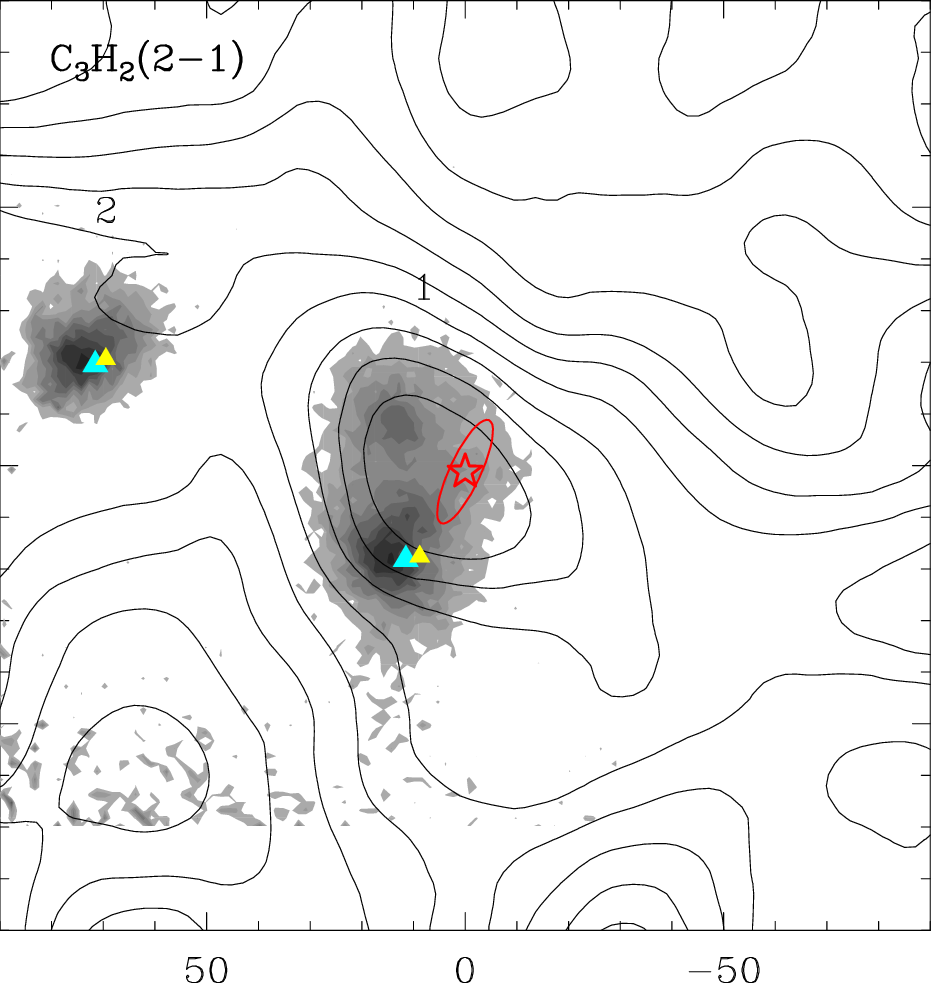}
\end{minipage}
\begin{minipage}[b]{0.3\textwidth}
    \includegraphics[width=\textwidth,angle=-0]{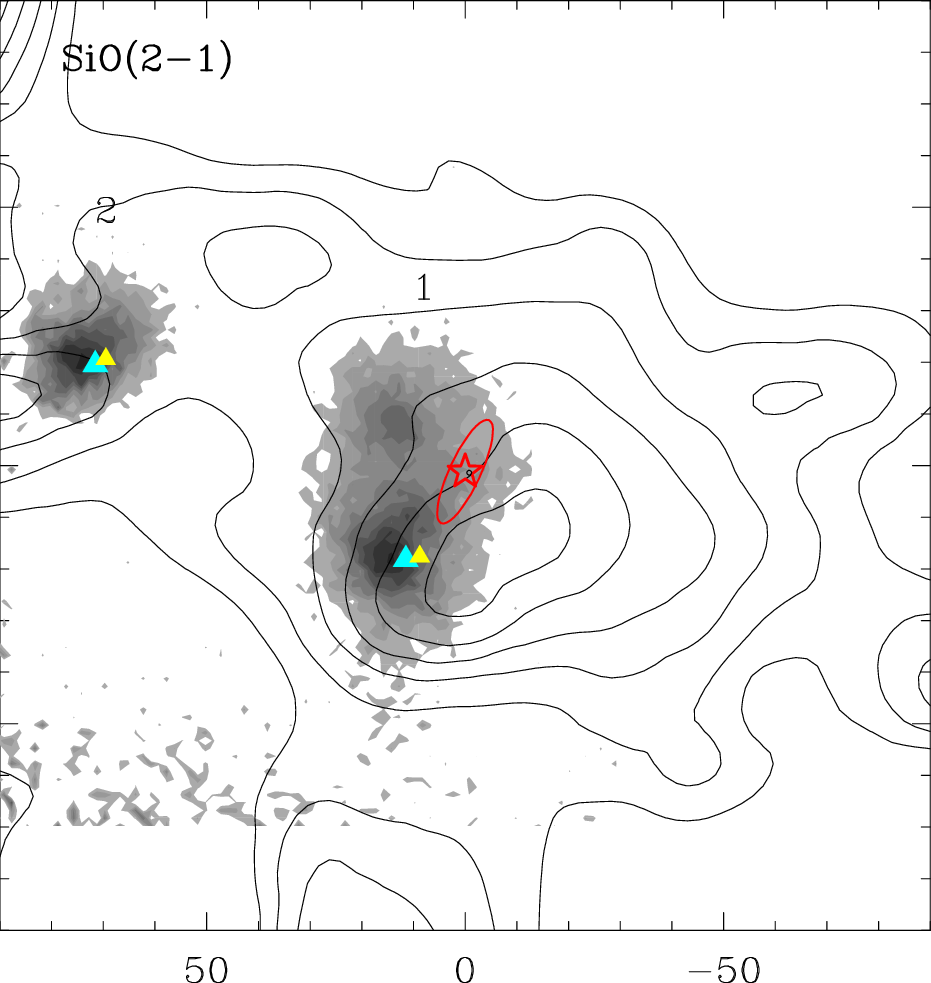}
\end{minipage}

\vskip 20mm

\caption{\scriptsize
Maps of molecular lines observed in G294.97--1.73.
Water masers \cite{Breen11} are indicated by blue triangles, class II methanol
masers (6.7 GHz) \cite{Caswell09,Green12} are indicated by yellow triangles.
The remaining symbols are the same as in Fig.~\ref{fig:G268}.
The numbers indicate cores whose parameters are calculated separately.
}
\label{fig:G294}
\end{figure}

Integrated intensity maps of various molecular
lines are shown in Figs.~\ref{fig:G268},\ref{fig:G269},\ref{fig:G270},\ref{fig:G285},\ref{fig:G291},\ref{fig:G294}.
The figures show the
IRAS sources and the masers of water, class II methanol
and hydroxyl which exist in the cores and are indicators
of the regions of massive star formation.
The figures also show maps of dust emission in the continuum
at a wavelength of 350~$\mu$m \cite{Pir22}, obtained with significantly
better angular resolution and allowing
the internal structure of cores to be estimated in more detail.

In most cases, the emission regions in different
lines are spatially correlated with each other.
The best correlation is observed between
the HCO$^+$(1--0), HCN(1--0), HNC(1--0), and HC$_3$N(10--9) maps
as well as between the maps of the isotopes
H$^{13}$CO$^+$(1--0), H$^{13}$CN(1--0). The maps in the
c-C$_3$H$_2$(2$_{1,2}$--1$_{0,1}$) and SiO(2--1) lines are less correlated
with the others. Continuum maps at 350~$\mu$m
are in most cases within half-intensity molecular
emission contours. The IRAS sources are located both
near the centers of the cores and at the periphery (in
G269.11 and G270.26). Maser sources are observed in
all cores, indicating star formation.

To estimate the sizes of emission regions, the convolution
of a two-dimensional Gaussian elliptical
function with unknown parameters and a two-dimensional
circular Gaussian with a width equal to the
width of the main beam of the telescope radiation pattern
is fitted into integrated intensity maps \cite{Pir03}.
The sizes of the cores are determined as the geometric
mean of the sizes of the fitted elliptic two-dimensional
Gaussian. The centers of the emission regions
are generally close to each other within the size of the
telescope main beam. Their morphology
generally correlates with the morphology of the
350~$\mu$m continuum maps (with the exception of
G291.27), although it should be noted that the latter
are significantly smaller in size. The ratios of the axes
of the fitted ellipses, the angular sizes of the emission
regions in various lines at half the maximum
intensity level are given below in Table~\ref{table:physpar}
along with the errors in their determination.

\newpage

\newpage

\newpage

\newpage

\newpage

\subsection{Line parameters}

We processed spectral data using standard methods
with a help of the GILDAS package
\footnote{http://iram.fr/IRAMFR/GILDAS} and our original programs.
After subtracting baseline from the spectral
subranges with lines, Gaussian functions (one or
more) are fitted (approximated) into the observed
spectra to determine intensity, velocity along
the line of sight corresponding to the center of the line,
and line width at half the maximum intensity (FWHM).
The parameters of the observed molecular
lines in the direction of the positions of the emission
peaks are given in Table~\ref{table:parln}.
When processing the CH$_3$C$_2$H(5--4) and CH$_3$CN(5--4) spectra,
the distances between lines with different quantum numbers $K$
are fixed, and their widths are considered identical.
The same procedure is used to process the spectra of
H$^{13}$CN(1--0), consisting of three close hyperfine components.
For the lines with large optical depth (HCN(1--0), HCO$^+$(1--0),
HNC(1--0)), the profiles of which
are in many cases non-Gaussian, Table~\ref{table:parln} shows only
integrated intensities. The uncertainties of integrated
intensities are calculated as $\Delta T_{MB}\sqrt{N_{ch}}\delta V_{ch}$,
where $\Delta T_{MB}$ is the noise level in the channel without a
line (calculated after subtracting the baseline), $N_{ch}$ is
the number of channels with the line, and $\delta V_{ch}$ is the
velocity resolution (channel width).

\newpage

\begin{table}[p]

\centering
\caption{Parameters of the observed lines}
\vskip 2mm
\scriptsize
\begin{tabular}{l|l|l|l|l|l|l|l|l|}
\noalign{\smallskip}\hline\noalign{\smallskip}
Линия & $I$ & $T_{\rm MB}$ & $V_{\rm LSR}$ & $\Delta V$
      & $I$ & $T_{\rm MB}$ & $V_{\rm LSR}$ & $\Delta V$ \\
      & (K km s$^{-1}$) & (K) & (km s$^{-1}$) & (km s$^{-1}$)
      & (K km s$^{-1}$) & (K) & (km s$^{-1}$) & (km s$^{-1}$) \\
\noalign{\hrule}\noalign{\smallskip}
       & \multicolumn{4}{|c|} {\bf G~268.42$-$0.85 (0$''$,--15$''$)}
       & \multicolumn{4}{|c|} {\bf G~269.11$-$1.12 (15$''$,30$''$)} \\
\noalign{\smallskip}\hline\noalign{\smallskip}

CH$_3$OH(5$_{-1}$--4$_0$~E)
                             & 1.67(0.12) & 0.51(0.04) &2.95(0.12)  & 3.21(0.30)
                             & 4.47(0.15) & 1.38(0.05) &10.25(0.05) & 2.86(0.12) \\
c-C$_3$H$_2$(2$_{1,2}$--1$_{0,1}$)
                             & 1.28(0.14) & 0.37(0.04) & 2.60(0.17) & 3.24(0.42)
                             & 1.72(0.15) & 0.44(0.04) &10.37(0.17) & 3.88(0.42) \\
CH$_3$C$_2$H(5$_0$--$4_0$)
                             & 1.46(0.16) & 0.46(0.05) &3.00(0.09)  &1.94(0.22)
                             & 1.44(0.19) & 0.23(0.03) &10.10(0.18) &2.98(0.40) \\
H$^{13}$CN(1--0~$F$=2--1)
                             & 3.36(0.21) & 0.61(0.04) &2.98(0.08)  &2.81(0.18)
                             & 4.12(0.20) & 0.75(0.04) &10.40(0.08) &3.06(0.17) \\
H$^{13}$CO$^+$(1--0)
                             & 2.53(0.11) & 0.96(0.05) &2.99(0.06)  &2.73(0.16)
                             & 2.65(0.11) & 0.75(0.04) &10.29(0.08) &3.40(0.20) \\
SiO(2--1)
                             & 1.00(0.13) & 0.21(0.03) &3.02(0.34)  &4.56(0.90)
                             & 0.77(0.12) & 0.31(0.04) &10.54(0.18) &2.67(0.45) \\
HN$^{13}$C(1--0)
                             & 0.91(0.11) & 0.34(0.05) &3.22(0.16)  &2.40(0.38)
                             & 1.10(0.14) & 0.39(0.05) &10.32(0.15) &2.68(0.38) \\
HCN(1--0~$F$=2--1)
                             & 38.6(0.2)  &            &            &
                             & 34.6(0.3)  &            &            &           \\
HCO$^+$(1--0)
                             & 20.5(0.2)  &            &            &
                             & 26.9(0.2)  &            &            &           \\
HNC$^+$(1--0)
                             & 15.0(0.1)  &            &            &
                             & 16.5(0.2)  &            &            &           \\
HC$_3$N(10--9)
                             & 3.53(0.10) & 1.40(0.04) &2.82(0.03)  &2.43(0.08)
                             & 5.21(0.12) & 1.68(0.04) &10.21(0.03) &2.90(0.08) \\
CH$_3$CN(5$_0$--4$_0$)
                             & 0.75(0.09) & 0.11(0.02) &3.05(0.30)  &4.11(0.58)
                             & 1.04(0.13) & 0.16(0.02) &10.55(0.23) &3.83(0.46) \\
\noalign{\smallskip}\hline\noalign{\smallskip}
       & \multicolumn{4}{|c|} {\bf G~270.26$+$0.83 (--15$''$,30$''$)}
       & \multicolumn{4}{|c|} {\bf  G~291.27$-$0.71 (--30$''$,--45$''$)} \\
\noalign{\smallskip}\hline\noalign{\smallskip}

CH$_3$OH(5$_{-1}$--4$_0$~E)
                             & 6.78(0.10) & 2.22(0.04) &10.04(0.02) & 2.74(0.06)
                             & 2.77(0.13) & 0.91(0.04) &--23.58(0.06)&2.72(0.15)  \\
c-C$_3$H$_2$(2$_{1,2}$--1$_{0,1}$)
                             & 1.02(0.10) & 0.31(0.04) &10.01(0.17) & 3.13(0.43)
                             & 2.57(0.18) & 0.48(0.03) &--23.06(0.16)&5.00(0.39)  \\
CH$_3$C$_2$H(5$_0$--$4_0$)
                             & 0.84(0.13) & 0.14(0.03) &9.20(0.26)  &3.36(0.54)
                             & 3.35(0.21) & 0.42(0.04) &--23.96(0.09)&2.80(0.22)  \\
H$^{13}$CN(1--0~$F$=2--1)
                             & 3.12(0.16) & 0.50(0.03) &9.66(0.10)  &3.29(0.19)
                             & 4.08(0.22) & 0.61(0.03) &--23.95(0.09)&3.31(0.16)  \\
H$^{13}$CO$^+$(1--0)
                             & 2.06(0.10) & 0.67(0.04) &9.62(0.07)  &2.85(0.18)
                             & 3.66(0.11) & 1.11(0.04) &--23.72(0.05)& 3.09(0.12) \\
SiO(2--1)
                             & 1.33(0.10) & 0.38(0.03) &10.28(0.16) &4.24(0.42)
                             & 2.42(0.14) & 0.50(0.03) &--23.42(0.13)& 4.42(0.32) \\
HN$^{13}$C(1--0)
                             & 0.75(0.10) & 0.21(0.03) &9.64(0.28)  &3.81(0.73)
                             & 1.41(0.13) & 0.46(0.04) &--23.57(0.11)& 2.82(0.28) \\
HCN(1--0~$F$=2--1)
                             & 28.9(0.2)  & 7.5(0.1)   &10.02(0.01) &2.15(0.03)
                             & 54.4(0.3)  & 5.50(0.07) &--23.37(0.03)& 4.33(0.05) \\
HCO$^+$(1--0)
                             & 25.9(0.2)  &10.6(0.1)   & 9.92(0.01) &2.44(0.03)
                             & 45.2(0.2)  & 8.20(0.05) &--23.10(0.01)& 5.04(0.03) \\
HNC$^+$(1--0)
                             & 13.0(0.1)  & 4.6(0.1)   & 9.86(0.01) &2.55(0.03)
                             & 22.4(0.1)  & 5.00(0.03) &--23.46(0.01)& 4.09(0.03) \\
HC$_3$N(10--9)
                             & 2.58(0.10) & 0.75(0.03) &9.70(0.07)  &3.46(0.18)
                             & 5.73(0.12) & 2.01(0.04) &--23.97(0.03)& 2.59(0.06) \\
CH$_3$CN(5$_0$--4$_0$)
                             & 1.16(0.12) & 0.16(0.02) &9.54(0.17)  &3.30(0.35)
                             & 1.62(0.14) & 0.23(0.03) &--24.03(0.12)& 3.00(0.27) \\
\noalign{\smallskip}\hline\noalign{\smallskip}
\end{tabular}
\end{table}

\newpage

\addtocounter{table}{-1}

\begin{table}[p]

\centering
\caption{Parameters of the observed lines (continued)}
\vskip 2mm
\scriptsize
\begin{tabular}{l|l|l|l|l|l|l|l|l|l}
\noalign{\smallskip}\hline\noalign{\smallskip}
Линия  & $I$ & $T_{\rm MB}$ & $V_{\rm LSR}$ & $\Delta V$
       & $I$ & $T_{\rm MB}$ & $V_{\rm LSR}$ & $\Delta V$ \\
       & (K km s$^{-1}$) & (K) & (km s$^{-1}$) & (km s$^{-1}$)
       & (K km s$^{-1}$) & (K) & (km s$^{-1}$) & (km s$^{-1}$) \\
\noalign{\smallskip}\hline\noalign{\smallskip}
       & \multicolumn{4}{|c|} {\bf G~285.26$-$0.05 (0$''$,--15$''$)}
       & \multicolumn{4}{|c|} {\bf G~285.26$-$0.05 (--45$''$,--75$''$)} \\
\noalign{\smallskip}\hline\noalign{\smallskip}

CH$_3$OH(5$_{-1}$--4$_0$~E)
                             & $<$0.4     &            &            &
                             & 1.92(0.18) & 0.50(0.05) &1.37(0.21)  &4.53(0.55)  \\
c-C$_3$H$_2$(2$_{1,2}$--$1_{0,1}$)
                             & 1.28(0.15) & 0.22(0.03) &2.28(0.45)  &7.19(1.34)
                             & 1.62(0.19) & 0.44(0.05) &2.84(0.21)  &3.64(0.53)  \\
CH$_3$C$_2$H(5$_0$-$4_0$)
                             & $<0.9$     &            &            &
                             & $<1.1$     &            &            &            \\
H$^{13}$CN(1--0~$F$=2--1)
                             & 0.92(0.22) & 0.12(0.04) &2.68(1.11)  &5.26(2.05)
                             & 1.19(0.31) & 0.34(0.06) &2.64(0.19)  &2.26(0.46)  \\
H$^{13}$CO$^+$(1--0)
                             & 0.57(0.13) & 0.25 (0.05)&3.45(0.24)  & 2.31(0.58)
                             & 1.33(0.14) & 0.45(0.05) &2.53(0.16)  & 2.84(0.40) \\
SiO(2--1)
                             & 0.97(0.13) & 0.1 (0.03) &4.23(0.48)  & 6.47(1.38)
                             & $<$0.5 \\
HN$^{13}$C(1--0)
                             & $<0.4$     &            &            &
                             & $<0.5$     &            &            &            \\
HCN(1--0)
                             & 32.9(0.2)  & 3.82(0.04) & 2.93(0.03)  & 4.37(0.05)
                             & 31.0(0.3)  & 4.01(0.05) & 2.55(0.03)  & 3.92(0.05) \\
HCO$^+$(1--0)
                             & 27.0(0.2)  & 5.17(0.05) & 3.05(0.02)  & 4.78(0.05)
                             & 25.2(0.2)  & 5.55(0.05) & 2.55(0.02)  & 4.29(0.05) \\
HNC$^+$(1--0)
                             &  6.8(0.2)  & 1.52(0.03) & 3.14(0.05)  & 4.28(0.12)
                             & 10.5(0.2)  & 2.78(0.05) & 2.51(0.03)  & 3.47(0.07) \\
HC$_3$N(10--9)
                             & 0.65(0.11) & 0.19(0.03) & 3.21(0.34)  & 4.48(0.88)
                             & 1.63(0.14) & 0.64(0.05) & 2.07(0.10)  & 2.56(0.26) \\
CH$_3$CN(5$_0$--4$_0$)
                             & $<0.9$     &            &            &
                             & $<0.9$     &            &            &            \\

\noalign{\hrule}\noalign{\smallskip}
       & \multicolumn{4}{|c|} {\bf G~294.97$-$1.73 (15$''$,--15$''$)}
       & \multicolumn{4}{|c|} {\bf G~294.97$-$1.73 (60$''$,15$''$)} \\
\noalign{\smallskip}\hline\noalign{\smallskip}

CH$_3$OH(5$_{-1}$--4$_0$~E)
                             & 1.42(0.12) & 0.63(0.05) &--8.08(0.08) &2.10(0.20)
                             & 0.51(0.12) & 0.25(0.05) &--8.47(0.18) &1.79(0.44) \\
c-C$_3$H$_2$(2$_{1,2}$--$1_{0,1}$)
                             & 1.61(0.14) & 0.51(0.04) &--8.83(0.12) &2.82(0.29)
                             & 1.07(0.14) & 0.37(0.04) &--9.35(0.16) &2.92(0.41) \\
CH$_3$C$_2$H(5$_0$-$4_0$)
                             & 1.09(0.14) & 0.26(0.04) &--8.14(0.11) &1.89(0.26)
                             & $<0.6$ \\
H$^{13}$CN(1--0~$F$=2--1)
                             & 1.23(0.19) & 0.41(0.05) &--8.31(0.09) &1.73(0.23)
                             & 1.19(0.19) & 0.28(0.04) &--8.89(0.17) &2.41(0.44) \\
H$^{13}$CO$^+$(1--0)
                             & 1.25(0.11) & 0.64(0.05) &--8.39(0.07) &1.82(0.17)
                             & 1.29(0.10) & 0.51(0.05) &--9.00(0.11) &2.49(0.27) \\
SiO(2--1)
                             & 0.47(0.11) & 0.14(0.04) &--8.17(0.38) &3.23(0.98)
                             & $<$0.4 \\
HN$^{13}$C(1--0)
                             & 0.72(0.13) & 0.46(0.06) &--8.26(0.10) &1.57(0.25)
                             & 0.48(0.12) & 0.35(0.06) &--8.53(0.12) &1.44(0.30) \\
HCN(1--0)
                             & 31.5(0.2)  & 5.45(0.05) &--8.49(0.01) &3.22(0.03)
                             & 25.5(0.2)  & 4.36(0.04) &--9.22(0.01) &3.18(0.03) \\
HCO$^+$(1--0)
                             & 24.9(0.2)  & 6.73(0.05) &--8.48(0.01) &3.38(0.03)
                             & 20.9(0.2)  & 6.14(0.05) &--9.16(0.01) &3.15(0.03) \\
HNC$^+$(1--0)
                             & 12.2(0.1)  & 3.91(0.04) &--8.43(0.01) &2.90(0.03)
                             &  9.4(0.1)  & 3.02(0.04) &--9.18(0.02) &2.83(0.05) \\
HC$_3$N(10--9)
                             & 2.76(0.10) & 1.32(0.05) &--8.30(0.03) &1.89(0.08)
                             & 1.81(0.11) & 0.89(0.05) &--8.90(0.05) &1.89(0.12) \\
CH$_3$CN(5$_0$--4$_0$)
                             & $<0.5$     &            &            &
                             & $<0.5$     \\

\noalign{\smallskip}\hline\noalign{\smallskip}
\end{tabular}

\flushleft
{\scriptsize
The CH$_3$C$_2$H(5$_0$--4$_0$) parameters are calculated for the spectra
averaged over the following regions:
$\Delta\alpha$: [$0''\div 30''$], $\Delta\delta$: [$15''\div 45''$] (G269.11);
$\Delta\alpha$: [$-15''\div 0''$], $\Delta\delta$: [$0''\div 30''$] (G270.26);
$\Delta\alpha$: [$-15''\div 15''$], $\Delta\delta$: [$-30''\div 0''$] (G294.97).

The CH$_3$CN(5$_0$--4$_0$) parameters
are calculated for the spectra averaged over the following regions:
$\Delta\alpha$: [$-30''\div 30''$] and $\Delta\delta$: [$-30''\div 30''$] (G268.42);
$\Delta\alpha$: [$0''\div 30''$], $\Delta\delta$: [$15''\div 45''$] (G269.11);
$\Delta\alpha$: [$-15''\div -30''$] and$\Delta\delta$: [$-30''\div 45''$] (G291.27).

The integrated intensities of CH$_3$C$_2$H(5--4),
H$^{13}$CN(1--0), HCN(1--0) and CH$_3$CN(5--4)
are calculated using all components of the spectra.
}
\label{table:parln}
\end{table}

\section{Physical parameters of cores}

\subsection{Kinetic temperatures}

The methyl acetylene (CH$_3$C$_2$H) and methyl cyanide
(CH$_3$CN) lines can be used to determine kinetic
temperatures in dense cores. For molecules of the
symmetric top type, such as these molecules,
transitions with different quantum numbers consist
of several lines with different values of the quantum
number $K$, which determines the projection of the
angular momentum on the axis of symmetry of the
molecule. Approximating the dependence of the integrated
intensities of lines with different $K$ on energy
levels with a straight line, the rotational temperature can
be determined by the slope coefficient and used as an
estimate of the kinetic temperature (see, for example, \cite{Mal}).
Kinetic temperatures are calculated by this
method for five cores (see Table~\ref{table:tkin}), in which at least
two CH$_3$C$_2$H(5--4) lines with different $K$ values
are reliably detected. Three cores are selected for
estimates using the CH$_3$CN(5--4) lines (see Table~\ref{table:tkin}).
All these lines appear to be optically thin. This is confirmed
by the fact that their intensities lie in the intensity
ranges of the corresponding lines observed in
regions of formation of massive stars, the optical depth
of which is small (see, for example, \cite{Alakoz02,Araya05}).
The product of the $T_{MB}$ and $\Delta V$ values,
which are determined from the approximation, is taken as the
integrated intensities of lines with different $K$.
The kinetic temperature errors are calculated by the method of
error propagation from approximation errors. Due to
the relatively low signal-to-noise ratios in the observations
of CH$_3$C$_2$H(5--4) and CH$_3$CN(5--4) ($\la 3$), the
temperature errors turned out to be quite high
(of the order of the values themselves). To reduce them, temperatures
in four of the five cores are calculated from
spectra averaged over several nearby points (averaging
ranges are indicated in the 2nd and 3rd columns of
Table~\ref{table:tkin}).

Table~\ref{table:tkin} shows the calculated kinetic temperatures.
The obtained values are $\sim 30-40$~K, the estimates from
CH$_3$C$_2$H(5--4) and CH$_3$CN(5--4) coincide within the
error for G269.11 and G270.26. For G291.27, the temperature
estimate from CH$_3$C$_2$H(5--4) reaches $\sim 50$~K,
while the estimate from CH$_3$CN(5--4) is $\sim 35$~K.
This discrepancy is discussed in Section 7. Thermal linewidths
corresponding to the temperature estimates for
H$^{13}$CO$^+$ are $\sim 0.2-0.3$~km/s for $T_{\rm KIN}=30-50$~K.

\begin{table}[h]
\centering
\caption{Kinetic temperatures}
\vskip 2mm
\small
\begin{tabular}{l|r|r|r|r|}
\noalign{\hrule}\noalign{\smallskip}
Source & $\Delta\alpha ('')$  & $\Delta\delta ('')$
& $T_{\rm KIN}$ (K) & $T_{\rm KIN}$ (K) \\
&&& (CH$_3$C$_2$H) & (CH$_3$CN) \\
 \noalign{\hrule}\noalign{\smallskip}
G~268.42$-$0.85  & 0             & --15         & 32(8)  &        \\
G~269.11$-$1.12  & $0\div 30$    & $15\div 45$  & 33(12) & 44(13) \\
G~270.26$+$0.83  & $-15\div 0$   & $15\div 30$  & 27(15) & 56(34) \\
G~291.27$-$0.71  & $-30\div -15$ & $-45\div-30$ & 51(8)  & 36(7) \\
G~294.97$-$1.73  & $-15\div 15$  & $-30\div 0$  & 48(17) &        \\
\noalign{\hrule}\noalign{\smallskip}
\end{tabular}
\flushleft
{\small
Kinetic temperatures for all the objects except G268.42--0.85 are calculated
from spectra averaged in the indicated
$\Delta\alpha$ and $\Delta\delta$ ranges.}
\label{table:tkin}
\end{table}

\subsection{Column densities and molecular abundances}

Using integrated intensities of optically thin
lines, it is possible to estimate the number of
molecules along the line of sight in the local thermodynamic
equilibrium (LTE) approximation \cite{MangumShirley,Pir16}.
We calculated the H$^{13}$CN,
H$^{13}$CO$^+$, HN$^{13}$C, HC$_3$N, c-C$_3$H$_2$, SiO, CH$_3$C$_2$H, and
CH$_3$CN column densities towards the positions of maximum intensity
in the objects. The excitation temperature is taken to
be 10~K for all lines except CH$_3$C$_2$H(5--4) and
CH$_3$CN(5-4). With this value, the calculated values of $N_{\rm MOL}$
are close to the minimum. If the excitation temperature
is different from 10 K, column densities can be
up to several times higher. For CH$_3$C$_2$H(5--4)
and CH$_3$CN(5--4), estimates of kinetic temperatures
are taken as excitation temperatures (see Section~5.1).
The calculated values of molecular column
densities are given in Table~\ref{table:colden}.
Their analysis indicates that
$N$(H$^{13}$CN), $N$(H$^{13}$CO$^+$) and $N$(HC$_3$N)
in G285.26 and G294.97 are several times lower compared
to the rest. The column densities of c-C$_3$H$_2$ and
SiO do not vary much from object to object.

To estimate molecular abundances
($X=N_{\rm MOL}/N($H$_2)$), we used the values of molecular hydrogen
column densities, calculated from
the 1.2~mm continuum observations \cite{Pir07}
which were obtained with an angular resolution of 24$''$.
Dust temperatures are taken equal to 30 K for all
the objects except G285.26(2) (20 K) and G268.42 (35 K),
which is consistent with the estimates made in \cite{Pir07,Pir22}
for the value of $\beta$, the power-law index of the dependence
of the dust emissivity on frequency, equal to 2.
The resulting column densities of molecular hydrogen
are given in Table~\ref{table:colden}. Their errors are associated mainly
with uncertainty in $\beta$, which affects the estimates
of dust temperatures and for most objects can be $\sim 20$\%
for the temperature variations of $\pm 5$~K (an effect of $\beta$
on dust temperature estimates is discussed, for example, in \cite{Pir22}).
For G285.26(2) at the dust temperature of 15~K,
the calculated column density of hydrogen is higher
by 50\% compared to the column density
calculated for 20~K. The molecular abundance values
are given in Table~\ref{table:colden}.
They lie in the following ranges:
$X$(C$_3$H$_2)\simeq (0.2-1.2)\,10^{-10}$,
$X$(CH$_3$C$_2$H$)\simeq (0.6-1.5)\,10^{-9}$,
$X$(CH$_3$CN$)\simeq (1.2-4.6)\,10^{-11}$,
$X$(H$^{13}$CN$)\simeq (0.8-9.2)\,10^{-11}$,
$X$(H$^{13}$CO$^+)\simeq (0.3-3.5)\,10^{-11}$,
$X$(H$_3$CN$)\simeq (2.3-45.9)\,10^{-11}$,
$X$(HN$^{13}$C$)\simeq (0.6-2.3)\,10^{-11}$,
$X$(SiO)$\simeq (0.7-3.3)\,10^{-11}$.
The lowest abundances of H$^{13}$CN, H$^{13}$CO$^+$ and H$_3$CN
are observed in the G285.26(1) core.

\begin{table}[h]
\centering
\caption{Column densities and molecular abundances}
\vskip 2mm
\scriptsize
\begin{tabular}{l|l|l|l|l|l|l|l|l|}
\noalign{\hrule}\noalign{\smallskip}
Molecule  & $N$, cm$^{-2}$ & $X$ & $N$, cm$^{-2}$ & $X$
          & $N$, cm$^{-2}$ & $X$ & $N$, cm$^{-2}$ & $X$ \\
              & ($\times 10^{12}$) & ($\times 10^{-11}$)  & ($\times 10^{12}$) & ($\times 10^{-11}$)
              & ($\times 10^{12}$) & ($\times 10^{-11}$)  & ($\times 10^{12}$) & ($\times 10^{-11}$) \\
\noalign{\smallskip}\hline\noalign{\smallskip}
  &\multicolumn{2}{|c|}{\bf G268.42($0'',-15''$)}
  &\multicolumn{2}{|c|}{\bf G269.11($15'',30''$)}
  &\multicolumn{2}{|c|}{\bf G270.26($-15'',30''$)}
  &\multicolumn{2}{|c|}{\bf G291.27($-30'',-45''$)} \\
\noalign{\smallskip}\hline\noalign{\smallskip}
H$^{13}$CN     & 6.6  & 2.2  & 8.1  & 9.2  & 6.1 & 5.6 & 8.0 & 5.0 \\
H$^{13}$CO$^+$ & 2.9  & 1.0  & 3.1  & 3.5  & 2.4 & 2.2 & 4.2 & 2.6 \\
SiO            & 2.3  & 0.7  & 1.7  & 1.9  & 2.9 & 2.6 & 5.3 & 3.3 \\
HN$^{13}$C     & 1.7  & 0.6  & 2.1  & 2.3  & 1.4 & 1.3 & 2.6 & 1.7 \\
HC$_3$N        &27.4  & 9.1  &40.0  & 45.9 &20.0 &18.2 &44.5 &27.8 \\
CH$_3$C$_2$H   & 460  & 153  &94.2  & 107  &64.6 &58.7 & 162 &101  \\
CH$_3$CN       & 3.5  & 1.2  & 4.0  & 4.6  & 3.5 & 3.2 & 4.5 & 2.8 \\
c-C$_3$H$_2$   & 6.3  & 2.1  & 8.5  & 9.6  & 5.0 & 4.6 &12.6 & 7.9 \\
H$_2^1$   & 30   &      & 8.8  &      & 11  &     & 16  &     \\
\noalign{\smallskip}\hline\noalign{\smallskip}
  &\multicolumn{2}{|c|}{\bf G285.26($0'',-15''$)}
  &\multicolumn{2}{|c|}{\bf G285.26($-45'',-75''$)}
  &\multicolumn{2}{|c|}{\bf G294.97($15'',-15''$)}
  &\multicolumn{2}{|c|}{\bf G294.97($60'',15''$)}  \\
\noalign{\smallskip}\hline\noalign{\smallskip}
H$^{13}$CN     & 1.8  & 0.8  & 2.3  & 2.9  & 2.4 & 3.6 & 2.3 & 3.6 \\
H$^{13}$CO$^+$ & 0.6  & 0.3  & 1.5  & 1.9  & 1.4 & 2.1 & 1.5 & 2.3 \\
SiO            & 2.1  & 1.0  &      &      & 1.0 & 1.5 &     &     \\
HN$^{13}$C     &      &      &      &      & 1.4 & 2.0 & 0.9 & 1.4 \\
HC$_3$N        & 5.0  & 2.3  &12.6  & 15.8 &21.4 &31.5 & 6.9 &10.6 \\
CH$_3$C$_2$H   &      &      &      &      &67.5 &99.3 &     &     \\
c-C$_3$H$_2$   & 6.3  & 2.9  & 8.0  & 1.0  & 7.9 &11.6 & 5.3 & 8.1 \\
H$_2^1$   & 22   &      & 8.0  &      & 6.8 &     & 6.5 &     \\
\noalign{\smallskip}\hline\noalign{\smallskip}
\end{tabular}

\flushleft
{\scriptsize
$^1$ The H$_2$ column density values should be multiplied by $10^{22}$.}

\label{table:colden}
\end{table}

\subsection{Sizes, velocity dispersions and virial masses
of the cores}

Table~\ref{table:physpar} shows the axes ratio of the fitted ellipses,
the angular and linear dimensions of the emission
regions at half the maximum intensity level, obtained
from the approximation of the maps. The emission
regions in the G270.26 core are, on average, closest to
a spherically symmetrical shape; the emission regions
in G294.97(2) are the most elongated. On average, the
HC$_3$N(10--9) and CH$_3$OH(5$_{-1}$--4$_0$~E) emission
regions are the most compact. Thus, in G270.26,
G285.26(2), G294.97(1) their sizes are $\sim 1.5-2$ times
smaller than the emission regions in other lines (see
Table~\ref{table:physpar}). In G268.42, a compact emission region of
SiO(2--1) with a size of 0.2~pc is observed.

Table~\ref{table:physpar} shows the line widths averaged over
areas within the half-maximum intensity contour. The
line widths (in most cases $\sim 2-3$~km/s) are significantly
higher than thermal ones. The SiO(2--1) lines,
which are indicators of shock waves in the shells
surrounding young stars (e.g., \cite{Harju98}), are the most
broad ($\sim 3-5$~km/s).

The virial masses given in the last column of
Table 6, are determined from the values of the
emission region sizes and velocity dispersions (averaged line widths),
as $M_{\rm VIR}(M_{\odot})=105\,\langle\Delta V\rangle^2\,d$, where
$\Delta V$ and $d$ are taken in km/s and in pc, respectively
(see, for example, \cite{Pir03}). This expression is valid for
spherically symmetric cores with constant density in
the absence of external pressure and magnetic field.
Virial masses of the cores, calculated from the lines of
rare isotopes, the optical depth of which is apparently
less than unity (H$^{13}$CN(1--0), H$^{13}$CO$^+$(1--0),
HN$^{13}$C(1--0)), vary from hundreds of solar masses
(G270.26, G294.97) to $\ga 1000~M_{\odot}$ (G285.26, G291.27).

\begin{table}[p]
\centering
\caption{Physical parameters of the cores}
\vskip 2mm
\scriptsize
\begin{tabular}{l|l|l|l|l|l}
\noalign{\hrule}\noalign{\smallskip}
Line   & Axes & $\Delta\Theta$ & $d$ & $\langle\Delta V\rangle$ & $M_{\rm VIR}$ \\
       &  ratio     & ($''$)         & (pc) & (km s$^{-1}$)  & ($M_{\odot}$) \\
\noalign{\smallskip}\hline\noalign{\smallskip}
\multicolumn{6}{c}{\bf G268.42--0.85} \\
CH$_3$OH      &  1.9(0.2) & 61.3(3.8)  & 0.51(0.04)  & 2.8(0.1) & 420(46)   \\
c-C$_3$H$_2$  &  1.4(0.2) & 103.9(8.2) & 0.86(0.08)  & 3.1(0.2) & 860(140)  \\
H$^{13}$CN    &  1.2(0.1) & 73.8(3.6)  & 0.61(0.05)  & 2.6(0.1) & 430(47)  \\
H$^{13}$CO$^+$&  2.0(0.1) & 67.5(1.8)  & 0.56(0.04)  & 2.0(0.1) & 234(28)  \\
SiO           &  2.1(1.5) & 27(10)     & 0.22(0.08)  & 4.9(0.2) & 560(215)  \\
HN$^{13}$C    &  2.7(0.4) & 98.4(6.9)  & 0.81(0.07)  & 2.1(0.2) & 380(80)  \\
HCN           &  1.12(0.04) & 74.0(1.3) & 0.61(0.04) &          &     \\
HCO$^+$       &  1.04(0.03) & 66.2(1.1) & 0.55(0.03) &          &   \\
HNC           &  1.4(0.1) & 69.7(1.6)  &  0.57(0.04) &          &   \\
HC$_3$N       &  1.4(0.1) & 67.1(1.9)  &  0.55(0.04) & 2.0(0.1) &  230(28) \\
\noalign{\smallskip}\hline\noalign{\smallskip}
\multicolumn{6}{c}{\bf G269.11--1.12} \\
CH$_3$OH      &  2.2(0.3) & 41.5(2.7)  & 0.52(0.07)  & 2.5(0.1) & 340(53)   \\
H$^{13}$CN    &  2.1(0.3) & 51.2(3.5)  & 0.65(0.09)  & 2.7(0.1) & 490(76)  \\
H$^{13}$CO$^+$&  2.8(0.7) & 38.7(4.7)  & 0.49(0.08)  & 2.4(0.1) & 300(55)  \\
SiO           &  1.8(0.5) & 38.0(4.9)  & 0.48(0.08)  & 3.2(0.5) & 515(180)  \\
HN$^{13}$C    &  3.8(1.6) & 32.9(7.1)  & 0.41(0.10)  & 2.4(0.1) & 250(65)  \\
HCN           &  1.7(0.2) & 73.3(3.3)  & 0.92(0.11)  &          &   \\
HCO$^+$       &  2.4(0.2) & 66.1(1.8)  & 0.83(0.10)  &          &   \\
HNC           &  2.0(0.1) & 60.1(1.9)  & 0.76(0.09)  &          &   \\
HC$_3$N       &  2.5(0.2) & 60.0(2.4)  & 0.76(0.09)  & 2.7(0.1) & 580(83)  \\
\noalign{\smallskip}\hline\noalign{\smallskip}
\end{tabular}
\label{table:physpar}
\end{table}

\newpage

\addtocounter{table}{-1}

\begin{table}[p]
\centering
\caption{Physical parameters of the cores (continued)}
\vskip 2mm
\scriptsize
\begin{tabular}{l|l|l|l|l|l}
\noalign{\hrule}\noalign{\smallskip}
Line   &  Axes   & $\Delta\Theta$ & $d$ & $\langle\Delta V\rangle$ & $M_{\rm VIR}$ \\
       &  ratio  & ($''$)         & (pc) & (km s$^{-1}$)  & ($M_{\odot}$) \\
\noalign{\smallskip}\hline\noalign{\smallskip}
\multicolumn{6}{c}{\bf G270.26$+$0.83} \\
CH$_3$OH      &  1.3(0.2) & 27.6(1.9)  & 0.17(0.03)   & 2.7(0.1) & 133(25)  \\
c-C$_3$H$_2$  &  1.3(0.3) & 121(13)    & 0.76(0.14)   & 2.0(0.1) & 320(68)  \\
H$^{13}$CN    &  1.2(0.2) & 40.6(3.4)  & 0.26(0.04)   & 2.8(0.2) & 210(48)  \\
H$^{13}$CO$^+$&  1.1(0.1) & 46.3(2.6)  & 0.29(0.05)   & 2.0(0.1) & 123(24)  \\
SiO           &  1.6(0.5) & 43.0(6.1)  & 0.27(0.06)   & 3.4(0.4) & 330(100) \\
HN$^{13}$C    &  1.8(0.2) & 69.3(4.6)  & 0.44(0.07)   & 3.2(0.2) & 470(100) \\
HCN           &  1.4(0.1) & 64.7(1.5)  & 0.41(0.06)   &          &   \\
HCO$^+$       &  1.4(0.1) & 59.6(1.3)  & 0.38(0.06)   &          &   \\
HNC           &  1.2(0.1) & 60.3(1.4)  & 0.38 0.06    &          &   \\
HC$_3$N       &  1.1(0.2) & 31.5(2.8)  & 0.20(0.04)   & 2.5(0.2) & 130(30) \\
\noalign{\smallskip}\hline\noalign{\smallskip}
\multicolumn{6}{c}{\bf G291.27--0.71} \\
CH$_3$OH      &  1.3(0.3) & 74(9)     & 1.0(0.2)  & 3.0(0.1  & 950(170)  \\
c-C$_3$H$_2$  &  1.8(0.1) & 103(18)   & 1.4(0.3)  & 5.6(0.2) & 4600(1000)   \\
H$^{13}$CN    &  1.9(0.4) & 46.0(4.3) & 0.6(0.1)  & 3.3(0.1) & 710(110)  \\
H$^{13}$CO$^+$&  1.7(0.1) & 76.2(2.5) & 1.0(0.1)  & 3.0(0.1) & 980(130)  \\
SiO           &  1.9(0.2) & 74.9(4.3) & 1.0(0.1)  & 4.9(0.1) & 2600(330)  \\
HN$^{13}$C    &  1.8(0.2) & 78.4(5.0) & 1.1(0.1)  & 2.8(0.1) & 880(130)  \\
HCN           &  1.7(0.2) & 120(6)    & 1.6(0.2)  &          &   \\
HCO$^+$       &  1.7(0.1) & 128(5)    & 1.7(0.2)  &          &   \\
HNC           &  1.9(0.2) & 97.7(4.1) & 1.3(0.2)  &          &   \\
HC$_3$N       &  1.4(0.2) & 70.2(3.8) & 1.0(0.1)  & 2.5(0.1) & 630(90)  \\
\noalign{\smallskip}\hline\noalign{\smallskip}
\end{tabular}
\label{table:physpar}
\end{table}

\newpage

\addtocounter{table}{-1}

\begin{table}[p]
\centering
\caption{Physical parameters of the cores (continued)}
\vskip 2mm
\scriptsize
\begin{tabular}{l|l|l|l|l|l}
\noalign{\hrule}\noalign{\smallskip}
Line   &  Axes   & $\Delta\Theta$ & $d$ & $\langle\Delta V\rangle$ & $M_{\rm VIR}$ \\
       &  ratio  & ($''$)         & (pc) & (km s$^{-1}$)  & ($M_{\odot}$) \\
\noalign{\smallskip}\hline\noalign{\smallskip}
\multicolumn{6}{c}{\bf G285.26--0.05 (1)} \\
H$^{13}$CN    &  2.0(0.6) & 84(13)    & 1.9(0.4)  & 3.6(0.4) & 2600(760)   \\
SiO           &  1.6(0.4) & 41.7(5.6) & 1.0(0.2)  & 4.7(1.0) & 2200(1000)  \\
HCN           &  2.1(0.3) & 50.6(4.2) & 1.2(0.2)  &          &   \\
HCO$^+$       &  2.3(0.5) & 36.6(4.2) & 0.8(0.1)  &          &   \\
\noalign{\smallskip}\hline\noalign{\smallskip}
\multicolumn{6}{c}{\bf G285.26--0.05 (2)} \\
CH$_3$OH      &  2.0(0.3) & 54.3(3.9)  & 1.2(0.2) & 3.4(0.3) & 1500(330)   \\
c-C$_3$H$_2$  &  1.0(0.4) & 68(12)     & 1.5(0.3) & 4.0(0.1) & 2600(550)  \\
HCN           &  1.9(0.7) & 135(24)    & 3.1(0.6) &          &   \\
HCO$^+$       &  1.6(0.3) & 107(9)     & 2.4(0.3) &          &   \\
HNC           &  1.5(0.8) & 120(30)    & 2.7(0.7) &          &   \\
HC$_3$N       &  4.2(1.4) & 36.4(6.2)  & 0.8(0.2) & 2.4(0.1) & 500(110)   \\
\noalign{\smallskip}\hline\noalign{\smallskip}
\multicolumn{6}{c}{\bf G294.97--1.73 (1)} \\
CH$_3$OH      &  1.8(0.3) & 34.5(3.3)  & 0.20(0.05)  & 2.1(0.1) & 90(30)  \\
c-C$_3$H$_2$  &  2.1(0.3) & 75.1(5.4)  & 0.4(0.1)    & 2.5(0.1) & 290(80)   \\
H$^{13}$CN    &  2.0(0.7) & 86(16)     & 0.5(0.2)    & 1.9(0.1) & 190(60) \\
H$^{13}$CO$^+$&  1.1(0.5) & 44(10)     & 0.3(0.1)    & 2.0(0.1) & 110(40)  \\
SiO           &  1.4(1.3) & 84(10)     & 0.5(0.1)    & 3.2(0.1) & 530(150)  \\
HN$^{13}$C    &  1.6(0.3) & 59.3(5.4)  & 0.3(0.1)    & 1.4(0.1) & 70(20)   \\
HCN           &  1.5(0.1) & 90.9(3.2)  &  &          &   \\
HCO$^+$       &  1.5(0.1) & 78.2(3.2)  &  &          &   \\
HNC           &  1.5(0.1) & 65.7(2.7)  &  &          &   \\
HC$_3$N       &  1.6(0.2) & 49.1(2.5)  & 0.3(0.1)    & 1.88(0.04) & 110(30)   \\
\noalign{\smallskip}\hline\noalign{\smallskip}
\multicolumn{6}{c}{\bf G294.97--1.73 (2)} \\
H$^{13}$CO$^+$&  4.0(1.1) & 70(10)    & 0.4(0.1)    & 2.6(0.1) & 290(90)   \\
HCO$^+$       &  3.2(1.2) & 124(24)   &  &          &   \\
HNC           &  3.4(1.2) & 95(16)    &  &          &   \\
HC$_3$N       &  2.7(0.7) & 80(10)    & 0.5(0.1)    & 1.93(0.04) & 180(50)   \\
\noalign{\smallskip}\hline\noalign{\smallskip}
\end{tabular}
\label{table:physpar}
\end{table}


\subsection{Radial velocity dispersion profiles}

Line widths obtained from Gaussian fitting the spectra significantly
exceed thermal widths (see Table~\ref{table:physpar}),
so they can be approximately used as an estimate of the
dispersions of non-thermal velocities in sources.
This parameter is used in theoretical models that include
non-thermal motions of gas in the cores (e.g., \cite{MT03,vazquez}),
in the equation of state of gas in turbulent cores (e.g., \cite{Lai03,Pir09})
as well as for calculating virial masses and
virial parameters used to estimate the stability of cores
and clouds (e.g., \cite{BM92,Camacho20}).

To estimate spatial variations of velocity dispersions,
the dependences of averaged line widths on
impact parameters are calculated. When averaging,
the width values higher than $3\sigma$ are used, where $\sigma$
is the error obtained from the Gaussian fitting
(see Table~\ref{table:parln}). The impact parameter ($b$) is calculated
as a square root of $A/\pi$, where $A$ is the area of the
region that includes all points with intensity above a
given level, the value of which varied from $I_{MAX}$ to $0.05\,I_{MAX}$
with increments of $0.01\,I_{MAX}$. If the difference $b_i-b_{i-1}$
becomes greater than half the mapping step
(7.5$''$), then the average line width and its error
are calculated for this area (the calculation method
is described in more detail in \cite{Caselli02,Pir03}).

\begin{figure}[h]

\begin{minipage}[b]{0.24\textwidth}
    \includegraphics[width=\textwidth]{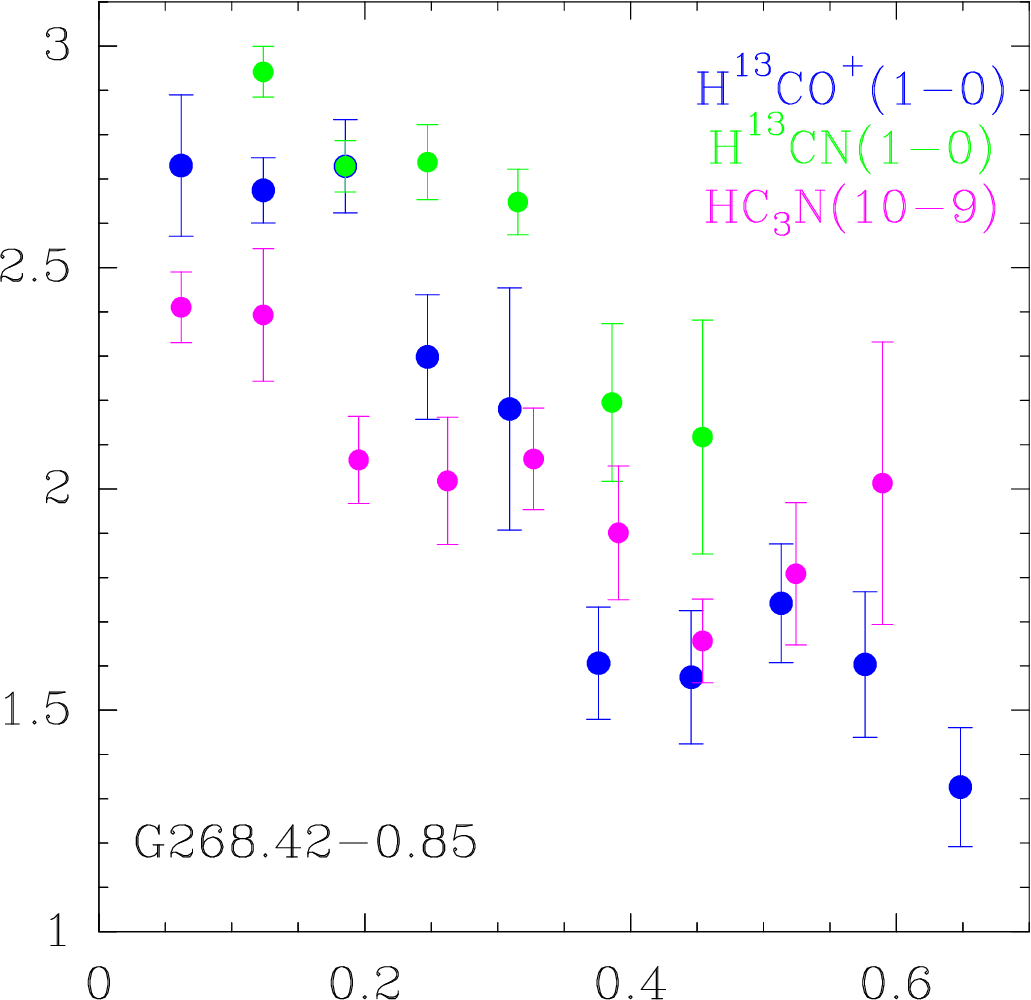}
\end{minipage}
\begin{minipage}[b]{0.24\textwidth}
    \includegraphics[width=\textwidth]{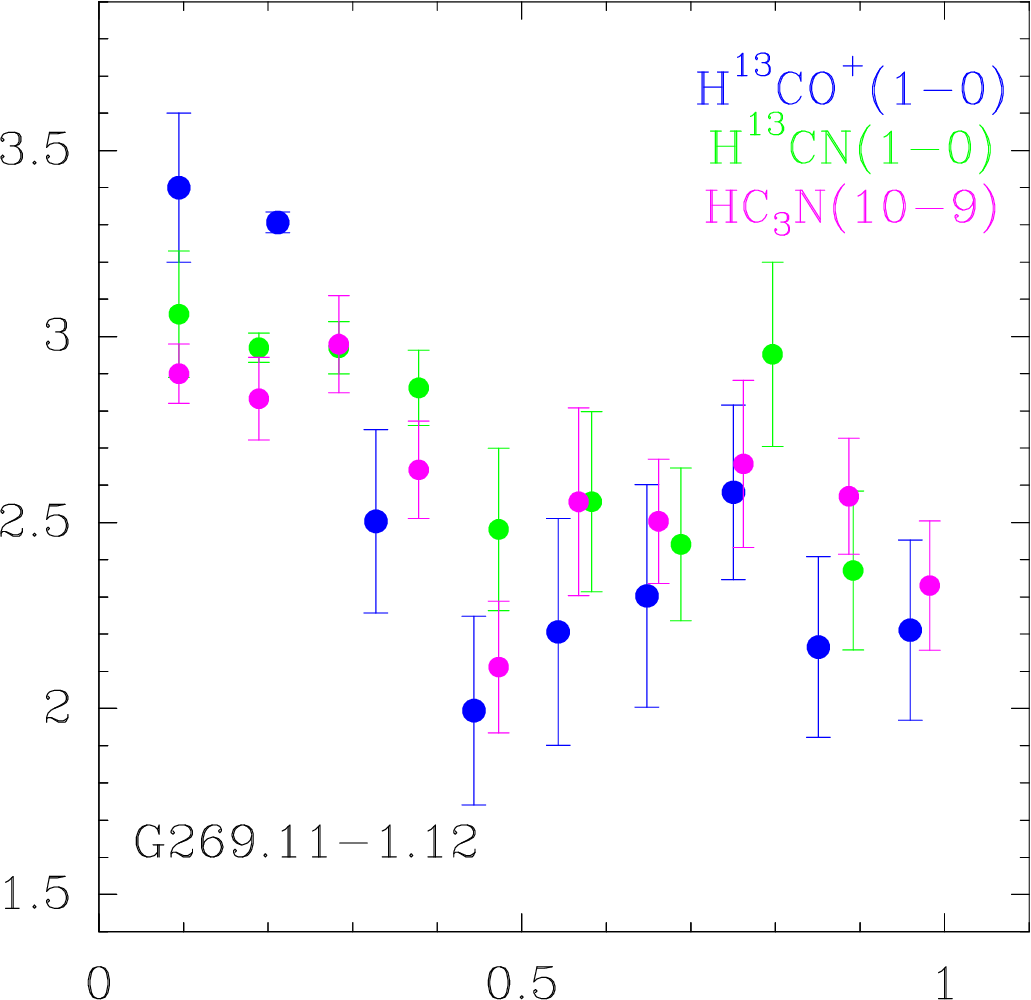}
\end{minipage}
\begin{minipage}[b]{0.24\textwidth}
    \includegraphics[width=\textwidth]{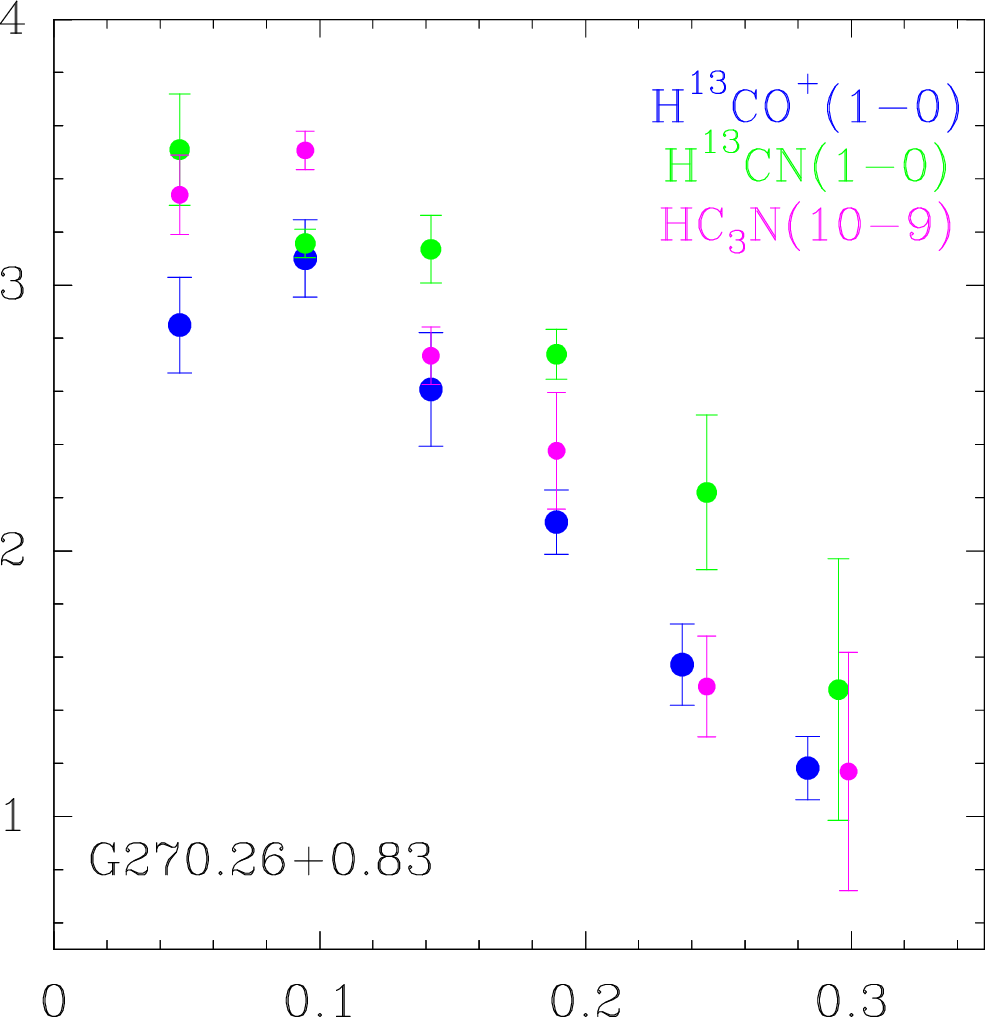}
\end{minipage}
\begin{minipage}[b]{0.24\textwidth}
    \includegraphics[width=\textwidth]{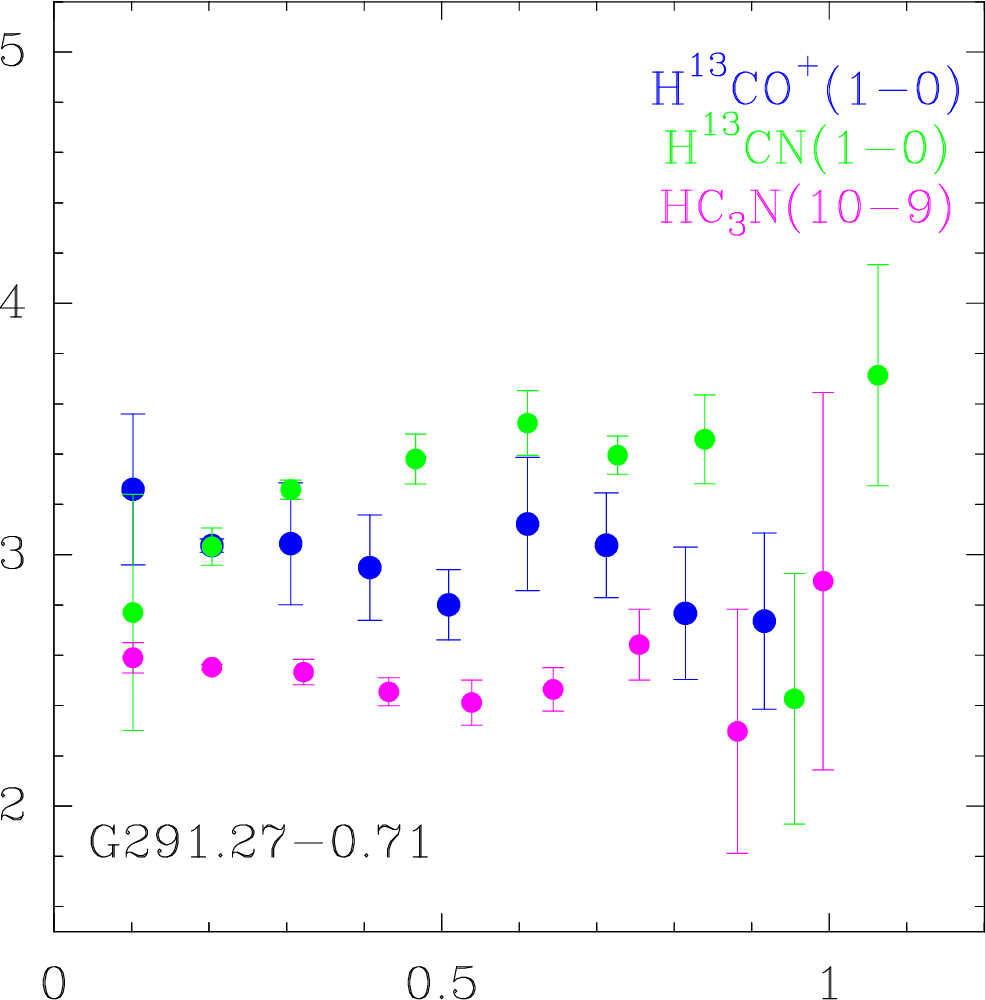}
\end{minipage}

\vskip 2mm

\begin{minipage}[b]{0.24\textwidth}
    \includegraphics[width=\textwidth]{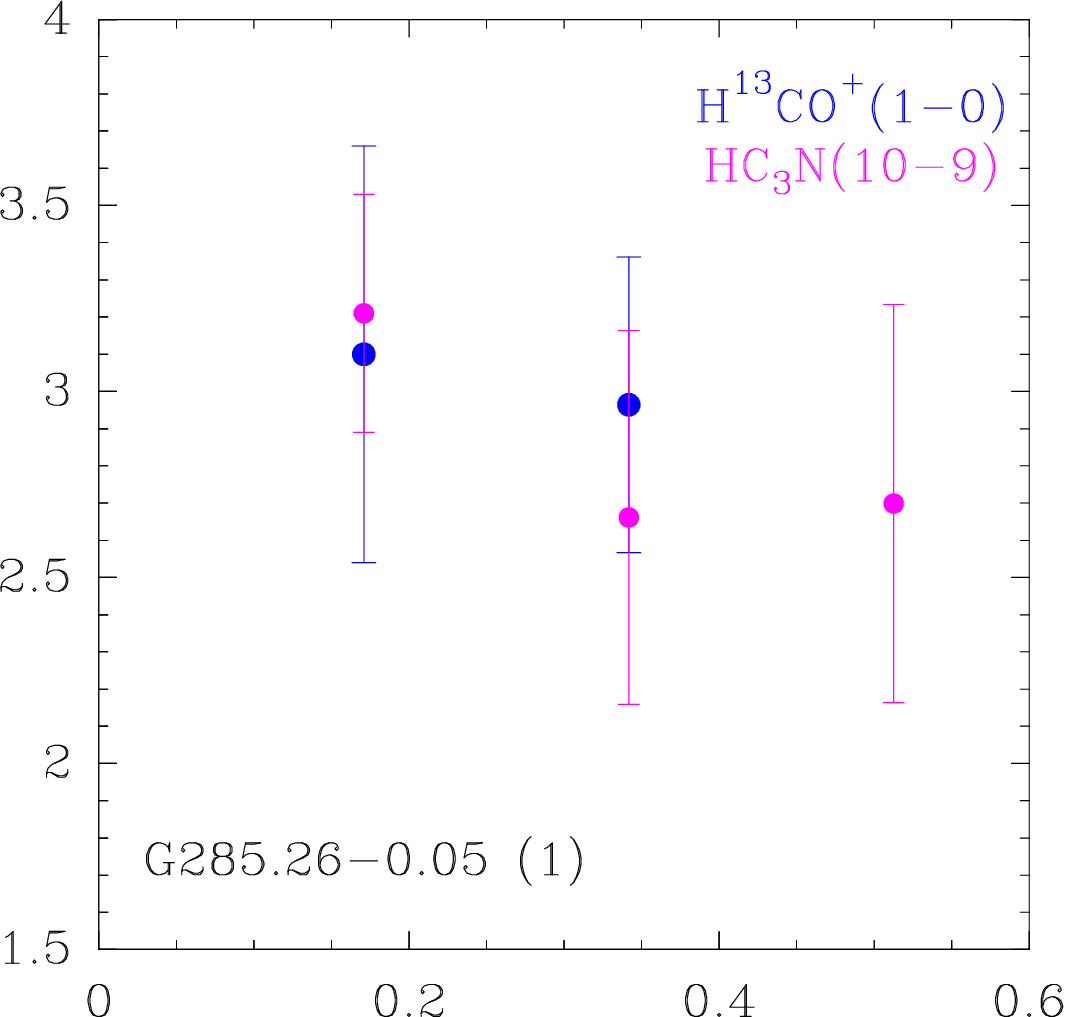}
\end{minipage}
\begin{minipage}[b]{0.24\textwidth}
    \includegraphics[width=\textwidth]{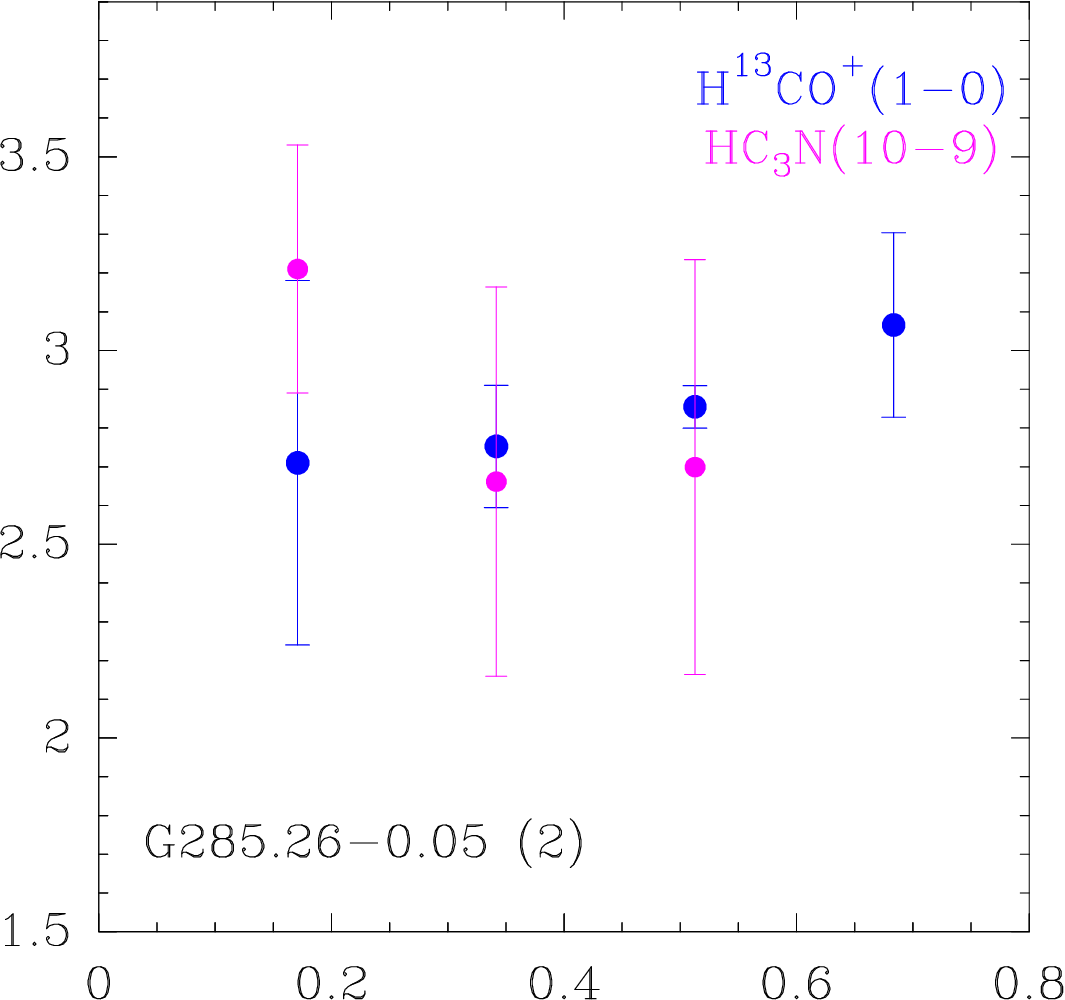}
\end{minipage}
\begin{minipage}[b]{0.24\textwidth}
    \includegraphics[width=\textwidth]{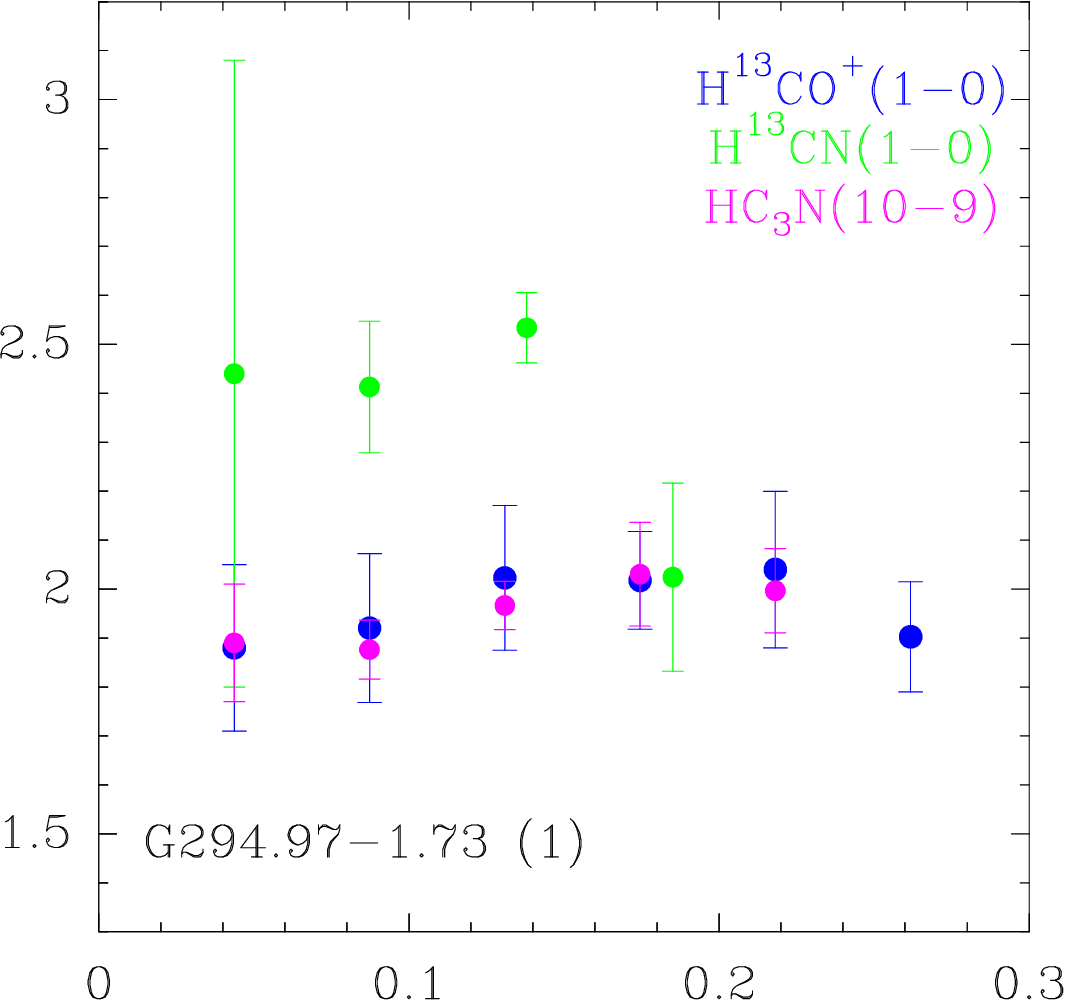}
\end{minipage}
\begin{minipage}[b]{0.24\textwidth}
    \includegraphics[width=\textwidth]{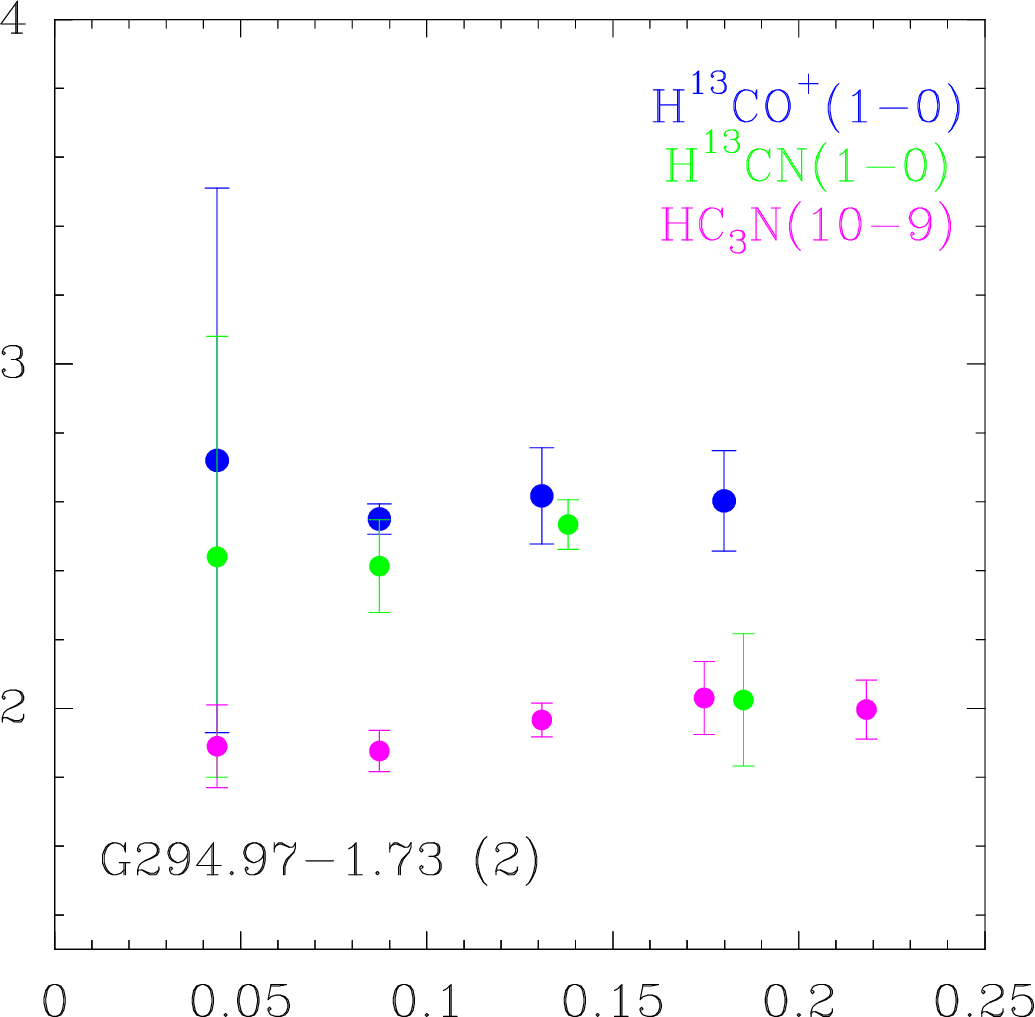}
\end{minipage}

\vskip 20mm

\caption{\scriptsize
Profiles of the widths of optically thin lines of H$^{13}$CO$^+$(1--0),
H$^{13}$CN(1--0) and HC$_3$N(10--9), which are a measure of the
dispersion of gas velocities in objects. The vertical axis shows
the values of average widths in km/s (see text for details).
The horizontal axis shows the impact parameters ($b$) in parsecs.
}
\label{fig:dvprof}
\end{figure}

Figure~\ref{fig:dvprof} shows the dependences of the average
widths of the optically thin H$^{13}$CO$^+$(1--0),
H$^{13}$CN(1--0) lines, as well as the HC$_3$N(10--9) lines.
The emission regions in these lines are spatially correlated
with each other and are relatively extended. The number
of averaged line width values increases with
increasing $b$ from 1 in the center to $\sim 10-15$ at the edge
(for the cores in the G285.26 and G294.97 regions, the
number of averaged values at the edge is $\sim 5-10$).
In three cases (G268.42, G269.11 and G270.26), the line
widths decrease with increasing impact parameter
from the direction to the center to $b\sim 0.3-0.6$~pc. The
difference between the line widths in the center and at
the edges in these cores reaches $\sim 1.5-2$~km/s. No similar
trends are found in G285.26(1, 2), G291.27, and G294.97(1, 2).

The broadening of lines in the center of the cores
may be partly due to an increase in the optical depth.
Indeed, optical depth $\sim 1$ can lead to broadening of
Gaussian profiles by a factor of $\sim 1.2$ due to saturation
(e.g., \cite{Phillips}). However, it is hardly possible to completely
explain the found trends in G268.42, G269.11
and G270.26 by optical depth effects. Note, that the
widths of the H$^{13}$CN(1--0) lines exceed the widths of
other lines in G268.42, G270.26, G291.27 and
G294.97(1), while the widths of the HC$_3$N(10--9)
lines are close to the corresponding widths of
H$^{13}$CO$^+$(1--0), or less than them (G291.27, G294.97(2)).

A possible explanation for the obtained trends is an
increase in the degree of dynamic activity (turbulence,
systematic motions, outflows) in the direction of the
centers of the cores, associated with star formation
processes. In the G268.42--0.85 and
G269.11--1.12 cores, the main factors causing line broadening
can be both turbulence and systematic motions of
the contraction type, the existence of which is indicated
by a specific type of asymmetry in the profiles of
optically thick lines (see Fig.~\ref{spectra}).

\section{Model estimates of parameters of radial density and velocity
profiles in cores}

An absorption dip is observed in the profiles of the optically
thick HCO$^+$(1--0) and HCN(1--0) lines in the G268.42--0.85
and G269.11--1.12 cores (Fig.~\ref{spectra}). Its position is close to
the peak position of H$^{13}$CO$^+$(1--0) and other lines with
a small optical depth. The red and blue wings of the
lines, separated by the dip, have different amplitudes,
with the blue peak exceeding the red one over significant
part of the observed maps. The presence of asymmetry
of this type for optically thick lines and symmetrical,
close to Gaussian, profiles of optically thin lines,
the maximum of which is close to the position of the
dip in optically thick lines, indicates a probable contraction
of the core \cite{Evans}. Comparison of such spectral
maps with maps calculated within non-LTE models
can provide information about the distribution of
physical parameters, including the contraction velocity.
This information can be used to select one or
another theoretical model. The difficulty of finding
optimal values when varying several parameters simultaneously
is due to the fact that the error function may
have more than one local minimum, and the parameters
themselves may correlate with each other, which
leads to dependence on initial conditions and poor
convergence. To overcome these difficulties, an algorithm
was proposed in \cite{PZ21} based on the use of statistical
methods of principal components (PC) and
k-nearest neighbors (kNN) for fitting model spectral
maps into observed ones.

The HCO$^+$(1--0) spectra with a dip and asymmetry
were observed in the central part of the
G268.42--0.85 core and in the northern part of the
G269.11--1.12 core. To estimate radial profiles of
physical parameters, a multilayer spherically symmetric
model is used, the parameters of which (density,
turbulent, systematic velocity and kinetic temperature)
depend on the radial distance according to the
law: $P=P_0/(1+(r/R_0)^{\alpha_p})$, where $R_0$ is the radius of
the central layer, taken equal to $2\times 10^{16}$~cm. The
description of the model is given in \cite{PZ21}.
The morphology of molecular emission regions, as well
as the structure of inner regions of the cores
according to the 350~$\mu$m data,
shows differences from spherical symmetry
(see Figs.~\ref{fig:G268},~\ref{fig:G269}).
The use of a spherically symmetric model
in this case can provide information about the averaged
radial profiles of physical parameters in the outer
regions of the cores.

The model parameters are the values of $P_0$ for
density, turbulent and systematic velocity radial profiles
($n_0, V_{turb}, V_{sys}$, respectively), the corresponding
power-law indices $\alpha_p$ ($\alpha_n, \alpha_{turb}, \alpha_{sys}$),
as well as the relative abundances of HCO$^+$ and H$^{13}$CO$^+$,
considered to be independent of the radial distance,
the outer radius of the model cloud ($R_{max}$) and the source
Doppler velocity ($V_{LSR}$). Note, that in accordance with
the given type of radial profiles, the values of physical
parameters in the central layer are equal to half of
the corresponding $n_0$, $V_{turb}$ and $V_{sys}$ values.
The kinetic temperature profile has the form
$T=80\,K/(1+(r/R_0)^{0.3})$ and is not changed during
the calculations. We used the values of
the HCO$^+$--H$_2$ collision probabilities from \cite{Flower}, calculated
for fixed temperatures with increments of 10~K.
Accordingly, the temperature in each layer is
rounded to a multiple of 10~K. Thus, the kinetic temperature
in the model is equal to 40~K in central layers
(which is close to the estimates of kinetic temperatures
in G268.42 and G269.11, see Table~\ref{table:tkin}) and dropped
to $\sim 10$~K at the periphery. For the calculations, 14 layers
are taken, the width of which increased according
to a power law with increasing distance from the
center. After calculating model spectra, convolution
with the telescope radiation pattern is carried
out and the error function is calculated, depending
on the difference between the model and observed
spectra at different points. To reduce information loss
when calculating the error function, maps with a step
of 6$''$ in both coordinates are used, obtained as a
result of processing the data with the $Gridzilla$ package.

The methodology for applying the algorithm to
find the global minimum of the error function corresponded
to that given in \cite{PZ21} with some additions
described in the Appendix. As part of the algorithm, a
set of precedents is generated and probable model
parameters and their confidence regions are estimated.
The PC method is used to reduce the model
dimension (from 10 to 6) and uniformly fill the parameter
space. The kNN method is used to find optimal
parameter values corresponding to the global minimum
of the error function determined by the difference
between the model and observed spectra (for more details, see \cite{PZ21}).

\begin{figure}[h]

\begin{minipage}[b]{0.48\textwidth}
    \includegraphics[width=\textwidth]{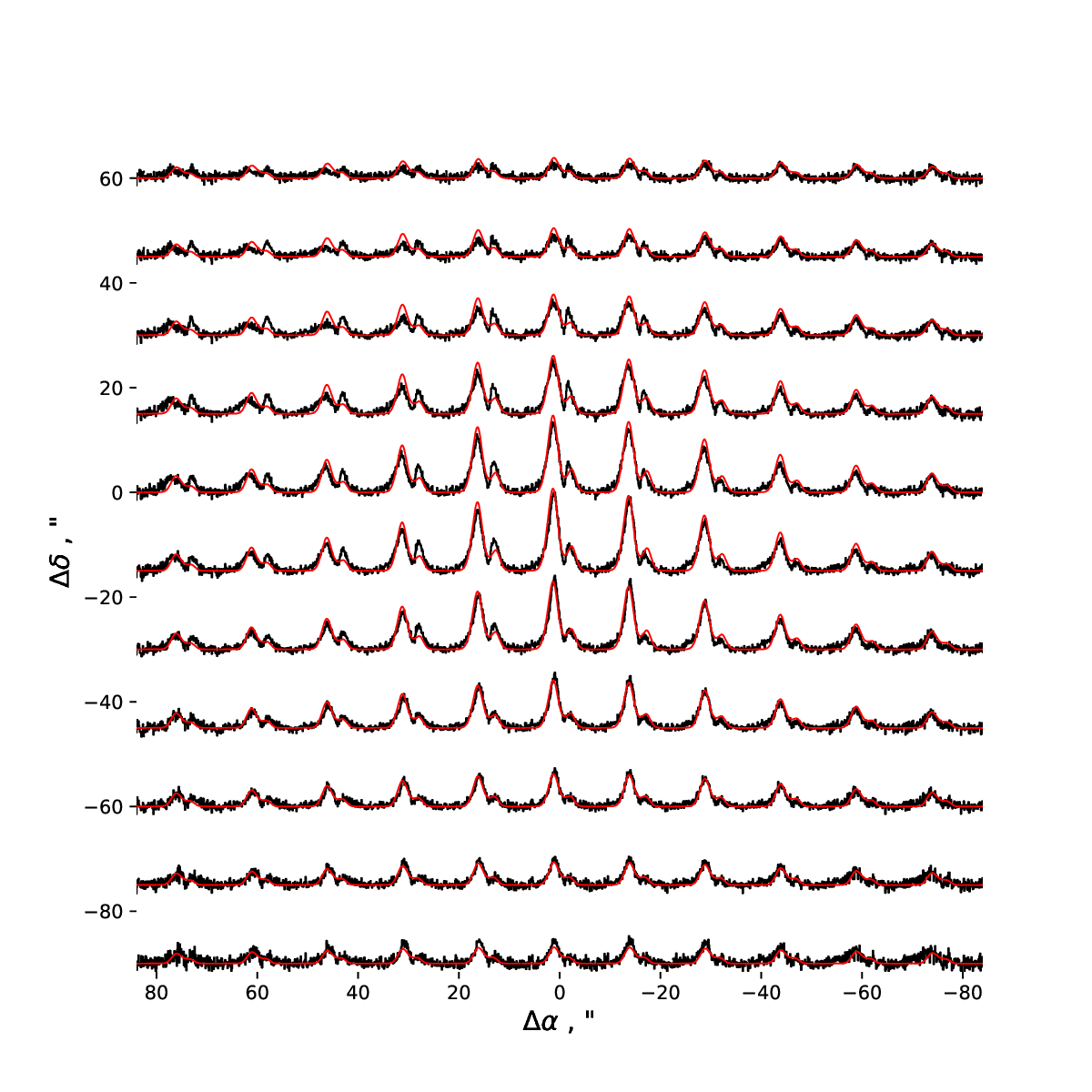}
\end{minipage}
\begin{minipage}[b]{0.48\textwidth}
    \includegraphics[width=\textwidth]{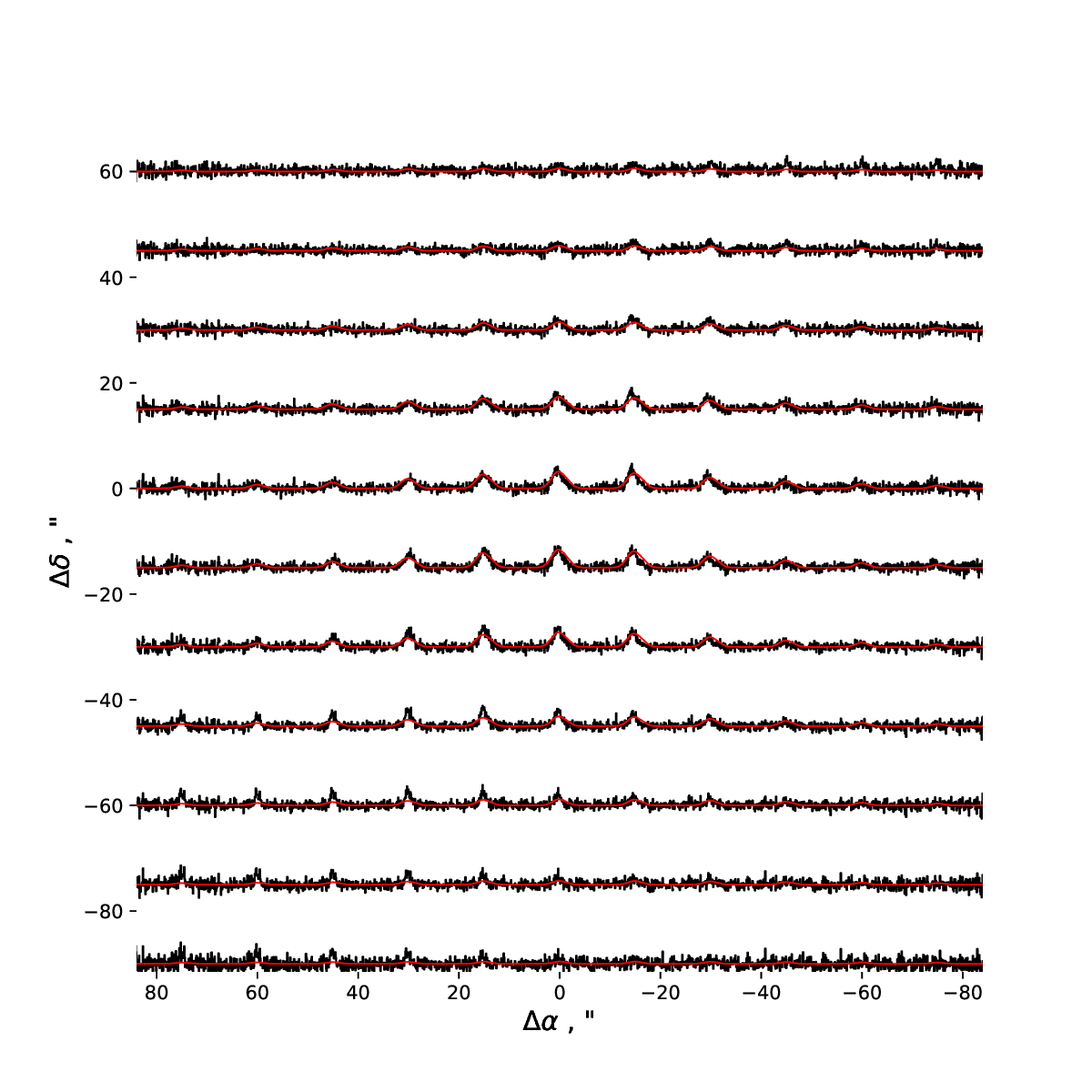}
\end{minipage}



\vskip 20mm

\caption{\scriptsize
Maps in the HCO$^+$(1--0) (left) and H$^{13}$CO$^+$(1--0) (right) lines in the G268.42--0.85 core. The results of observations are
shown in black. In red we show the results of model calculations corresponding to the obtained estimates of the physical parameters
from Table~\ref{table:modelpar}.
}
\label{fig:model}
\end{figure}

Model HCO$^+$(1--0) and H$^{13}$CO$^+$(1--0) line maps
are fitted into the central region of G268.42--0.42
with a size of $\sim 1.5'\times 1.5'$, which corresponds to
linear size $\sim 0.7$~pc, as well as in the map of the
northern part of G269.11--1.12 with a size of $\sim 1'\times 1.3'$
or $\sim 0.7\times 1$pc. Since, within the framework of the
model used, we are unable to achieve a satisfactory
fit of the model HCO$^+$(1--0) maps into those
observed in G269.11--1.12 (see Section~\ref{sec:discussion}),
below we present the results of calculations only for the
G268.42--0.42 core. Model and observed HCO$^+$(1--0) and
H$^{13}$CO$^+$(1--0) line maps
for this core are shown in Fig.~\ref{fig:model}.
The values of the physical parameters for G268.42--0.42, corresponding
to the minimum of the error function, as well as
uncertainties of these estimates, corresponding to the
boundaries of the confidence ranges obtained from
the analysis of projections of the multidimensional
error function on the plane of various pairs of parameters
(Fig.~\ref{fig:corner}), are given in Table~\ref{table:modelpar}.

\begin{table}[h]
\centering
\caption{Model values of physical parameters of the G268.42--0.85 core}
\vskip 2mm
\begin{tabular}{l|l|l|l}
\noalign{\hrule}\noalign{\smallskip}
Parameter & Initial range & Value & \\
\noalign{\hrule}\noalign{\smallskip}
$n_0$(cm$^{-3}$), 10$^7$      &$0.6\div 10$  & 5.5$_{-4.3}^{+8.0}$  \\
$\alpha_n$                    &$1.3\div 2.0$ & 1.6$_{-0.2}^{+0.2}$  \\
V$_{turb}$ (km/s)             &$3\div 9$     & 6.7$_{-2.3}^{+1.4}$  \\
$\alpha_{turb}$               &$-0.2\div 2.1$& 0.2$_{-0.1}^{+0.1}$  \\
V$_{sys}$(km/s)               &$-1.5\div-0.1$&--1.2$_{-0.5}^{+0.2}$ \\
$\alpha_{sys}$                &$-0.6\div 0.6$&--0.5$_{-0.2}^{+0.3}$ \\
R$_{max}$(pc)                 &$0.2\div 4$   & 4.6$_{-3.2}^{+2.1}$  \\
X(HCO$^+$), 10$^{-10}$        &$0.2\div 20$  & 1.5$_{-0.5}^{+3.3}$  \\
X(H$^{13}$CO$^+$), 10$^{-11}$ &$0.2\div 10$   & 1.0$_{-0.2}^{+0.3}$ \\
\noalign{\smallskip}\hline\noalign{\smallskip}
\end{tabular}
\flushleft
{\small
$n_0$, V$_{turb}$ and V$_{sys}$ represent double values
of the parameters in the central layer}
\label{table:modelpar}
\end{table}

\section{Discussion}
\label{sec:discussion}

The studied regions contain dense cores, associated with the
regions of formation of massive stars and stars in clusters,
as indicated by the presence of high-luminosity
IRAS sources, compact H~II regions, methanol and
water masers, near-infrared sources, and other sources
(see Section~\ref{sec:sources}). From observations, it follows that the
studied regions have rich molecular composition
(see Table~\ref{table:colden}), which is typical for the cores associated
with regions of formation of high-mass stars or stars in clusters.
Core 1 from the G285.26--0.05 region is poorer in
chemical composition than the others.

The calculated kinetic temperatures (see Table~\ref{table:tkin})
demonstrate both closeness and discrepancies with
independent estimates. Thus, for G268.42, the calculated
kinetic temperature is close to the estimate from
\cite{Araya05} (33~K), obtained from observations of methyl cyanide
lines in the $J=6-5$ transition, as well as to the
dust temperature (35~K \cite{Pir07}).
The kinetic temperature estimate in G270.26 from methyl acetylene,
taking into account the error, practically coincides
with the dust temperature estimate (29~K \cite{Pir07}),
but is slightly lower than the peak CO(1--0) line  temperature
(44~K \cite{Zin95}), which in the case of
large optical depth can serve as a measure of kinetic
temperature. The kinetic temperature estimate in
G294.97, taking into account the error, exceeds the
estimate of the dust temperature in the direction of
the IRAS source (27~K \cite{Pir07}) and is close to the value
of the peak temperature of CO(1--0) (41 K \cite{Zin95}). The
difference in kinetic temperatures in G291.27, calculated
using methyl acetylene and methyl cyanide lines,
is beyond the error limits (see Table~\ref{table:tkin}). Observations
in ammonia lines with an angular resolution of $\sim 11''$
\cite{Purcell09} showed that the kinetic temperature in
the core can vary from $\sim 30$~K to $\sim 50$~K in the region
where we calculated the average kinetic temperatures.
It is possible that the spatial distributions of emission in
the methyl acetylene and methyl cyanide lines on
scales smaller than the size of the MOPRA-22m beam
in a given region differ from each other, which leads to
differences in temperature estimates. New observations
in these lines with better angular resolution (for
example, $\sim 10''$) could be useful in answering the question
of the reason for the discrepancy in the estimates
obtained.

Virial masses calculated from different lines
(Table~\ref{table:physpar}) differ from each other, often exceeding the
error limits ($\sim 10-50$\%), which may be due to differences
in the sizes of emission regions, line widths,
excitation conditions for various spices, as well as variations
in the chemical composition in the cores. For
the cores of G268.42, G269.11, and G291.27, the
masses calculated from dust emission data \cite{Pir22} lie in
the range of virial estimates. For the remaining cores
(especially for G294.97), the virial masses exceed the
masses calculated from dust. It should be noted that
the estimates of $M_{\rm VIR}$ may be overestimated. An increase
in line widths in the inner regions of the
G268.42, G269.11 and G270.26 cores ($b\la 0.3-0.6$~pc)
(Fig.~\ref{fig:dvprof}) is apparently associated with the influence of
internal sources, which cause an increase in the dispersion
of non-thermal velocities and increase the
average values of $\Delta V$ and $M_{\rm VIR}$.
If the cores are inhomogeneous,
then virial masses will be lower. Thus, for
the density profile $r^{-2}$, the virial masses will be 40\%
less than the values given in Table~\ref{table:physpar}.

The cores differ in their internal kinematics, as
indicated by different profiles of the optically thick
lines HCO$^+$(1--0) and HCN(1--0) (Fig.~\ref{spectra}), as well as
the different shapes of the radial profiles of non-thermal
velocity dispersion (Fig.~\ref{fig:dvprof}). In two cores
(G268.42--0.85, G269.11--1.12), an asymmetry of the
HCO$^+$(1--0) profiles indicates probable gas contraction.
Using a detailed analysis of the
HCO$^+$(1--0) and H$^{13}$CO$^+$(1--0) line maps with a spherically
symmetric model, the parameters of the radial dependences
of density, turbulent velocity dispersion,
and contraction velocity in the G268.42--0.85 core
are estimated (Table~\ref{table:modelpar}). The results of
model calculations of HCO$^+$(1--0) and H$^{13}$CO$^+$(1--0)
generally describe the observed spectra in G268.42--85
quite well, although there are some discrepancies between model and observed
HCO$^+$(1--0) line profiles in the northeastern part of the
core (see Fig.~\ref{fig:model}). For G269.11-1.12, the model and
observed spectra are consistent only in the center of
the core, while in the northeastern and southwestern
parts the discrepancies in both the peak temperature
values and the form of line profiles turned out to be very
significant. These discrepancies may be associated
with the difference in the shape of the cores from
spherical symmetry, as well as with an influence of
rotation, which can lead to differences in the form of
the HCO$^+$(1--0) line profiles from those expected in the
case of radial contraction. The G269.11-1.12 core is
more elongated (compare the ratios of the axes of the
fitted ellipses for G268.42 and G269.11, Table~\ref{table:physpar}).
In addition, model estimates of physical parameters in
G269.11-1.12 differ from independent estimates.
It is possible that the model used is too simplified to
describe this core.

For the G268.42--0.85 core, the calculated model
parameters (see Table~\ref{table:modelpar}) are consistent with independent
estimates, taking into account uncertainties
of the compared quantities. Note, that the obtained
outer radius value ($\sim 4$~pc) represents the boundary of
the model cloud and cannot be directly compared with
the observed sizes of emission regions in molecular
lines (see Table~\ref{table:physpar}). In addition, an uncertainty of this
value, calculated from the confidence range, reaches
$\sim 70$\%( see Table~\ref{table:modelpar}).
Due to a sharp decrease in density,
the outer layers of the model cloud with low density
($< 10^4$~cm$^{-3}$ for $r>1$~pc) affect primarily the magnitude
of the dip in the profiles of optically thick lines,
absorbing radiation from the central layers, and practically
do not contribute when estimating the sizes of
emission regions at half the maximum level.
Taking into account large uncertainty in the density
value in the central layer ($\sim 100$\%, Table~\ref{table:modelpar}),
the model column densities of molecular hydrogen averaged
for the region of $\sim 24''$ ($\sim 0.2$~pc) and the masses
for the HCO$^+$ and H$^{13}$CO$^+$ emission regions ($\sim 0.6$~pc,
Table~\ref{table:modelpar}) are easy to reconcile with the corresponding
values $N$(H$_2$) and $M_{\rm VIR}$ from Tables~\ref{table:colden},~\ref{table:physpar}.
A power-law index of the radial density profile obtained from the
model coincides with the value obtained from the
1.2~mm observations \cite{Pir09}. The estimate of the
power-law index of the radial density profile for G269.11-1.12,
obtained from model calculations, is significantly
lower than the index calculated from observations of
dust emission (1.2 and 2 \cite{Pir09}, respectively).

The ratio of model abundances [HCO$^+$]/[H$^{13}$CO$^+$]
in G268.42--0.85 is lower than the range of probable
values of the [$^{12}$C]/[$^{13}$C] ratio ($\sim 35-55$), calculated
from the dependence from \cite{Yan19} for a galactocentric
distance $R_G\sim 8.7$~kpc. The uncertainty in the
model abundance ratio is, however, rather high (Table~\ref{table:modelpar}).
The model estimate of $X$(H$^{13}$CO$^+$) for G268.42--0.85
coincides with the estimate obtained from LTE calculations
(Table~\ref{table:physpar}). For G269.11-1.12 these estimates differ.

Turbulent velocity decreases with radial distance
from $\sim 3.3$~km/s in the central layer to 1.5~km/s
in the outer layer. Similar dependence was obtained
previously for the L1287 core \cite{PZ21}. The contraction
velocity in the central layer of G268.42--0.85 is
$\sim -0.6$~km/s, which is close to the estimate of the two-layer
model \cite{Myers}, obtained from the parameters of the
observed HCO$^+$(1--0) line profile for the point
($0'',-15''$). The model power-law index of the radial profile
of the contraction velocity, taking into account
uncertainties, turned out to be less than zero
($\alpha_{sys}\sim -0.5^{+0.3}_{-0.2}$), which for a given type of model
dependence means an increase in the contraction
velocity in absolute value with increasing distance
from the center.

This result differs both from the case of gas collapse
onto a protostar in the free fall mode ($\alpha_{sys}=0.5$) \cite{Shu,MT03,vazquez},
and from the case of a non-equilibrium
core in the mode of global hierarchical collapse with a
constant velocity ($\alpha_{sys}=0$) \cite{naranjo,vazquez}.
However, the fact that for a part of the HCO$^+$(1--0) map in G268.42--0.85
we are unable to reproduce completely the form of
the line profiles (Fig.~\ref{fig:model}) indicates the likelihood
of the existence of a more complex spatial distribution of systematic
velocities in the core compared to the radial contraction.
Taking into account the core rotation, indications
of which were reported in \cite{Pir03}, could possibly be
useful for model calculations in this case. Contribution
of rotation to the overall systematic velocity
along the line of sight will obviously be different for
different parts of the core, which should change the
ratio of the intensities of the ``blue” and ``red” peaks of
the profiles and, possibly, more accurately reproduce
the features of the observed spectra in different parts of
the map and correct the parameters of the radial contraction
velocity profile.

Thus, an analysis of the results of
model calculations showed that the used spherically
symmetric model with single power-law indices of the
physical parameter radial profiles is simplified
for the cores under consideration. Nevertheless, with
the help of this model, using the developed algorithm
for fitting model spectral maps into observed ones
\cite{PZ21}, we are able to reproduce the
HCO$^+$(1--0) and H$^{13}$CO$^+$(1--0) line maps observed
in the G268.42--0.85 core and to estimate radial profiles of
physical parameters. As in the L1287 core \cite{PZ21}, the
calculated value of $\alpha_{sys}$ differs from that expected in
the case of free fall of gas onto a protostar \cite{Shu,MT03}.
In this case, however, the found value does not agree with the
prediction of the global collapse model \cite{naranjo,vazquez}.
To verify the obtained result, it is necessary to carry out
observations of this core with better sensitivity and
higher spectral resolution both in 1--0
and in higher transitions of the rotational spectrum of
HCO$^+$ and H$^{13}$CO$^+$ molecules, as well as to use for analysis
the models that take into account both radial
motions and gas rotation, as well as the difference from
spherical symmetry. To obtain statistically significant
conclusions about the nature of gas contraction in the
cores associated with the regions of high-mass stars
and star clusters formation, it is important to further expand
the number of objects of analysis, where profiles of
molecular lines, indicators of high density (HCO$^+$,
HCN, CS, etc.), have indications of contraction.

\section{Conclusions}

Using the 22-m radio telescope of the MOPRA
observatory (Australia), spectral observations of six
regions of formation of massive stars in the southern
sky were carried out in the $\sim 84-92$~GHz frequency
range in the lines
CH$_3$OH(5$_{-1}$--4$_0$~E), c-C$_3$H$_2$(2$_{1,2}$--1$_{0,1}$), CH$_3$C$_2$H(5--4),
H$^{13}$CN(1--0), H$^{13}$CO$^+$(1--0), SiO(2--1), HN$^{13}$C(1--0),
HCN(1--0), HCO$^+$(1--0), HNC(1--0), HC$_3$N(10--9) and CH$_3$CN(5--4).
These regions have dense cores and are associated with bright
IRAS point sources, compact H~II regions, maser
sources and near-infrared sources, and have been
observed previously in continuum and molecular
lines. The following results are obtained:

1. The physical parameters of cores are determined,
including kinetic temperatures ($\sim 30-50$~K),
sizes of emission regions in various lines ($0.2-3.1$~pc)
and virial masses ($\sim 70-4600~M_{\odot}$). The line widths
significantly exceed the thermal widths ($\sim 2-5$~km/s).
Within the framework of LTE, the column densities as
well as the abundances of the
H$^{13}$CN, H$^{13}$CO$^+$, HN$^{13}$C, HC$_3$N, c-C$_3$H$_2$, SiO,
CH$_3$C$_2$H and CH$_3$CN molecules are calculated. The lowest abundances of
H$^{13}$CN, H$^{13}$CO$^+$ and H$_3$CN are observed in the
G285.26(1) core. The widths of various molecular
lines, the optical depth of two of which
is apparently small, increase in three cores with decreasing impact
parameter, indicating the influence of the internal
source on the dynamic activity of the gas. The difference
between the line widths in the center and at the
periphery in these objects reaches $\sim 1.5-2$~km/s.

2. An asymmetry of the
optically thick HCO$^+$(1--0) and HCN(1--0) line profiles
indicates the probable contraction of gas in the
G268.42--0.85, G269.11--1.12 cores. Within the framework
of a spherically symmetric model, using an algorithm
for minimizing the error function when fitting
model maps into observed ones \cite{PZ21}, the optimal values
of the parameters of density,
turbulent velocity and contraction velocity radial profiles
in the G268.42--0.85 core are calculated and their uncertainties
are determined. Density decreases with distance
from the center as $r^{-1.6}$, turbulent velocity
decreases as $r^{-0.2}$, and contraction velocity
increases as $r^{0.5}$. The radial profile of the contraction
velocity differs from that expected both in the case of
free fall of gas onto a protostar ($r^{-0.5}$), and in the
case of global core collapse (contraction velocity does
not depend on distance). A possible reason for this
discrepancy could be rotation of the G268.42--0.85 core,
which is not taken into account in the model.

\newpage

{\flushright\bf\large
Appendix
}

{\flushright\large
APPLICATION OF AN ALGORITHM
FOR FINDING THE GLOBAL MINIMUM
OF THE ERROR FUNCTION
AND ESTIMATING CONFIDENCE
INTERVALS OF MODEL PARAMETERS
}

In \cite{PZ21}, to analyze the maps of molecular lines in
the L1287 core, the excitation of molecules was simulated
in a spherically symmetric model and model
spectral maps corresponding to the observed ones
were calculated. The optimal values of the model
parameters and their confidence ranges were calculated
using an original algorithm for finding the global
minimum of the error function between the model and
observed spectra at each point, including the PC
method used for the statistical procedure of reducing
the model dimension, and the kNN method for finding
the optimal values.

This approach is applied to analyze the maps
G268.42--0.85 and G269.11--1.12 in the HCO$^+$(1--0),
H$^{13}$CO$^+$(1--0) lines. Compared to the approach
described in \cite{PZ21}, the generation of model precedents
before the application of the dimensionality reduction
procedure is changed: when generating the sample,
precedents are randomly excluded in accordance
with the criterion:

\begin{equation}
	(\sum_{k=1}^b r_k/\chi^2)/b < \alpha ,
\label{eq:puge}
\end{equation}

\noindent{where}

\begin{equation}
		r_k=\sum_{i=1}^{n}\sqrt{(p^{new}_i-p_i)^2/\sigma_i^2}
\end{equation}

\noindent{is the normalized distance to previously generated
model precedents of the parameter $p_i$ (density, systematic
and turbulent velocities).} Normalization is
carried out to the standard deviation $\sigma_i$ of the
parameter $p_i$ for the previously generated data set. Here $n$
is the number of parameters, $p^{new}_i$ is the value of the
parameter $p_i$ to which criterion (\ref{eq:puge}) is applied. When
calculating using formula (\ref{eq:puge}), $b$ precedents are
used, selected according to the criterion of the minimum
error function. The value $p^{new}_i$ is randomly
selected within the given ranges. The coefficients $b$
and $\alpha$ are chosen empirically and are equal to 7 and
0.1, respectively. The precedents for which relation
(\ref{eq:puge}) is satisfied are discarded. For the rest, spectral
maps are calculated, the error function is calculated,
and the precedent is added to the total sample.
Thus, we avoided calculating model parameters if
a precedent with a larger value of the error function
was previously calculated nearby.

The sample is produced in several generations
(1000 implementations), some of which are discarded
according to the criterion (\ref{eq:puge}), except for the first
generation. Next, the sample is randomly divided
into two parts. For each, the transformation kernel of
the nonlinear principal component method (with
exponential nonlinearity) is calculated. If the conversion
coefficients calculated for the full sample and
for the parts did not differ by more than 10\%, it is
accepted that the dimensionality reduction procedure
could be applied. Then, similarly to \cite{PZ21}, precedents
are calculated at the nodes of a regular grid for a
space of reduced dimensionality. Generations are
produced until doubling the number of precedents led
to changes in estimates of the probable values of the
parameters and their confidence ranges. Although
these changes have little effect on filling the parameter
space with precedents after applying dimensionality
reduction procedures, they lead to an acceleration of
the process of accumulating precedents before the
reduction procedure. This fact is important, since with
the accumulation of statistics, the percentage of discarded
precedents is close to $\sim 95$\% in the original version
of the algorithm and $\sim 70$\% in the modified version,
which makes it possible to speed up preliminary
calculations by $\sim 6$ times. This is achieved due to the
fact that parameters for which $k$ nearest implementations
have a larger error function value can be
discarded without carrying out model calculations.
Although this modification is not critical when analyzing
individual sources, it is necessary when analyzing
multiple objects, as it speeds up calculations without
affecting the confidence of the final estimates.

Based on the modified algorithm, parameters
of the G268.42--0.85 and G269.11--1.12 cores are
estimated. A priori assumptions about initial
ranges of parameter values and the obtained most
probable values with the confidence ranges at the $1\sigma$
level for G268.42--0.85 are given in Table~\ref{table:modelpar}.
The initial ranges of values, however, did not strictly define the
boundaries of the areas filled by model precedents.
When calculating the principal
components, these areas could be automatically
expanded (as in the case of $R_{max}$). The corresponding
spectral maps are shown in Fig.~\ref{fig:model}.

\begin{figure}[h]


    \includegraphics[width=1.\textwidth]{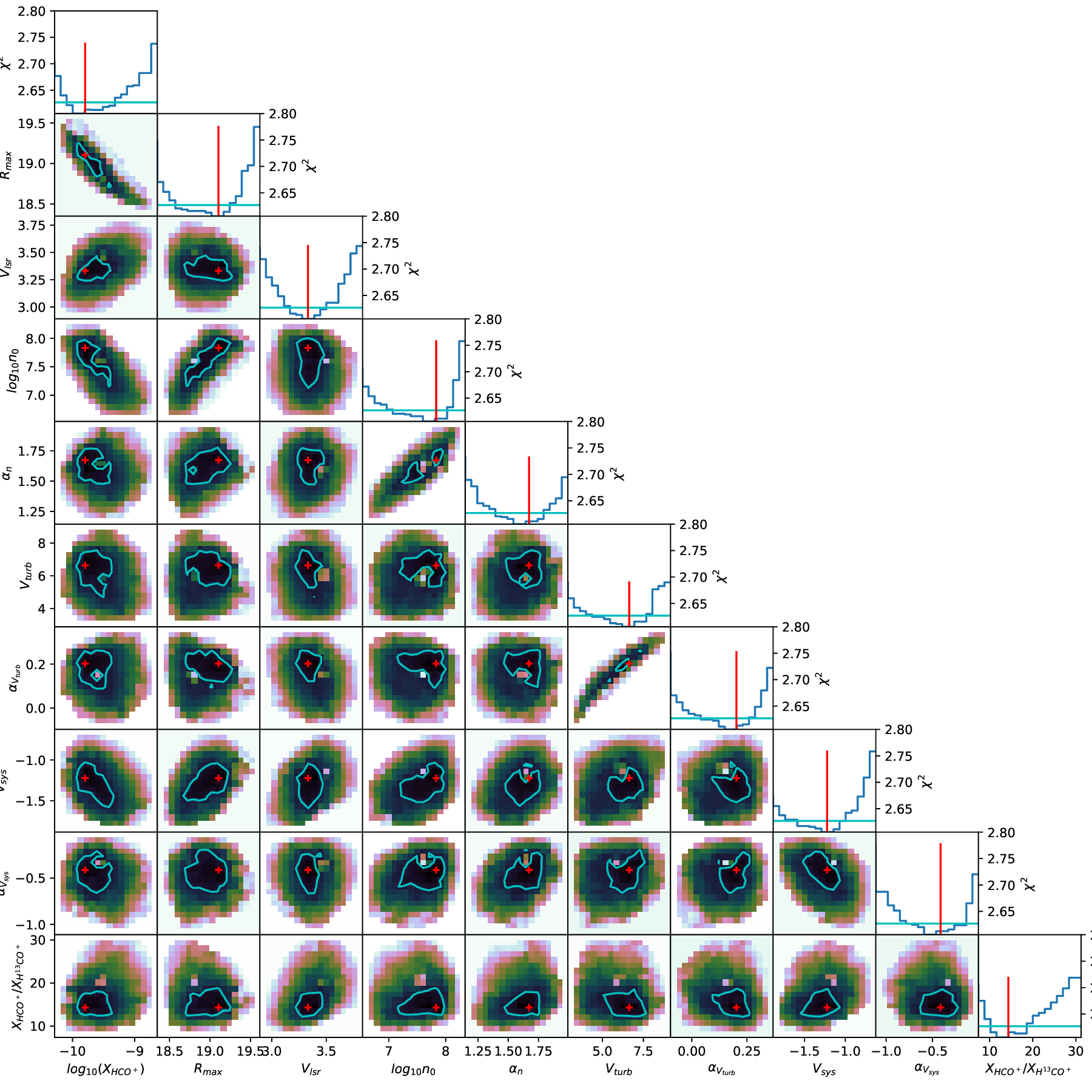}

\vskip 25mm

\caption{\scriptsize
Projections of the 10-dimensional error function ($\chi^2$)
onto the plane of various pairs of parameters, calculated from
fitting model spectral maps in the HCO$^+$(1--0) and H$^{13}$CO$^+$(1--0)
lines into the observed maps in the G268.42--0.85 core.
Above each column there are graphs of the error function depending
on an individual parameter. The red dots in the diagrams
and the red vertical lines in the top plots correspond to the global minimum
of the error function obtained from the kNN method.
Confidence regions for optimal parameter values, calculated from the $\chi^2_{\sigma}$
hyperplane section of the error function, are
shown as blue contours and horizontal lines in the two-dimensional projections
and one-dimensional plots, respectively.
}
\label{fig:corner}
\end{figure}

Figure~\ref{fig:corner} presents sets of two-dimensional projections
of a 10-dimensional error function between the
model and observed spectral maps of HCO$^+$(1--0) in
G268.42--0.85 on the plane of various pairs of parameters.
This figure (at the top of each column) shows
graphs of the projection of the error function
on each of the parameters. In the diagrams, there is
a global minimum of the error function.
For each parameter the area where $\chi^2$ is less than the confidence
threshold is marked. Several parameters
correlate with each other. A negative correlation is
observed between $R_{max}$ and $X_{\rm HCO^+}$. Positive correlations
are observed between $\alpha_n$ and $n_0$, as well as
between the turbulent velocity in the central layer and
$\alpha_{turb}$. Weaker correlations exist between $n_0$ and $R_{max}$
(positive) and $n_0$ and $X_{\rm HCO^+}$ (negative). The most
probable parameter values are estimated using the
kNN method. They are marked by red crosses on
two-dimensional projections and by red vertical lines on
graphs of the $\chi^2$  projections on individual
parameters. Confidence regions are calculated using
the $\chi^2_{\sigma}$ hyperplane cross-section of the error function.
The projections of the sections of the error function by the $\chi^2_{\sigma}$
hyperplane are contours on the planes of pairs of
parameters; they correspond to horizontal lines on the
graphs (Fig.~\ref{fig:corner}).

\section{Acknowledgements}

The authors express their gratitude to the reviewer for
valuable comments and instructions, which significantly improved
the text of the article.
The work is supported by the Russian Science Foundation
grant no. 23-22-00139.

{}

\end{document}